\documentclass[fleqn]{2022AMSE}
\usepackage{bm}
\usepackage{xcolor}
\usepackage{times}
\usepackage{algorithmic}
\usepackage{algorithm}
\usepackage{amssymb}
\usepackage{amsmath}
\usepackage{amsthm}
\usepackage{mathrsfs}
\DeclareMathAlphabet{\mathcal}{OMS}{cmsy}{m}{n}
\DeclareSymbolFont{largesymbols}{OMX}{cmex}{m}{n}
\usepackage{etoolbox}
\usepackage{subfig}
\usepackage{graphicx}
\usepackage{dcolumn}
\usepackage{bm}

\begin{document}

\ensubject{Fluid Dynamics}

\ArticleType{RESEARCH PAPER}

\title{Integrating Fourier Neural Operator with Diffusion Model for Autoregressive Predictions of Three-dimensional Turbulence}

\author[1,2,3]{Yuchi Jiang}{}%
\author[1,2,3]{Yunpeng Wang}{}
\author[1,2,3]{Huiyu Yang}{}%
\author[1,2,3,]{Jianchun Wang}{wangjc@sustech.edu.cn}



\address[1]{Department of Mechanics and Aerospace Engineering, Southern University of Science and Technology, Shenzhen 518055, China}
\address[2]{Shenzhen Key Laboratory of Complex Aerospace Flows, Southern University of Science and Technology, Shenzhen 518055, China}
\address[3]{Guangdong Provincial Key Laboratory of Turbulence Research and Applications, Southern University of Science and Technology, Shenzhen 518055, China}

\contributions{Executive Editor:}

\abstract{Accurately autoregressive prediction of three-dimensional (3D) turbulence has been one of the most challenging problems for machine learning approaches. Diffusion models have demonstrated high accuracy in predicting two-dimensional (2D) turbulence, but their applications in 3D turbulence are relatively limited. To achieve reliable autoregressive predictions of 3D turbulence, we propose the DiAFNO model which integrates the implicit adaptive Fourier neural operator (IAFNO) with diffusion model. IAFNO can effectively capture the global frequency and structural features, which is crucial for global consistent reconstructions of the denoising process in diffusion models. Furthermore, based on conditional generation from diffusion models, we design an autoregressive framework in DiAFNO to achieve long-term stable predictions of 3D turbulence. \textcolor{red}{The proposed DiAFNO model is systematically trained and tested separately with fixed hyperparameters in several types of 3D turbulence, including forced homogeneous isotropic turbulence (HIT) at Taylor Reynolds number $Re_{\lambda}\approx100$, decaying HIT at initial Taylor Reynolds number at $Re_{\lambda}\approx100$ and turbulent channel flow at friction Reynolds numbers $Re_{\tau}\approx395$ and $Re_{\tau}\approx590$ with case-specific training at each Reynolds number. The results in the \textit{a posteriori} tests demonstrate that DiAFNO exhibits a significantly higher prediction accuracy in most of the analyzed statistics (such as the velocity spectra, the root-mean-square (RMS) values of both velocity and vorticity, and Reynolds stresses), as compared to the elucidated diffusion model (EDM) and the traditional large-eddy simulation (LES) using dynamic Smagorinsky model (DSM). Although DiAFNO is not optimal in certain statistics, its overall performance is substantially better than all baseline models (EDM and DSM). Meanwhile, we record the time taken by machine learning models and DSM during the inference stage. Ignoring training costs, the well-trained DiAFNO achieves higher inference efficiency than EDM and LES with DSM.}}




\setlength{\textheight}{23.6cm}
\thispagestyle{empty}

\maketitle
\setlength{\parindent}{1em}

\vspace{-1mm}
\begin{multicols}{2}

\section{Introduction}
The modeling and prediction of turbulence has long been at the forefront of research across various fields \cite{pope2000turbulent}. With advancements in computer science and numerical method, computational fluid dynamics (CFD) has emerged as a vital tool for studying turbulence. However, due to very large scale range of turbulence at high Reynolds numbers, direct numerical simulation \Authorfootnote (DNS) becomes challenging and even impractical \cite{moin1998direct,ishihara2009study}. Therefore, Reynolds-averaged Navier–Stokes (RANS) and large-eddy simulation (LES) methods have been developed to achieve higher efficiency by sacrificing accuracy within certain flow scales on coarser grids, and have thus gained widespread industrial applications \cite{meneveau2000scale,motegi2025reynolds,songyue2025prediction}.

With the explosive development of artificial intelligence in recent years, various machine learning methods have been proposed for turbulent flow prediction tasks \cite{duraisamy2019turbulence,brunton2020machine,brunton2021applying}. Physics-informed neural networks (PINNs) was developed by Raissi et al., which incorporates physical laws as soft constraints into the neural network to directly predict the solutions of nonlinear partial differential equations \cite{raissi2019physics}. Based on the PINN framework, Jin et al. later proposed the Navier-Stokes flow nets (NSFnets) to simulate turbulent channel flow at friction Reynolds number $Re_{\tau}=999$ \cite{jin2021nsfnets}. Lu et al. proposed the deep operator network (DeepONet), which comprises a branch network dedicated to encoding input functions and a trunk network responsible for capturing spatial location information \cite{lu2021learning}. In 2020, Li et al. proposed the Fourier neural operator (FNO) via parameterization of the integral kernel in Fourier space, which enables the reconstruction of information in infinite-dimensional spaces \cite{li2020fourier}. Zhao et al. embedded LES equations into the FNO framework and proposed the LESnets, which can be trained without using label data and maintain the efficiency of data-driven neural operators \cite{zhao2025lesnets}. Wen et al. developed the U-FNO model by incorporating the U-Net structure to FNO and achieved high accuracy in solving multiphase flow problems \cite{wen2022u}. You et al. proposed an implicit Fourier neural operator (IFNO), which improves upon FNO by adopting an implicit iterative method, enabling stable training when networks grow deeper \cite{you2022learning}. Inspired by the above mentioned FNO-based models, Li et al. developed an implicit U-Net enhanced FNO (IUFNO) for the long-term prediction of 3D turbulence including homogeneous isotropic turbulence and turbulent mixing layer \cite{li2023long}. Wang et al. further applied the IUFNO network to predict 3D turbulent channel flows across different friction Reynolds numbers \cite{wang2024prediction}. The adaptive Fourier neural operator (AFNO) was developed by Guibas et al. as an efficient token mixer that learns to mix in the Fourier domain \cite{guibas2021adaptive}. Jiang et al. proposed an implicit AFNO (IAFNO) for fast and accurate long-term predictions of 3D turbulence \cite{jiang2026animplicit}.

Since Vaswani et al. introduced the attention-based transformer model in 2017, it has demonstrated outstanding performance across numerous domains \cite{vaswani2017attention}. Subsequently, transformers were applied to data-driven solutions for various types of partial differential equations including turbulence \cite{chattopadhyay2020deep,li2022transformer,dang2022tnt,yousif2023transformer,li2024transformer}. In 2021, Cao et al. introduced the concept of a learnable Petrov-Galerkin projection and proposed the Galerkin transformer to learn partial differential equations (PDEs) \cite{cao2021choose}. Li et al. developed a novel transformer-based convolutional neural network (TransCNN) method to effectively model the inverse energy cascade in two dimensional (2D) turbulence \cite{li2025transformer}. Hu et al. proposed the physics-informed transformer (PI-Transformer) by incorporating the Navier-Stokes and continuity equations into the loss function and achieved a significant reduction in prediction errors with enhanced robustness at high Reynolds numbers \cite{hu2025physics}. However, transformer typically requires substantial memory, making direct application to high-dimensional PDEs challenging. Li et al. introduced the factorized transformer (FactFormer) based on axial decomposition kernel integration, enabling efficient surrogate modeling \cite{li2024scalable}. Furthermore, Yang et al. introduced the implicit factorized transformer (IFactFormer) 3D turbulence, enabling stable deep training through implicit iterations \cite{yang2024implicit}. They further enhanced the model with a parallel decomposed attention mechanism (IFactFormer-m), demonstrating high accuracy in long-term prediction of 3D turbulent channel flows \cite{yang2026implicit}.

Recently, diffusion models have attracted widespread attention as a new type of generative model capable of achieving high accuracy in various complex tasks. In 2020, Ho et al. proposed the denoising diffusion probability model (DDPM), which progressively degrades signals with Gaussian noise before sequentially denoising them to recover signals from the same distribution \cite{ho2020denoising}. Song et al. improved training and sampling efficiency by accurately estimating the score (a noisy data distribution gradient field dependent on denoising time) via neural networks, and employing stochastic differential equation (SDE) solvers for sampling \cite{song2020score}. Karras et al. systematically analyzed various diffusion models within a unified framework, identifying key factors in model training and design \cite{karras2022elucidating}. Their proposed elucidated diffusion model (EDM) introduces a noise-weighted loss function and an improved sampling strategy based on preconditioning networks which successfully reduces the number of sampling steps \cite{karras2022elucidating}. Liu et al. proposed DiffFNO in 2025, a diffusion framework enhanced with weighted Fourier neural operator (WFNO) and attention-based neural operator (AttnNO) \cite{liu2025difffno}. Experimental validation demonstrated that the WFNO can effectively capture critical frequency essential for diffusion models to reconstruct high-frequency regions during image denoising \cite{liu2025difffno}. With the assistance of AttnNO, the DiffFNO framework was able to capture both global structures and local details of images \cite{liu2025difffno}.


The diffusion models have important applications in turbulence prediction. Its applications include but are not limited to: super-resolution of flow fields \cite{fan2025neural,sardar2024spectrally,guo2025physics,wang2025fundiff}, reconstructing complete flow fields from sparse data \cite{shu2023physics,li2024learning,li2024synthetic,li2025stochastic,li2026deterministic,du2024conditional}, generating realistic turbulent samples of high-quality \cite{lienen2023zero,whittaker2024turbulence,gao2024bayesian}, and high-precision temporal prediction of turbulent flows \cite{kohl2023benchmarking,tahmasebi2025using,sambamurthy2025lazydiffusionmitigatingspectral,mucke2025physics,liu2025confild,gao2024generative,oommen2025integrating}. Fan et al. employed a Bayesian conditional diffusion model to construct small-scale turbulence based on coarser 2D spatiotemporal large-scale turbulence\cite{fan2025neural}. Shu et al. proposed a physically constrained diffusion model, which can reconstruct high-precision data from regular low-precision or sparse samples of 2D turbulence \cite{shu2023physics}. Li et al. proposed a sparse-sensor-assisted score-based generative model ($\textrm{S}^3\textrm{GM}$), which employed diffusion models for reconstruction of spatiotemporal dynamics of 2D flow fields from sparse information \cite{li2024learning}. Li et al. presented a stochastic method that uses diffusion models to reconstruct missing velocity trajectory of objects passively affected by turbulence \cite{li2024synthetic,li2025stochastic,li2026deterministic}. Du et al. integrated conditional neural field encoding with latent diffusion processes and proposed the conditional neural field latent diffusion (CoNFiLD) model, which enabled a memory-efficient and robust generation of turbulence under diverse conditions in 3D domains \cite{du2024conditional,liu2025confild}. Lienen et al. proposed an approach that directly learns the manifold of all possible turbulent flow states without relying on any initial flow state \cite{lienen2023zero}. Gao et al. proposed generative learning of effective dynamics (G-LED), which embeds Bayesian diffusion models that map low-dimensional manifold onto its corresponding high-dimensional space to achieve efficient forecasts of turbulence \cite{gao2024generative}. Oommen et al. applied EDM to neural operators and successfully enhanced the autoregressive forecasting ability of various 2D turbulent flows \cite{oommen2025integrating}. However, much of the work is based on 2D turbulence, while the application of diffusion models in 3D turbulence remains relatively limited \cite{lienen2023zero,du2024conditional}.





To facilitate the application of diffusion models in the long-term continuous prediction of 3D turbulent flows, we propose the DiAFNO model, which integrates IAFNO with diffusion model based on stochastic sampling. The IAFNO model has demonstrated its ability to effectively capture global frequency and structural features in 3D turbulence \cite{jiang2026animplicit,guibas2021adaptive}, we consequently integrate it with a diffusion model to achieve more precise and continuous temporal predictions of 3D turbulence. Our proposed DiAFNO achieves a more accurate autoregressive long-term prediction of various turbulence with higher computational efficiency compared to the state-of-the-art (SOTA) EDM and traditional dynamic Smagorinsky model (DSM).


The rest of the paper is organized as follows. In Section \ref{method}, governing equations of the large-eddy simulation, and the architecture of EDM and DiAFNO are presented. We then present the results of DiAFNO, EDM and DSM for forced homogeneous isotropic turbulence (HIT) at Taylor Reynolds number $Re_{\lambda}\approx100$, decaying HIT at initial Taylor Reynolds number at $Re_{\lambda}\approx100$ and turbulent channel flow at friction Reynolds numbers $Re_{\tau}\approx395$ and $Re_{\tau}\approx590$ in Section \ref{numerical}. Moreover, in Section \ref{numerical}, we also compare the computational cost of DiAFNO with EDM and DSM. In Section \ref{conclusion}, conclusions are drawn.


\section{Methodology}
\label{method}
\subsection{Governing equations}

The governing equations of the 3D incompressible turbulence are given by \cite{pope2000turbulent,ishihara2009study}:

\begin{equation}
\frac{\partial u_i}{\partial x_i}=0 ~, \label{eq:NS1}
\end{equation}
\begin{equation}
\frac{\partial u_i}{\partial t}+\frac{\partial u_iu_j}{\partial x_j}=-\frac{\partial p}{\partial x_i}+\nu\frac{\partial^2 u_i}{\partial x_j\partial x_j}+\mathcal{F}_i ~, \label{eq:NS2}
\end{equation} where $u_i$ denotes the $i$th component of velocity, $p$ is the pressure divided by the constant density, $\nu$ represents the kinematic viscosity, and $\mathcal{F}_i$ stands for a large-scale forcing to the  momentum of the fluid in the $i$th coordinate direction. Throughout this paper, the summation convention is used unless otherwise specified.


In contrast to DNS, LES only resolves the dominant energy-containing large-scale motions on a coarse grid and approximates the effects of subgrid-scale (SGS) motions by SGS models \cite{smagorinsky1963general,deardorff1970numerical,germano1992turbulence}. A filtering approach is employed to decompose turbulent physical variables into distinct large-scale and subgrid-scale components, formally defined as \cite{lesieur1996new,meneveau2000scale}:

\begin{equation}
\bar{f}(\textbf{x})=\int_{D}f(\textbf{x}-\textbf{r})G(\textbf{r};\Delta)\mathrm{d}\textbf{r} ~, \label{eq:filter}
\end{equation} where $f$ represents any physical quantity of interest in physical space, and $D$ is the entire domain. $G$ and $\Delta$ are the filter kernel and filter width, respectively. As shown in Eq.~\ref{eq:filter}, filtering is essentially a convolution operator, hence in Fourier space a filtered quantity is given by $\bar{f}(\textbf{k})=\hat{G}(\textbf{k})f(\textbf{k})$, where $\hat{G}$ is the Fourier transform of $G$: $\hat{G}(\textbf{k})=\int_{-\infty}^{\infty}G(\textbf{x})e^{-i\textbf{k}\textbf{x}}\mathrm{d}\textbf{x}$. In the present study, a sharp spectral filter $\hat{G}(\textbf{k})=H(k_c-\textbf{k})$ is utilized in Fourier space \cite{pope2000turbulent}, where the Heaviside function $H(x)=1$ if $x\geq0$; otherwise $H(x)=0$. Here, the cutoff wavenumber $k_c=\pi/\Delta$.

Applying filtering to Eqs.~\ref{eq:NS1} and \ref{eq:NS2} yields:

\begin{equation}
\frac{\partial \bar{u}_i}{\partial x_i}=0 ~, \label{eq:fNS1}
\end{equation}
\begin{equation}
\frac{\partial \bar{u}_i}{\partial t}+\frac{\partial \bar{u}_i\bar{u}_j}{\partial x_j}=-\frac{\partial \bar{p}}{\partial x_i}-\frac{\partial \tau_{ij}}{\partial x_j}+\nu\frac{\partial^2 \bar{u}_i}{\partial x_j\partial x_j}+\bar{\mathcal{F}}_i ~, \label{eq:fNS2}
\end{equation} where the unclosed SGS stress $\tau_{ij}$ is defined by:

\begin{equation}
\tau_{ij}=\overline{u_iu_j}-\bar{u}_i\bar{u}_j ~, \label{eq:sgstress}
\end{equation} and it represents the nonlinear effects of SGS dynamics on the resolved flow structures.

To accurately solve the LES equations, the SGS stress must be modeled based on the resolved variables. One of the most widely adopted SGS models is the Smagorinsky model, formally given by \cite{smagorinsky1963general}:

\begin{equation}
\tau_{ij}^A=\tau_{ij}-\frac{\delta_{ij}}{3}\tau_{kk}=-2C^2_{\mathrm{Smag}}\bar{\Delta}^2|\bar{S}|\bar{S}_{ij} ~, \label{eq:sgstressA}
\end{equation} where $\bar{S}_{ij}$ is the filtered strain rate, and $|\bar{S}|=\sqrt{2\bar{S}_{ij}\bar{S}_{ij}}$ is the characteristic filtered strain rate. The classical value for the Smagorinsky coefficient is $C_{\mathrm{Smag}}=0.16$, which can be determined through theoretical arguments for isotropic turbulence \cite{pope2000turbulent,smagorinsky1963general}.

The Smagorinsky model exhibits an excessive dissipation in the non-turbulent regime and requires attenuation in the near-wall region \cite{wang2024prediction}. To address this limitation, the DSM has been proposed \cite{germano1992turbulence}. In the DSM, the Smagorinsky coefficient is determined appropriately via the Germano identity \cite{meneveau1999dynamic,lilly1992proposed,pope2000turbulent,germano1992turbulence}. Using a least-squares method, the coefficient $C^2_{\mathrm{Smag}}$ is computed as:

\begin{equation}
C^2_{\mathrm{Smag}}=\frac{\langle L_{ij}M_{ij}\rangle}{\langle M_{kl}M_{kl}\rangle} ~, \label{eq:Csmag}
\end{equation} where $L_{ij}=\widetilde{\bar{u}_i\bar{u}_j}-\tilde{\bar{u}}_i\tilde{\bar{u}}_j$, $\alpha_{ij}=-2\bar{\Delta}^2|\bar{S}|\bar{S}_{ij}$, $\beta_{ij}=-2\tilde{\bar{\Delta}}^2|\tilde{\bar{S}}|\tilde{\bar{S}}_{ij}$ and 
$M_{ij}=\beta_{ij}-\tilde{\alpha}_{ij}$. Here the over-bar denotes the filtering at scale $\bar{\Delta}$, and a tilde denotes a coarser filtering ($\tilde{\Delta}=2\bar{\Delta}$) \cite{wang2024prediction}.



For isotropic turbulence, the Kolmogorov length scale $\eta$, the Taylor length scale $\lambda$, and the Taylor-scale Reynolds number $Re_{\lambda}$ are defined, respectively, as \cite{pope2000turbulent,wang2022constant}:


\begin{equation}
\eta=\left ( \frac{\nu^3}{\epsilon}\right )^{\frac{1}{4}},~~\lambda=\sqrt{\frac{5\nu}{\epsilon}}u^{\textrm{rms}},~~Re_{\lambda}=\frac{u^{\textrm{rms}}\lambda}{\sqrt{3}\nu} ~, \label{eq:variousPhysq}
\end{equation} where $u^{\textrm{rms}}$ is root-mean-square value of velocity magnitude, $\epsilon=2\nu\langle S_{ij}S_{ij} \rangle$ denotes the average kinetic energy dissipation rate and $S_{ij}=(\partial u_i/\partial x_j+\partial u_j/\partial x_i)/2$ represents the strain rate tensor. Furthermore, the integral length scale $L_I$ and the large-eddy turnover time $\tau$ are respectively given by \cite{pope2000turbulent}:

\begin{equation}
L_I=\frac{3\pi}{2(u^{\textrm{rms}})^2}\int_{0}^{\infty}\frac{E(k)}{k}\mathrm{d}k,~~\tau=\frac{L_I}{u^{\textrm{rms}}} ~, \label{eq:turnoverT}
\end{equation} where $E(k)$ is the energy spectrum.

For turbulent channel flow, the friction Reynolds number is defined as \cite{pope2000turbulent}:


\begin{equation}
Re_{\tau}=\frac{u_{\tau}\delta}{\nu} ~, \label{eq:REtau}
\end{equation} where $\delta$ is the characteristic length scale and $u_{\tau}=\sqrt{\tau_{\omega}/\rho}$ is the wall shear velocity. Here, the wall-shear stress is calculated as $\tau_{\omega}=\mu\partial\langle u\rangle/\partial y$ at the wall ($y = 0$), with $\langle \cdot \rangle$ denoting a spatial average over the homogeneous streamwise and spanwise directions \cite{wang2024prediction}.


\subsection{The elucidated diffusion model}

Generative models including flow models and diffusion models aim to sample the desired output $y_{t}$ from the unknown target distribution $p_{data}$ by mapping the initial input $y_{0}\sim p_{init}=\mathcal{N}(0,\textbf{I})$ using neural networks \cite{lipman2022flow,ho2019flow++}. Here, $\mathcal{N}$ represents a normal distribution and $\textbf{I}$ represents the identity matrix. A vector field $v_t$ can be used to construct a time-dependent diffeomorphic map, called a flow, $\phi:[0,1]\times\mathbb{R}^d\rightarrow\mathbb{R}^d$, defined via ordinary differential equation (ODE) \cite{lipman2022flow}:

\begin{equation}
\frac{\mathrm{d}}{\mathrm{d}t}\phi_t(y)=v_t(\phi_t(y));~~\phi_0(y)=y_0 ~. \label{eq:flowmodel}
\end{equation}

By incrementally increasing $t$ by $\mathrm{d}t$ and solving the above equation, sampling (generating) procedure from $y_{0}\sim p_{init}=\mathcal{N}(0,\textbf{I})$ to $y_{t=1}\sim p_{data}$ can be achieved. However, the ODE path is unique and deterministic. Consequently, the stochastic differential equation (SDE) framework proposed by Song et al. demonstrates that the process of generating data from noise possesses an infinite number of diffusion paths \cite{song2020score}. This lays the foundation for designing more efficient and higher-quality samplers. Song et al. defined their forward SDE as \cite{song2020score}:

\begin{equation}
\mathrm{d}\textbf{x}=\textbf{f}(\textbf{x},t)\mathrm{d}t+g(t)\mathrm{d}\omega_t ~, \label{eq:songsde}
\end{equation} where $\omega_t$ is the standard Wiener process and $\textbf{f}(\cdot,t):\mathbb{R}^d\rightarrow\mathbb{R}^d$ and $g(\cdot):\mathbb{R}\rightarrow\mathbb{R}$ are the drift and diffusion coefficients, respectively, where $d$ denotes the dimensionality of the dataset \cite{karras2022elucidating, song2020score}. In the work of Karras et al., they introduced the Fokker-Planck PDE and the heat equation PDE into Eq.~\ref{eq:songsde} and derived the following SDE \cite{karras2022elucidating}:


\begin{align}
\mathrm{d}\textbf{x}_{\pm}=\underbrace{-\dot{\sigma}(t)\sigma(t)\nabla_{\textbf{x}}\mathrm{log}p(\textbf{x};\sigma(t))\mathrm{d}t}_{\textrm{Probability flow ODE}}\nonumber\\
\pm\underbrace{\beta(t)\sigma(t)^2\nabla_{\textbf{x}}\mathrm{log}p(\textbf{x};\sigma(t))\mathrm{d}t}_{\textrm{Deterministic noise decay}}\nonumber\\
+\underbrace{\sqrt{2\beta(t)}\sigma(t)\mathrm{d}\omega_t}_{\textrm{Noise injection}} ~, \label{eq:karrassde}
\end{align} where $\sigma(t)$ denotes the intensity of noise at the current sampling step, $\beta(t)$ expresses the relative rate at which existing noise is replaced with new noise and $\nabla_{\textbf{x}}\mathrm{log}p(\textbf{x};\sigma(t))$ is a score function which denotes the gradient of the logarithmic probability of the data distribution \cite{karras2022elucidating,song2020score}. Moreover, $\nabla_{\textbf{x}}\mathrm{log}p(\textbf{x};\sigma(t))$ is a tensor field pointing towards the direction of real data \cite{karras2022elucidating,song2020score}. In practice, we use a neural network $D_{\theta}(\textbf{x};\sigma)$ to approximate: $(D_{\theta}(\textbf{x};\sigma)-\textbf{x})/\sigma^2\approx\nabla_{\textbf{x}}\mathrm{log}p(\textbf{x};\sigma)$ \cite{song2020score}. The positive sign denotes a forward time advance of the separate SDE, while the negative sign denotes a backward advance \cite{karras2022elucidating}.


Furthermore, Karras et al. provided their definition of $D_{\theta}(\textbf{x};\sigma)$ as:

\begin{align}
D_{\theta}(\textbf{x};\sigma)=c_{\mathrm{skip}}(\sigma)\textbf{x}+c_{\mathrm{out}}(\sigma)F_{\theta}(c_{\mathrm{in}}(\sigma)\textbf{x};c_{\mathrm{noise}}(\sigma))~, \label{eq:karrasD}
\end{align} where $\textbf{x}=\textbf{x}_0+\sigma\epsilon$, $\textbf{x}_0$ is the desired image without noise, $\epsilon\sim\mathcal{N}(0,\textbf{I})$ is the random noise, $F_{\theta}$ denotes the neural network to be trained, $c_{\mathrm{skip}}(\sigma)$ modulates the skip connection, $c_{\mathrm{in}}(\sigma)$ and $c_{\mathrm{out}}(\sigma)$ scale the magnitudes of input and output, respectively, and $c_{\mathrm{noise}}(\sigma)$ maps the noise level $\sigma$ to a conditioning input for $F_{\theta}$ \cite{karras2022elucidating}.

In order to ensure that when $F_{\theta}=0$, $D_{\theta}$ can approximate the optimal Wiener filter, the elucidated diffusion model (EDM) set \cite{karras2022elucidating}:

\begin{equation}
c_{\mathrm{skip}}(\sigma)=\frac{\sigma^2_{\mathrm{data}}}{\sigma^2+\sigma^2_{\mathrm{data}}} ~, \label{eq:cskip}
\end{equation} where $\sigma_{\mathrm{data}}$ is the standard deviation of the data distribution. Ideally, $D_{\theta}$ should provide the distribution of the original image input, i.e., the noise-free $\textbf{x}_0$:

\begin{align}
\textbf{x}_0=c_{\mathrm{skip}}(\sigma)\textbf{x}+c_{\mathrm{out}}(\sigma)F_{\theta}(c_{\mathrm{in}}(\sigma)\textbf{x};c_{\mathrm{noise}}(\sigma)); \nonumber \\
F^*_{\theta}(c_{\mathrm{in}}(\sigma)\textbf{x};c_{\mathrm{noise}}(\sigma))=\frac{\textbf{x}_0-c_{\mathrm{skip}}(\sigma)\textbf{x}}{c_{\mathrm{out}}}~, \label{eq:ideal}
\end{align} where $F^*_{\theta}$ is the training target $F_{\textrm{target}}$.

Both the training input of $F_{\theta}(\cdot)$ and training target $F_{\textrm{target}}$ have unit variance \cite{karras2022elucidating}:

\begin{align}
\mathrm{Var_{\textbf{x}}}[c_{\mathrm{in}}(\sigma)(\textbf{x})]&=1,\nonumber \\
c_{\mathrm{in}}(\sigma)^2&=\frac{1}{\mathrm{Var}[\textbf{x}]},\nonumber\\
c_{\mathrm{in}}(\sigma)&=\sqrt{\frac{1}{\sigma^2+\sigma_{\textrm{data}}^2}},\\
\mathrm{Var_{\textbf{x}}}\left[\frac{\textbf{x}_0-c_{\mathrm{skip}}(\sigma)\textbf{x}}{c_{\mathrm{out}}}\right]&=1,\nonumber \\
c_{\mathrm{out}}(\sigma)^2&=\mathrm{Var}[\textbf{x}_0-c_{\mathrm{skip}}(\sigma)(\textbf{x}_0+\sigma\epsilon)],\nonumber\\
c_{\mathrm{out}}(\sigma)&=\sqrt{(1-c_{\textrm{skip}})^2\sigma_{\textrm{data}}^2\textcolor{red}{+}c_{\textrm{skip}}^2\sigma^2},\nonumber\\
c_{\mathrm{out}}(\sigma)&=\frac{\sigma\sigma_{\textrm{data}}}{\sqrt{\sigma^2+\sigma_{\textrm{data}}^2}}~. \label{eq:cincout}
\end{align} Therefore, the model's inputs must first undergo Max-Min normalization, and the model's final outputs require corresponding inverse normalization.

Additionally, based on previous experience, they set the following relation \cite{karras2022elucidating}:

\begin{equation}
c_{\textrm{noise}}=\frac{\textrm{ln}(\sigma)}{4} ~. \label{eq:cnoise}
\end{equation}

Since the model's training target has been explicitly defined ($F^*_{\theta}$), we can specify the training loss used to update the model parameters:

\begin{equation}
\textrm{Training loss}=\mathbb{E}_{\sigma,\textbf{x},\epsilon}[||F_{\theta}-\underbrace{\frac{\textbf{x}_0-c_{\mathrm{skip}}(\sigma)\textbf{x}}{c_{\mathrm{out}}}}_{F^*_{\theta}}||^2_2] ~. \label{eq:trainingloss}
\end{equation}

In practice, the training loss is calculated as the L2 error between the model’s one-step denoising output and the true value. Karras et al. use a 2D U-Net as $F_{\theta}$ in their original proposal of EDM \cite{karras2022elucidating}. After the neural network completes training, the sampling process is shown in Algorithm~\ref{alg:samplerEDM}. During the sampling process, the model removes part of the noise from the noisy image $\textbf{y}_t$ to obtain $\textbf{y}_{t+1}$, and then corrects it using the second-order Heun's method \cite{karras2022elucidating}. With this denoising approach, the pure random noise $\textbf{y}_0$ initially input into the model is sampled into a noise-free image $\textbf{y}_T$ through $T$ steps.

\end{multicols}
\begin{algorithm}[!h]
    \caption{Stochastic sampler of EDM \cite{karras2022elucidating}}
    \label{alg:samplerEDM}
    \renewcommand{\algorithmicrequire}{\textbf{Input:}}
    \renewcommand{\algorithmicensure}{\textbf{Output:}}
    
    \begin{algorithmic}[1]
        \REQUIRE $\textbf{y}_0\sim\mathcal{N}(0,\sigma^2_0\textbf{I})$~~~~~~\# Random noise 
        \ENSURE $\textbf{y}_T$~~~~~~\# $\textbf{y}_T\sim p_{data}$    
        
        \STATE  \textbf{for} $t\in\{0,...,T-1\}$ \textbf{do}:
            \STATE~~~~\textbf{sample} $\epsilon_t\in\mathcal{N}(0,S^2_{\textrm{noise}}\textbf{I})$~~~~~~\# Generating random noise $\epsilon_t$ with $S^2_{\textrm{noise}}=1.003$ based on experience
            \STATE~~~~$\hat{\sigma}_t=\sigma_t(1+\gamma_t)$~~~~~~\# $\gamma_t$: scaling factor which formulate as \textcolor{red}{$\gamma_i=\begin{cases}\displaystyle \min\left( \frac{S_{\rm churn}}{N},\sqrt{2}-1 \right), & t_i\in [S_{\rm tmin},S_{\rm tmax}]\\[6pt]0, & \text{otherwise}\end{cases}$ 
            where $S_{\textrm{churn}} = 80$, $S_{\textrm{tmin}} = 0.05$, $S_{\textrm{tmax}} = 50$, and $T =32$}
            \STATE~~~~$\hat{\textbf{y}}_t=\textbf{y}_t+\sqrt{\hat{\sigma}_t^2-\sigma_t^2}\epsilon_t$~~~~~~\# Adjustment of noise level
            \STATE~~~~$\textbf{d}_t=(\hat{\textbf{y}}_t-D_\theta(\hat{\textbf{y}}_t;\hat{\sigma}_t))/\hat{\sigma}_t$~~~~~~\# $\textbf{d}_t$: direction of update
            \STATE~~~~$\textbf{y}_{t+1}=\hat{\textbf{y}}_t+(\sigma_{t+1}-\hat{\sigma}_t)\textbf{d}_t$~~~~~~\# Update using Euler's method
            \STATE~~~~\textbf{if} $\sigma_{t+1}\neq0$ \textbf{do}:
            \STATE~~~~~~~~$\textbf{d}_{t+1}=(\textbf{y}_{t+1}-D_{\theta}(\textbf{y}_{t+1};\sigma_{t+1}))/\sigma_{t+1}$
            \STATE~~~~~~~~$\textbf{y}_{t+1}=\hat{\textbf{y}}_t+\frac{1}{2}(\sigma_{t+1}-\hat{\sigma}_t)(\textbf{d}_t+\textbf{d}_{t+1})$~~~~~~\# Update using second order Heun's method
        \STATE  \textbf{return}  $\textbf{y}_T$
    \end{algorithmic}
\end{algorithm}
\begin{multicols}{2}

\subsection{The DiAFNO model and the autoregressive prediction architecture}

\begin{figure*}[htb]
    \centering
    \includegraphics[width=16cm,height=9cm]{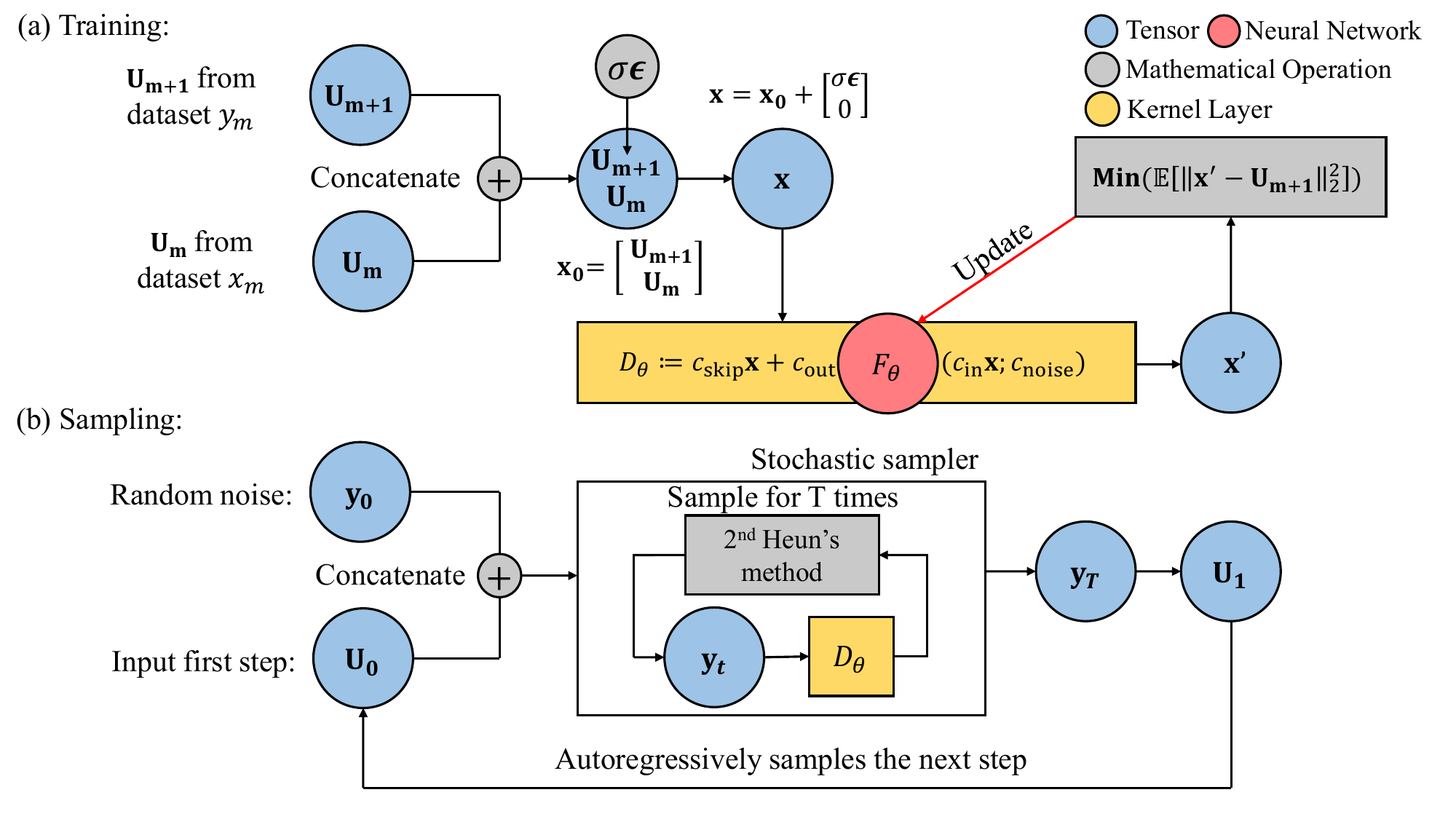}
    \caption{\label{fig:ModelStructureDiAFNO} The autoregressive prediction architecture of DiAFNO: (a) the training process; (b) the sampling process. Dataset $x_m$ contains flow fields from $U_0$ to $U_{N-1}$ and dataset $y_m$ contains flow fields from $U_1$ to $U_{N}$.}
\end{figure*}

The autoregressive prediction is achieved by the architecture shown in Fig.~\ref{fig:ModelStructureDiAFNO}. We start by adding noise of random intensity to $U_{m+1}$ and then concatenate the noised $U_{m+1}$ to $U_m$ along the channel dimension before input into $D_{\theta}$ for training. Here, $U_m$ is used as a condition to prompt the model to denoise the pure random noise to $U_{m+1}$ when sampling. Since $D_{\theta}(\textbf{x};\sigma)$ is an approximation of $\nabla_{\textbf{x}}\mathrm{log}p(\textbf{x};\sigma)$, the designed $D_{\theta}$ outputs a denoised $\textbf{x}'$. By minimizing the $L_2$ error between the output $\textbf{x}'$ and $U_{m+1}$, we can train the neural network $F_{\theta}$'s denoising capability. During the sampling process, we input the network with random noise $\textbf{y}_0$ and $U_0$. We then obtain the flow field for the next step by using the stochastic sampler demonstrated in Algorithm~\ref{alg:samplerEDM}. By taking the output flow field as input and predicting the flow field at the next time step, the autoregressive architecture is complete.



The IAFNO model has demonstrated its robust performance for 3D turbulent flow predictions in our previous study \cite{jiang2026animplicit}. Moreover, as an implicit Fourier neural operator based on self-attention mechanisms, IAFNO possesses the ability to effectively capture global frequency and structural features \cite{guibas2021adaptive}. This effective capture ability has been demonstrated to be capable of enhancing diffusion models for image denoising by Liu et al. \cite{liu2025difffno}. Therefore, we propose the DiAFNO model, which integrates IAFNO as the core denoising network with EDM as presented in Eq.~\ref{eq:IAFNOD}.

\end{multicols}
\begin{equation}
D^{\textrm{DiAFNO}}_{\theta}(\textbf{x};\sigma)=c_{\mathrm{skip}}(\sigma)\textbf{x}+c_{\mathrm{out}}(\sigma)\underbrace{\mathcal{L}^{\mathrm{IAFNO}}\circ\mathcal{L}^{\mathrm{IAFNO}}\circ\cdot\cdot\circ\mathcal{L}^{\mathrm{IAFNO}}}_{\textrm{Iterates}~L~\textrm{times as}~F_{\theta}}[c_{\mathrm{in}}(\sigma)\textbf{x};c_{\mathrm{noise}}(\sigma)]~. \label{eq:IAFNOD}
\end{equation}
\newline
\\
\newline
\begin{multicols}{2}

$\mathcal{L}^{\textrm{IAFNO}}$ denotes the implicit kernel layer of the IAFNO model. We provide a specific introduction to AFNO in Appendix.~\ref{app2}. Here, we will directly present the iterative formula for IAFNO used in DiAFNO as $F_{\theta}$ \cite{guibas2021adaptive,jiang2026animplicit}:

\end{multicols}
\begin{equation}
v(x,(l+1)\Delta t)=\mathcal{L}^{\mathrm{IAFNO}}[v(x,l\Delta t)]:=v(x,l\Delta t)+\Delta t\underbrace{\mathcal{K}_N\circ\cdot\cdot\circ\mathcal{K}_1}_{\textrm{N different layers}}[v(x,l\Delta t)],l\in[0,L-1] ~, \label{eq:iter1IAFNO}
\end{equation}
\begin{equation}
v_{n+1}(x,l\Delta t)=\mathcal{K}_{n+1}[v_n(x,l\Delta t)]:=\mathrm{MLP}\{ v(x,l\Delta t)+\mathcal{F}^{-1}\left[
R_{\mathrm{IAFNO}}\cdot\mathcal{F}(v_n(x,l\Delta t)) \right](x) \},n\in[0,N-1] ~, \label{eq:iter2IAFNO}
\end{equation}
\begin{equation}
R_{\mathrm{IAFNO}}\cdot\mathcal{F}(v_n(x,l\Delta t)):=S_{\lambda}\left[W_2\sigma\left(W_1\mathcal{F}(v_n(x,l\Delta t))+b_1\right)+b_2\right] ~. \label{eq:iter3IAFNO}
\end{equation}
\begin{multicols}{2}


Here, $v(x,(l+1)\Delta t)$ denotes the output of $\mathcal{L}^{\mathrm{IAFNO}}$ at the $(l+1)$th implicit iteration. $\mathcal{K}$ denotes the explicit kernel layer of IAFNO and multiple explicit layers act on the input $v(x,l\Delta t)$ within one implicit iteration. $v_n(x,l\Delta t))$ denotes the output of $\mathcal{K}$ at the $n$th explicit iteration. We write $v_0(x,l\Delta t))$ as $v(x,l\Delta t))$ for a concise notation. $\Delta t=1/(N\times L)$, where $L$,$N$ represent the total number of implicit iterations and explicit iterations, respectively. $\mathcal{F}$ and $\mathcal{F}^{-1}$ denote the Fourier transform and the inverse Fourier transform, respectively. MLP represents multilayer perception and $\sigma$ represents an activation layer. $W_1,W_2$ are weight matrix in Fourier space, $b_1,b_2$ are bias in Fourier space and $S_{\lambda}$ denotes a soft-thresholding (shrinkage) operation: $S_{\lambda}[x]=\mathrm{sign}(x)\mathrm{max}\{|x|-\lambda,0\}$, where $\lambda$ is a tuning parameter regulating the level of sparsity.



\begin{figure*}[!h]
    \centering
    \includegraphics[width=16cm,height=9cm]{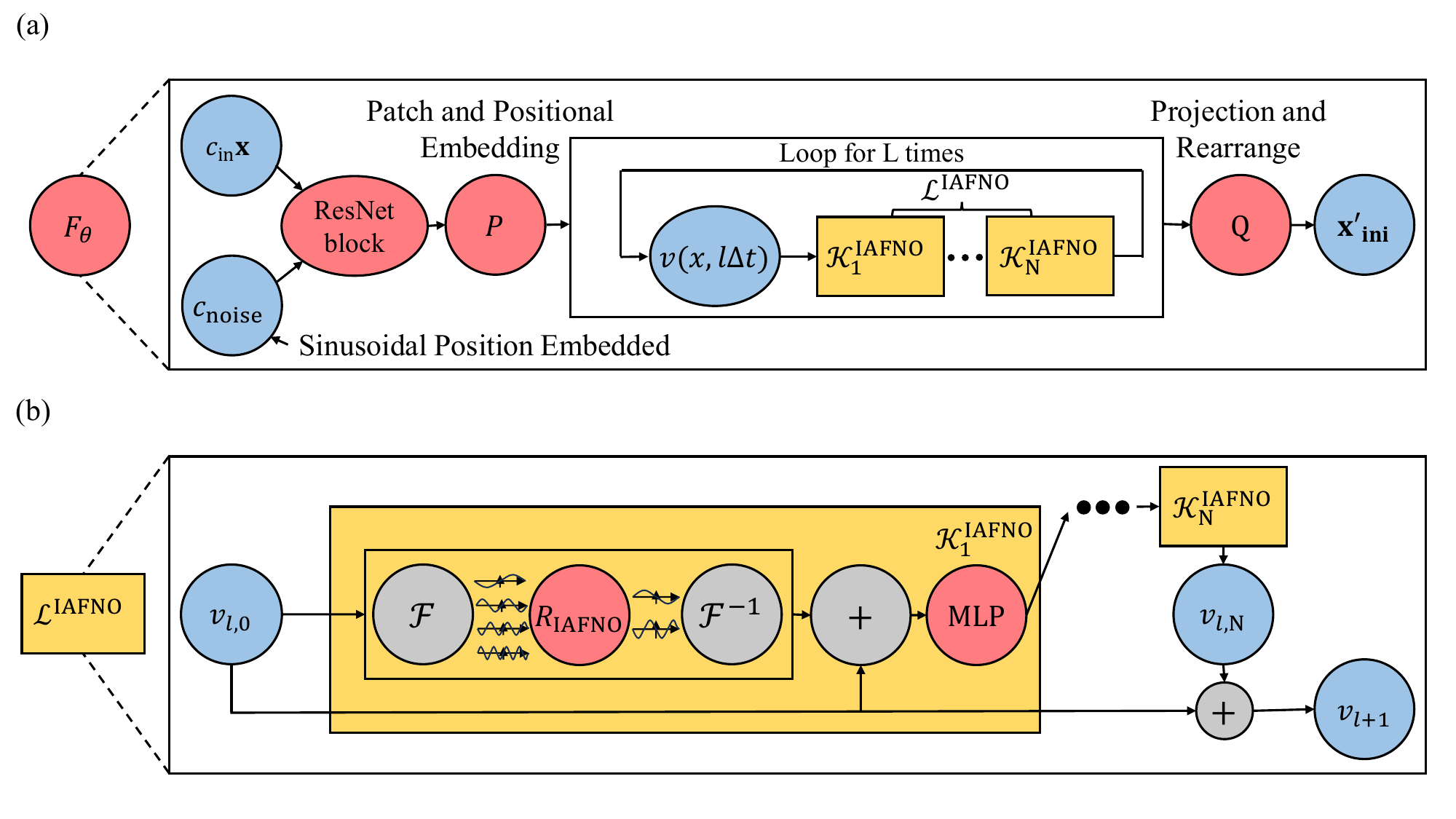}
    \caption{\label{fig:ModelStructureIAFNO} The architecture of IAFNO: (a) the macro architecture of IAFNO as $F_{\theta}$ in DiAFNO; (b) the architecture of $\mathcal{L}^{\textrm{IAFNO}}$.}
\end{figure*}

We demonstrate the architecture of IAFNO in Fig.~\ref{fig:ModelStructureIAFNO}. In Fig.~\ref{fig:ModelStructureIAFNO}(a), it can be observed that we first apply sinusoidal position embedding to $c_{\textrm{noise}}$ and feed the encoded tensor together with $c_{\textrm{in}}\textbf{x}$ into a ResNet block before entering the patch layer $P$, which implies:

\begin{equation}
v(x,l=0)=P[\mathrm{ResNet}(c_{\textrm{in}}\textbf{x};\textrm{SinuPosEmd}(c_{\textrm{noise}}))] ~. \label{eq:iter4IAFNO}
\end{equation} The input $v$ undergoes $n$ explicit iterations as described in Eq.~\ref{eq:iter2IAFNO}, followed by $l$ implicit iterations as described in Eq.~\ref{eq:iter1IAFNO}. Finally, the output $\textbf{x}'_{\textbf{ini}}$ is obtained through a fully connected layer on the feature dimension and a rearrange function after all iterations are complete. These reconstruction steps constitute the projection and rearrange module Q. The final output of IAFNO is $\textbf{x}'_{\textbf{ini}}$, and $\textbf{x}'$ in Fig.~\ref{fig:ModelStructureDiAFNO} is given by:

\begin{equation}
\textbf{x}'=c_{\textrm{skip}}\textbf{x}+c_{\textrm{out}}\textbf{x}'_{\textbf{ini}} ~. \label{eq:DiAFNO}
\end{equation} And $\textbf{x}'$ is an approximation of $U_{m+1}$. In Fig.~\ref{fig:ModelStructureIAFNO}(b), we present the architecture of $\mathcal{L}^{\textrm{IAFNO}}$ and $\mathcal{K}_{n}^{\textrm{IAFNO}}$, which is an intuitive illustration of Eq.~\ref{eq:iter1IAFNO}, Eq.~\ref{eq:iter2IAFNO} and Eq.~\ref{eq:iter3IAFNO} \cite{jiang2026animplicit}.

\end{multicols}
\begin{multicols}{2}
\section{Numerical Results}
\label{numerical}

In this section, flow fields from the filtered direct numerical simulation of three types of turbulent flows are employed to assess DiAFNO and EDM. It should be noted that since the EDM was originally applied to the generation of 2D images, we extended EDM to 3D and used the same autoregressive prediction framework as DiAFNO to ensure a fair comparison. These models are compared against the traditional large-eddy simulation (LES) incorporating the dynamic Smagorinsky model (DSM). The three turbulent flow configurations include forced homogeneous isotropic turbulence (HIT), decaying homogeneous isotropic turbulence (dHIT) and turbulent channel flow (CF).

\subsection{Data preparation}

The direct numerical simulated datasets for forced HIT and dHIT are generated by imposing periodic boundary conditions on a cube of size $(2\pi)^3$, implementing pseudo-spectral spatial discretization for the Navier-Stokes (NS) equations at a uniform grid resolution of $N=256^3$ \cite{ku1987pseudospectral}, and adopting a second-order two-step explicit Adams-Bashforth scheme for time integration \cite{chen1993statistical,he2007stability}. Notably, the adoption of periodic boundary conditions here enables efficient capture of global flow field information \cite{yuan2020deconvolutional,xie2020modeling,munters2016shifted} and facilitates compliance with the periodicity requirement for Fourier transforms. Aliasing errors arising from nonlinear convective terms are eliminated by truncating the high wavenumbers of Fourier modes via the two-third rule \cite{hussaini1986spectral}. The flow field has a kinematic viscosity $\nu \approx 0.00625$, which corresponds to a Taylor Reynolds number $Re_{\lambda}\approx100$. The time interval between neighboring DNS steps is $dt=0.001$. To ensure the statistical steady state of the flow field’s physical quantities, the forced HIT datasets are collected after 10 large-eddy turnover times. Here, the large-eddy turnover time $\tau$ is defined as $\tau \equiv L_I/u^{\textrm{rms}} \approx 1.0$. We use a statistically steady flow field of the forced HIT as an initial field, while removing the external force term for the generation of dHIT datasets.

The DNS data is filtered by a sharp spectral filter into a filtered DNS (fDNS) data with a uniform grid resolution of $32^3$ \cite{pope2000turbulent}. The truncation frequency $k_c = 10$ is employed during the filtering process \cite{pope2000turbulent}. Additionally, one time node is sampled every 200 DNS steps for forced HIT simulation, yielding a total of 200 time nodes. For the decaying HIT case, sampling is performed every 100 DNS steps, which yields 20 time nodes. The training dataset comprises 40 independent samples generated from 40 distinct initial conditions for forced HIT, and 320 independent samples from 320 different initial conditions for decaying HIT. We further generate another 5 independent sets of both forced HIT and dHIT for the purpose of validation. All details are summarized in Tab.~\ref{tab:hitfdns1}, Tab.~\ref{tab:hitfdns2} and Tab.~\ref{tab:hitfdns3} shown below.


\end{multicols}
\begin{table*}[!ht]
\centering
\caption{\label{tab:hitfdns1}Parameters and statistics for DNS and fDNS of forced HIT and decaying HIT at its initial state.}
\renewcommand{\arraystretch}{1.5}
\begin{tabular}{cccccccc}
\hline\hline
\mbox{Reso.(DNS)}&\mbox{Reso.(fDNS)}&\mbox{Domain}&\mbox{$Re_{\lambda}$}&\mbox{$\nu$}&\mbox{$\mathrm{d} t$}&\mbox{$k_c$}&\mbox{$\tau$}\\
\hline
\mbox{$256^3$}&\mbox{$32^3$}&\mbox{$(2\pi)^3$}&\mbox{100}&\mbox{0.00625}&\mbox{0.001}&\mbox{10}&\mbox{1.00}\\
\hline\hline
\end{tabular}
\end{table*}

\begin{table*}[!ht]
\centering
\caption{\label{tab:hitfdns2}Information of datasets of forced HIT.}
\renewcommand{\arraystretch}{1.5}
\begin{tabular}{cccc}
\hline\hline
\mbox{}&\mbox{Count of samples}&\mbox{Time nodes per sample}&\mbox{Interval between time nodes}\\
\hline
\mbox{training}&\mbox{40}&\mbox{200}&\mbox{200$\textrm{d}t$}\\
\mbox{validation}&\mbox{5}&\mbox{600}&\mbox{200$\textrm{d}t$}\\
\hline\hline
\end{tabular}
\end{table*}

\begin{table*}[!ht]
\centering
\caption{\label{tab:hitfdns3}Information of datasets of decaying HIT.}
\renewcommand{\arraystretch}{1.5}
\begin{tabular}{cccc}
\hline\hline
\mbox{}&\mbox{Count of samples}&\mbox{Time nodes per sample}&\mbox{Interval between time nodes}\\
\hline
\mbox{training}&\mbox{320}&\mbox{20}&\mbox{100$\textrm{d}t$}\\
\mbox{validation}&\mbox{5}&\mbox{60}&\mbox{100$\textrm{d}t$}\\
\hline\hline
\end{tabular}
\end{table*}
\begin{multicols}{2}


We use the filtered direct numerical simulation data generated by Xcompact3D in the case of turbulent channel flow at different friction Reynolds numbers \cite{laizet2009high,bartholomew2020xcompact3d,wang2024prediction}. Tab.~\ref{tab:cffdns1} presents the DNS setup for turbulent channel flow, featuring a three-dimensional computational domain with streamwise, transverse, and spanwise dimensions $(L_x,L_y,L_z)=(4\pi,2,4\pi/3)$. Uniform grid distributions are adopted in the streamwise and spanwise directions, while a non-uniform grid is employed in transverse direction to ensure finer mesh resolution near the walls. The normalized grid spacings in the streamwise and spanwise directions, and the normalized distance between the nearest grid point to the wall and the wall surface along the transverse direction are shown in Tab.~\ref{tab:cffdns1} with an order of $(\Delta X^+,\Delta Z^+,\Delta Y_w^+)$. The superscript ``+'' indicates normalization in viscous units, e.g., $\Delta X^+=\Delta X/\delta_{\nu}, \Delta Z^+=\Delta Z/\delta_{\nu},\Delta Y_{w}^+=\Delta Y_{w}/\delta_{\nu}$ where $\delta_{\nu}$ represents the viscous length scale and $(\Delta X,\Delta Z,\Delta Y_w)$ are the grid spacings \cite{wang2024prediction}. Although the mesh is non-uniform in the transverse direction, the DNS employs a structured mesh that can be transformed into a uniform grid \cite{li2023fourier}, enabling the application of FFT. Filtering is applied to the DNS data in the streamwise and spanwise directions where the grid distributions are homogeneous, with no filtering performed in the transverse direction \cite{wang2024prediction}. A linear interpolation is utilized for grid coarsening, resulting in a fDNS data.

Datasets construction of turbulent channel flow at two friction Reynolds numbers are identical. As shown in Tab.~\ref{tab:cffdns2}, we construct a sufficiently large training dataset with 20 different fDNS samples, each containing 200 time nodes that are saved every 200 DNS time steps. We generate a larger sample that contains 400 time nodes and each time node is saved every 200 DNS time steps for validation.

\end{multicols}
\begin{table*}[!ht]
\centering
\caption{\label{tab:cffdns1}Parameters and statistics for DNS and fDNS of turbulent channel flow at different friction Reynolds numbers.}
\renewcommand{\arraystretch}{1.5}
\begin{tabular}{ccccccc}
\hline\hline
\mbox{$Re_{\tau}$}&\mbox{Reso.(DNS)}&\mbox{Reso.(fDNS)}&\mbox{Domain}&\mbox{$\nu$}&\mbox{$\mathrm{d} t$}&\mbox{$(\Delta X^+,\Delta Z^+,\Delta Y_w^+)$}\\
\hline
\mbox{395}&\mbox{$256\times193\times128$}&\mbox{$64\times49\times32$}&\mbox{$4\pi\times2\times4\pi/3$}&\mbox{1/10500}&\mbox{0.005}&\mbox{$(19.1,12.8,1.4)$}\\
\mbox{590}&\mbox{$384\times257\times192$}&\mbox{$64\times65\times32$}&\mbox{$4\pi\times2\times4\pi/3$}&\mbox{1/16800}&\mbox{0.005}&\mbox{$(19.3,12.9,1.6)$}\\

\hline\hline
\end{tabular}
\end{table*}

\begin{table*}[!ht]
\centering
\caption{\label{tab:cffdns2}Information of datasets of turbulent channel flow at different friction Reynolds numbers.}
\renewcommand{\arraystretch}{1.5}
\begin{tabular}{cccc}
\hline\hline
\mbox{}&\mbox{Count of samples}&\mbox{Time nodes per sample}&\mbox{Interval between time nodes}\\
\hline
\mbox{training}&\mbox{20}&\mbox{200}&\mbox{200$\textrm{d}t$}\\
\mbox{validation}&\mbox{1}&\mbox{400}&\mbox{200$\textrm{d}t$}\\
\hline\hline
\end{tabular}
\end{table*}
\begin{multicols}{2}

\subsection{The \textit{a posteriori} tests}

We construct the input-output pairs with every two neighboring flow fields: $(U_m,U_{m+1})$. We use $80\%$ of the input-output pairs as the model's training set, and the remaining $20\%$ as the testing set. All data-driven models use the flow field $U_m$ as input and output the predicted flow field at the next time node $U^{\textrm{pre}}_{m+1}$. The loss function is defined as:

\begin{equation}
\mathrm{ Testing~Loss} =\frac{\vert\vert u^{\textrm{pre}}-u\vert\vert_2}{\vert\vert u \vert\vert_2},~~\vert\vert \textbf{u} \vert\vert = \frac{1}{n}\sqrt{\sum_{\textbf{k}=1}^{n}\vert \textbf{u}_{\textbf{k}} \vert^2} ~, \label{eq:lossfunction}
\end{equation} where $u^{\textrm{pre}}$ denotes the output of data-driven models ($U^{\textrm{pre}}_{m+1}$) and $u$ is the ground truth ($U_{m+1}$) \cite{li2023long,li2022fourier}. The autoregressive approach aiming at achieving long-term predictions of turbulent flows is given as follows:
\begin{align}
U_0&\rightarrow U_1^{\textrm{pre}},\nonumber \\
U_1^{\textrm{pre}}&\rightarrow U_2^{\textrm{pre}},\nonumber \\
&\dots ,\nonumber \\
U_{m}^{\textrm{pre}}&\rightarrow U_{m+1}^{\textrm{pre}}~. \label{eq:predict}
\end{align}

Since the Max-Min normalization is required for the input of the diffusion models, the output needs to undergo inverse normalization:

\begin{align}
\textrm{Input}:\hat{x}&=(x-x_{\textrm{min}})/(x_{\textrm{max}}-x_{\textrm{min}}),\nonumber \\
\textrm{Output}:y&=\hat{y}(y_{\textrm{max}}-y_{\textrm{min}})+y_{\textrm{min}}, \label{eq:max-min}
\end{align} where $\hat{x}$ and $\hat{y}$ are the flow fields that have been normalized and $x_{\textrm{max}},x_{\textrm{min}};y_{\textrm{max}},y_{\textrm{min}}$ denote the maximum and minimum values of the input and output datasets, respectively.



We present a detailed hyperparameters settings for EDM and DiAFNO in Tab.~\ref{tab:configs}. ``Dim\_mults'' refers to the dimension multipliers of 3D U-Net in EDM. ``Dim'' refers to the base channel dimension of 3D U-Net in EDM. ``Implicit layers'' refers to the number of implicit iterations of IAFNO in DiAFNO. ``Explicit layers'' refers to the number of explicit iterations of IAFNO in DiAFNO. ``Patchsize'' refers to the size of every patches in (x,y,z) directions of IAFNO in DiAFNO. ``Embed\_dim'' refers to the dimension of features of the tensor output by the patch layer of IAFNO in DiAFNO. ``Num\_blocks'' refers to the number of blocks in the block diagonal matrix of IAFNO in DiAFNO. ``Hidden\_size\_factor'' refers to the scaling factor of the dimension of features in a hidden layer of IAFNO in DiAFNO. ``Sample\_steps'' refers to the sampling steps in the stochastic sampler. The other hyperparameters in the stochastic sampler are kept the same as Karras et al. in their paper \cite{karras2022elucidating}.

During the model testing phase, we have drawn the following conclusions based on our experience: 1. As the number of retained Fourier modes increases, the model’s prediction accuracy improves. 2. As patch size and the number of blocks increase, the model’s prediction accuracy decreases. 3. The soft thresholding parameter has no significant effect on the model. 4. Increasing the number of explicit and hidden layers within a certain range can improve model performance, but further increases lead to numerical instability. The configuration of 2 explicit layers and 4 implicit layers used in this paper is the result of optimal hyperparameter tuning. 5. Increasing the number of sampling steps can improve the model’s prediction accuracy, but a balance between performance and efficiency must be considered. The current approach follows the common practice of 32 sampling steps. \textcolor{red}{Moreover, we have included a section on quantitative testing of hyperparameters in Appendix.~\ref{app34}.}

The minimum losses of two data-driven models with respect to the three types of turbulent flows are presented in Tab.~\ref{tab:losses}. By comparing the results shown in Tab.~\ref{tab:losses}, it can be found that during the 100-epoch training process, both losses of DiAFNO are lower than that of EDM.

\end{multicols}
\begin{table}[h]
\centering
\caption{Model configurations of EDM and DiAFNO.}
\renewcommand{\arraystretch}{1.5}
\label{tab:configs}
\begin{tabular}{lcc}
\hline\hline
Model & Configurations of denoising neural network & Trainer configurations\\
\hline
EDM & dim\_mults: (1,2,4,8) & learning rate: $1\times10^{-4}$ \\
            & dim: 16 & weight\_decay: $1\times10^{-4}$ \\
            &  & batchsize: 4 \\
            &  & optimizer: Adam \\
            &  & scheduler: CosineAnnealingLR \\
            &  & sample\_steps: 32 \\
\cline{1-3}
DiAFNO & implicit layers (L): 4 & learning rate: $1\times10^{-3}$ \\
& explicit layers (N): 2 & weight\_decay: 0 \\
& patch\_size: (2,2,2) & batchsize: 4 \\
& embed\_dim: 180 & optimizer: Adam \\
& num\_blocks: 1 & scheduler: CosineAnnealingLR \\
& hidden\_size\_factor: 4  & sample\_steps: 32\\
\hline\hline
\end{tabular}
\end{table}

\begin{table}[htb]
\centering
\caption{\label{tab:losses}Comparison of minimum training and testing loss of different data-driven models in three types of turbulent flows.}
\renewcommand{\arraystretch}{1.5}
\begin{tabular}{ccc}
\hline\hline
\multicolumn{3}{c}{(Training Loss: Eq.~\ref{eq:trainingloss},~~ Testing Loss: Eq.~\ref{eq:lossfunction}) in 100 epochs}
\\
\hline
\mbox{Turbulent Flows}&\mbox{EDM}&\mbox{DiAFNO}\\
\hline
\mbox{HIT}&\mbox{(0.0720,~0.0384)}&\mbox{(\textbf{0.0434},~\textbf{0.0288})}\\
\mbox{dHIT}&\mbox{(0.0448,~0.0235)}&\mbox{(\textbf{0.0211},~\textbf{0.0169})}\\
\mbox{CF395}&\mbox{(0.1091,~0.0501)}&\mbox{(\textbf{0.0816},~\textbf{0.0410})}\\
\mbox{CF590}&\mbox{(0.1220,~0.0555)}&\mbox{(\textbf{0.0962},~\textbf{0.0485})}\\
\hline\hline
\end{tabular}
\end{table}
\begin{multicols}{2}

\subsubsection{Forced homogeneous isotropic turbulence}

The data-driven models use ten different initial flow fields to predict 250 steps for each initial field, which cover 50 large-eddy turnover times. The physical statistics obtained from post-processing the ten sets of different prediction data are ensemble averaged.

\begin{figure*}[htb]
    \centering
    \subfloat{
    \includegraphics[width=6.4cm,height = 4.8cm]{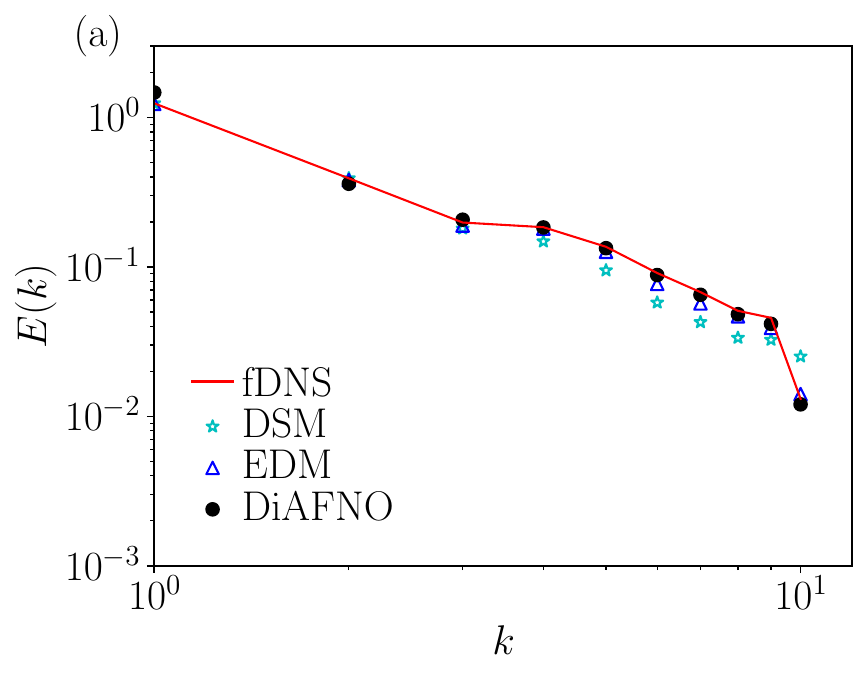}}
    \subfloat{
    \includegraphics[width=6.4cm,height = 4.8cm]{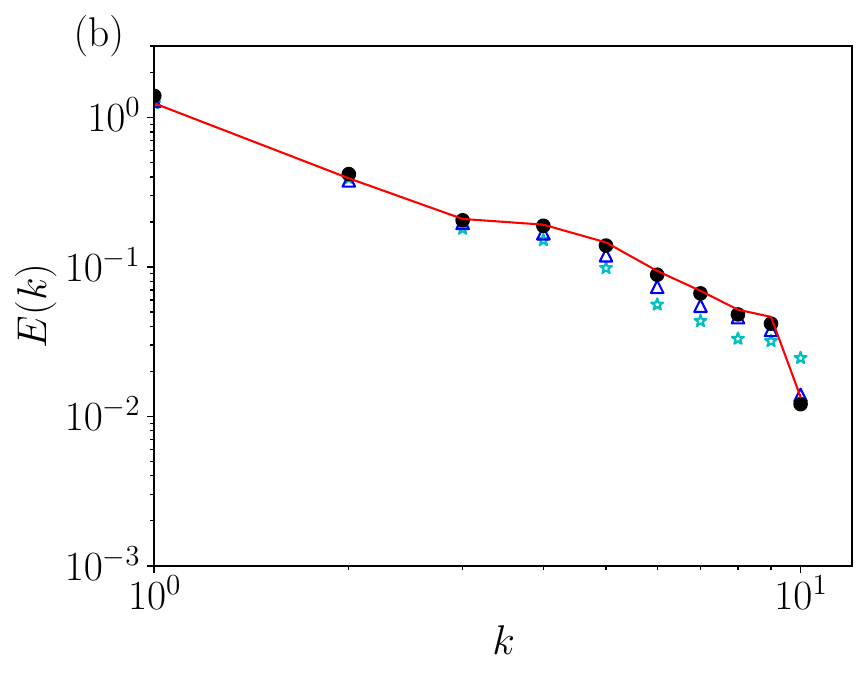}}
    \\
    \subfloat{
    \includegraphics[width=6.4cm,height = 4.8cm]{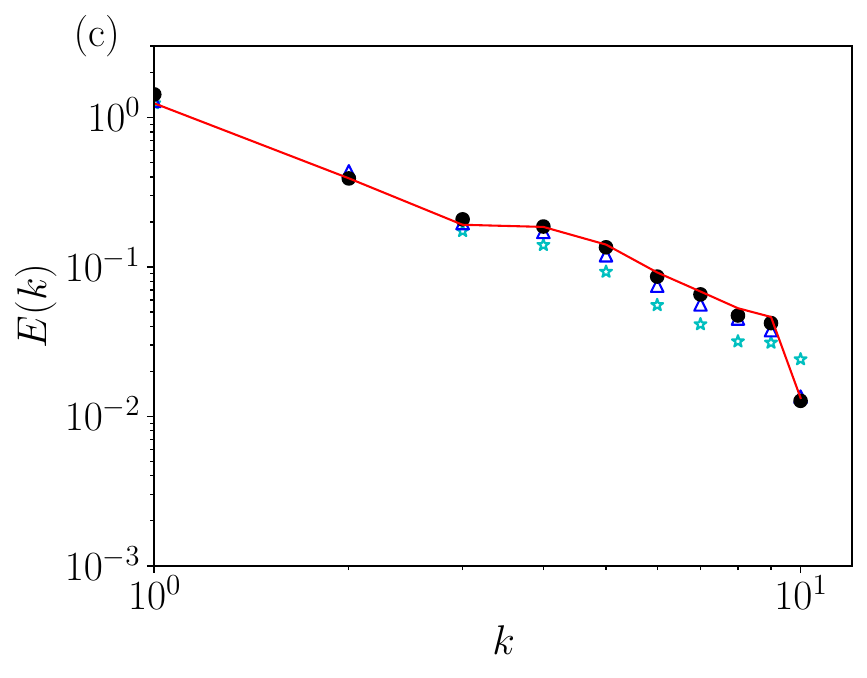}}
    \subfloat{
    \includegraphics[width=6.4cm,height = 4.8cm]{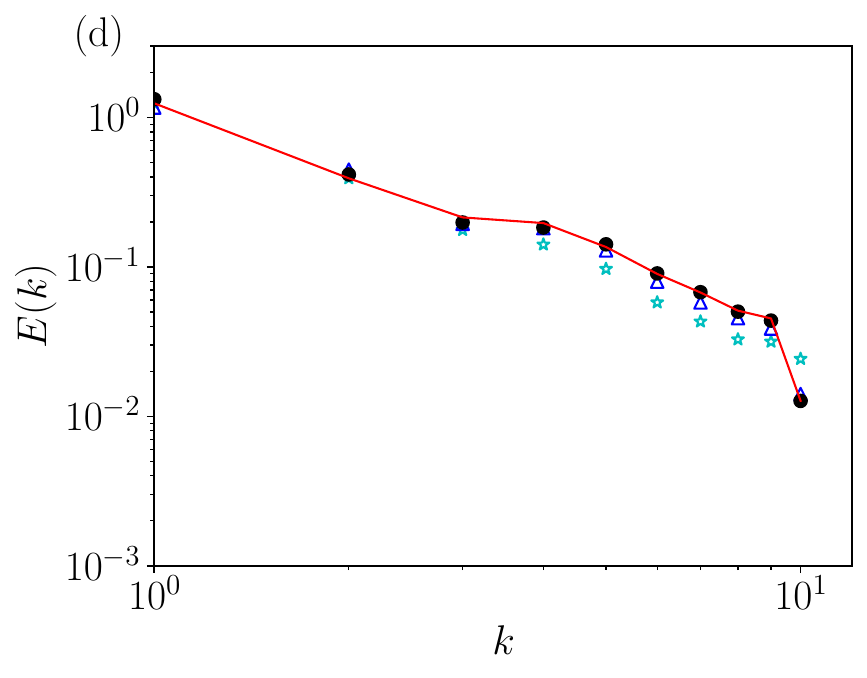}}
    \caption{\label{fig:HITspec} The velocity spectra of various models in the forced HIT at different time instants: (a) $t/\tau\approx 4.0$; (b) $t/\tau\approx 6.0$; (c) $t/\tau\approx 8.0$; (d) $t/\tau\approx 50.0$.}
\end{figure*}

The velocity spectra predicted by EDM, DiAFNO and the DSM in the forced HIT at different time instants are shown in Fig.~\ref{fig:HITspec}. The DiAFNO can accurately predict the velocity spectra at any prediction time. When $k\in[4,9]$, both EDM and DSM underestimate the energy spectrum, with DSM having a larger underestimation. When $k=10$, DSM gives an overestimated result, indicating that the result of DSM is not satisfactory.

\begin{figure*}[!h]
    \centering
    \subfloat{
    \includegraphics[width=6.4cm,height = 4.8cm]{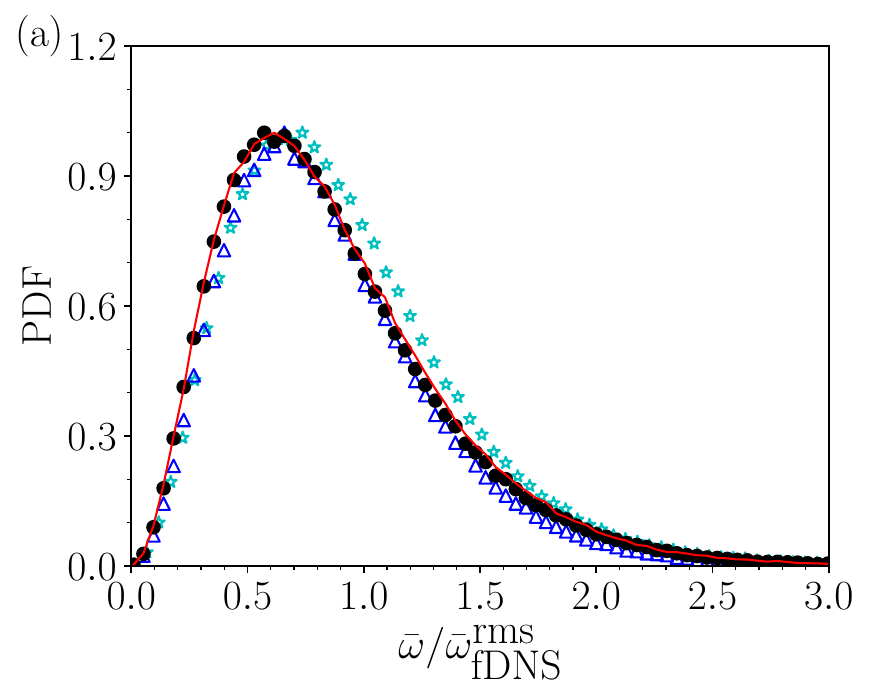}}
    \subfloat{
    \includegraphics[width=6.4cm,height = 4.8cm]{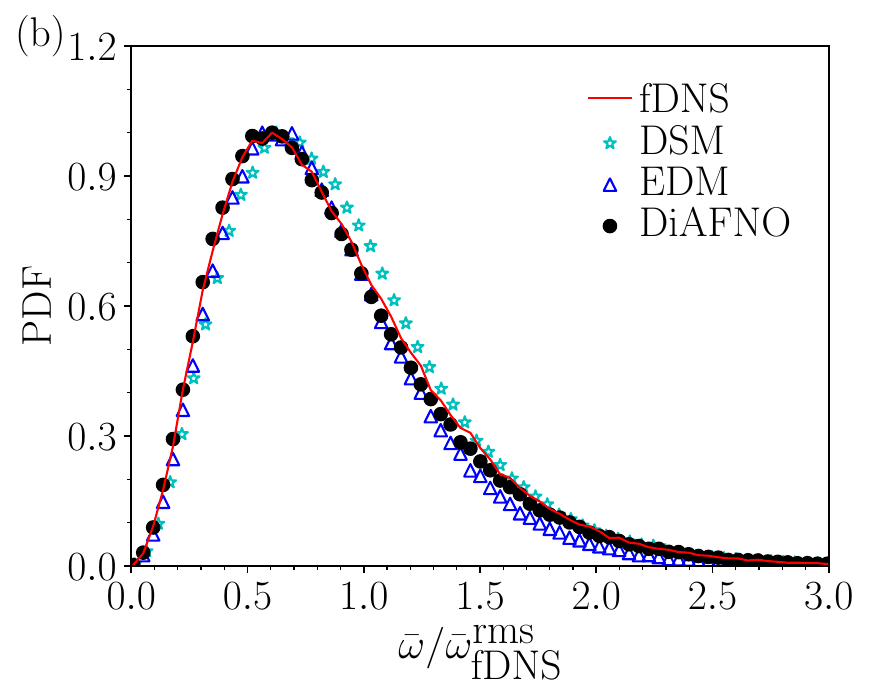}}
    \\
    \subfloat{
    \includegraphics[width=6.4cm,height = 4.8cm]{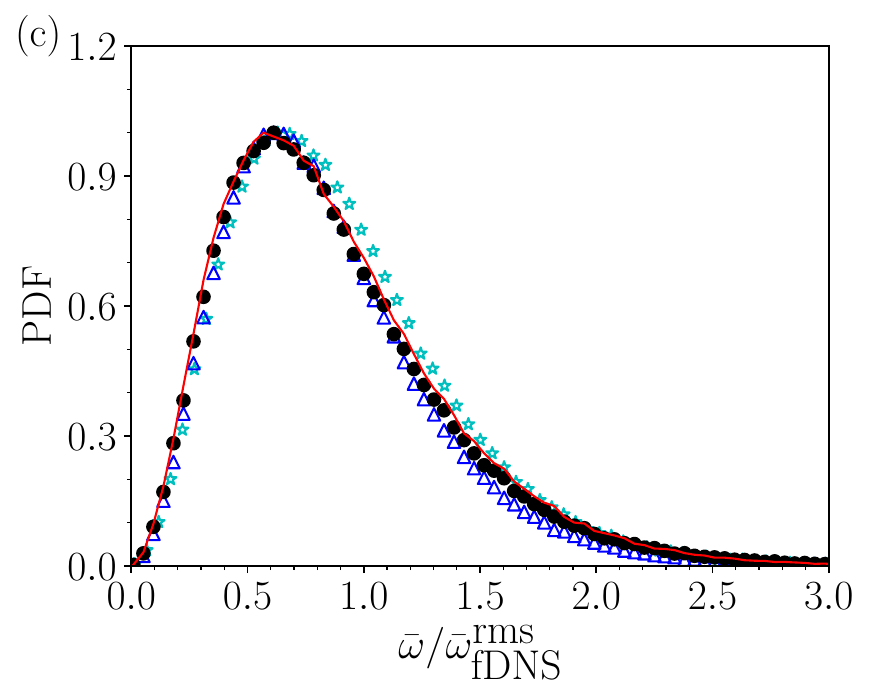}}
    \subfloat{
    \includegraphics[width=6.4cm,height = 4.8cm]{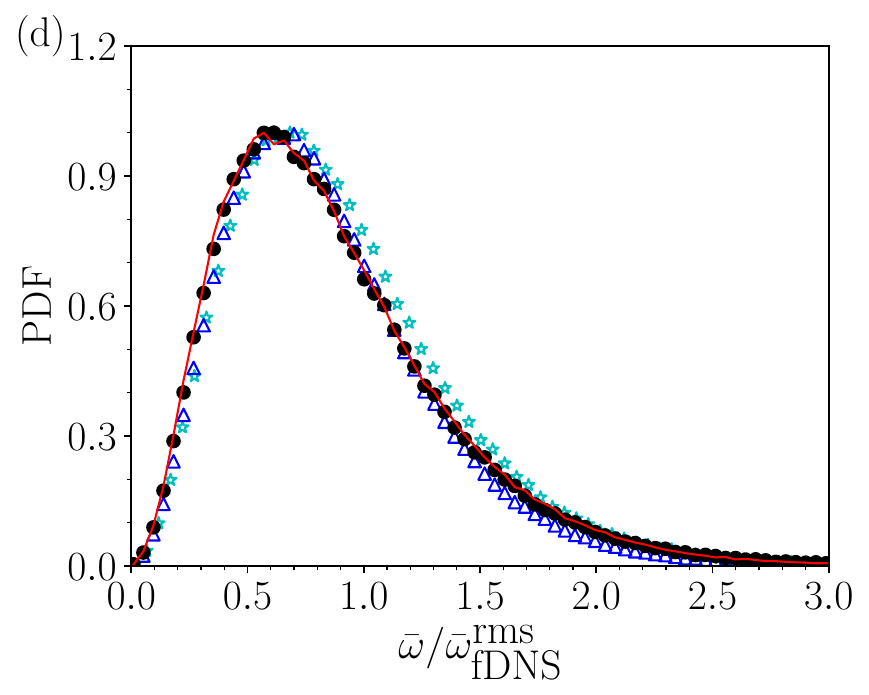}}
    \caption{\label{fig:HITvor} The PDFs of the normalized vorticity $\bar{\omega} / \bar{\omega}^{\textrm{rms}}_{\textrm{fDNS}}$ of various models in the forced HIT at different time instants: (a) $t/\tau\approx 4.0$; (b) $t/\tau\approx 6.0$; (c) $t/\tau\approx 8.0$; (d) $t/\tau\approx 50.0$.}
\end{figure*}

We compare the PDFs of the normalized vorticity magnitude $\bar{\omega} / \bar{\omega}^{\textrm{rms}}_{\textrm{fDNS}}$ predicted by various models at different time instants in Fig.~\ref{fig:HITvor}. Here, the vorticity is normalized by the rms values of the vorticity calculated by the fDNS data. The results of EDM are slightly deviated from fDNS results on both side of the peak ($\bar{\omega} / \bar{\omega}^{\textrm{rms}}_{\textrm{fDNS}}=0.6$). However, the overall PDF is shifted to the right for DSM as compared to the fDNS results. It can be seen that the DiAFNO is in a good agreement with fDNS and outperforms the EDM and DSM.

\begin{figure*}[!h]
    \centering
    \subfloat{
    \includegraphics[width=6.4cm,height = 4.8cm]{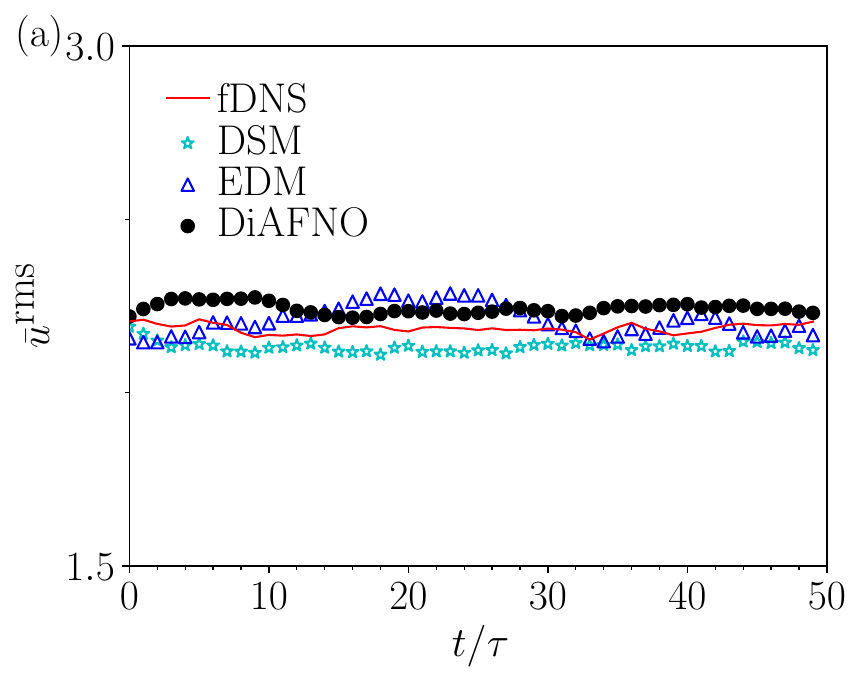}}
    \subfloat{
    \includegraphics[width=6.4cm,height = 4.8cm]{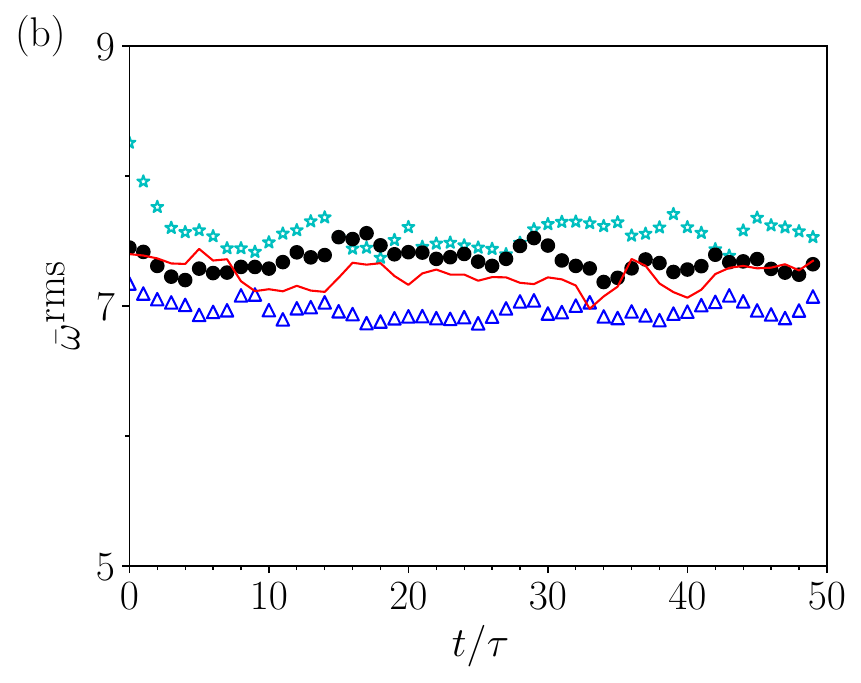}}
    \caption{\label{fig:HITrms} Temporal evolutions of (a) the velocity rms value and (b) vorticity rms value of various models in the forced HIT.}
\end{figure*}

The temporal evolutions of the root-mean-square (rms) values of velocity and vorticity predicted by various models are shown in Fig.~\ref{fig:HITrms}. In Fig.~\ref{fig:HITrms}(a), the DSM gives a relatively lower value while the DiAFNO gives a relatively higher value. The result given by EDM oscillates up and down, and is slightly better than the other two models for $t/\tau\in[4,10]$ and $t/\tau\in[30,50]$. In Fig.~\ref{fig:HITrms}(b), it is observed that DiAFNO's performance is better than EDM and DSM.

\begin{figure*}[!h]
    \centering
    \subfloat{
    \includegraphics[width=3cm,height=3cm]{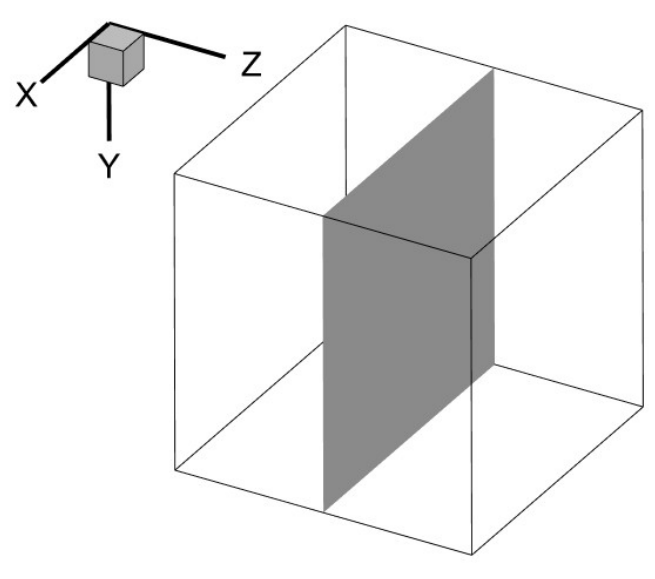}}
    \subfloat{
    \includegraphics[width=12cm,height=8cm]{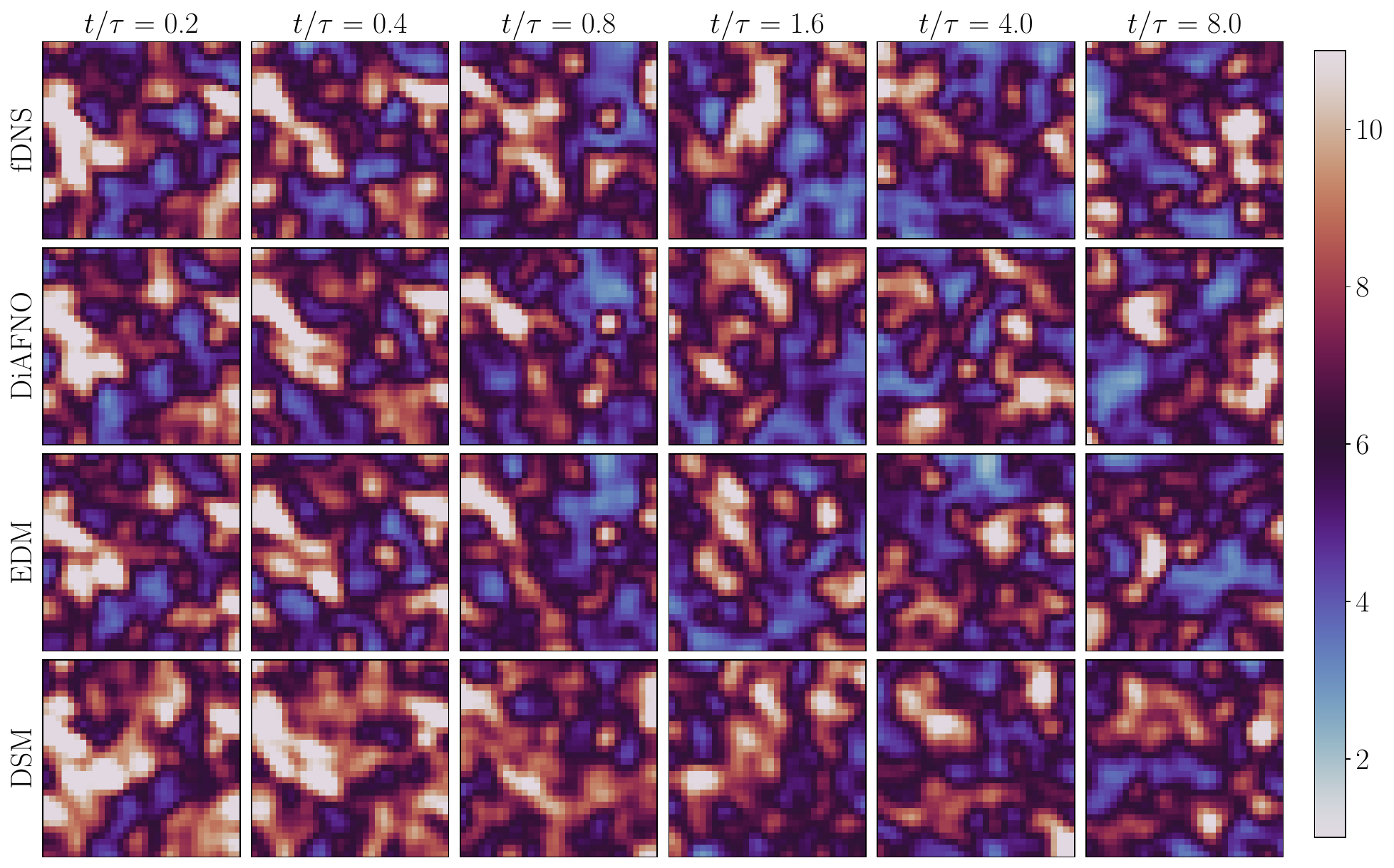}}
    \caption{\label{fig:HITContour} Contour of vorticity $\bar{\omega}$ on the xy-plane in the middle of the z-axis at different time instants for forced HIT.}
\end{figure*}

Moreover, we demonstrate the contour of vorticity $\bar{\omega}$ on the xy-plane in the middle of the z-axis at different time instants for HIT in Fig.~\ref{fig:HITContour}. By comparison, DSM shows a significant overestimation at the first few steps, which is consistent with the results shown in Fig.~\ref{fig:HITrms}(b). Thus, we can conclude that both DiAFNO and EDM are more accurate than DSM at this specific plane. However, there is no inherent difference between the performance of DiAFNO and EDM.

\end{multicols}

\begin{figure*}[!h]
    \centering
    \subfloat{
    \includegraphics[width=6.4cm,height = 4.8cm]{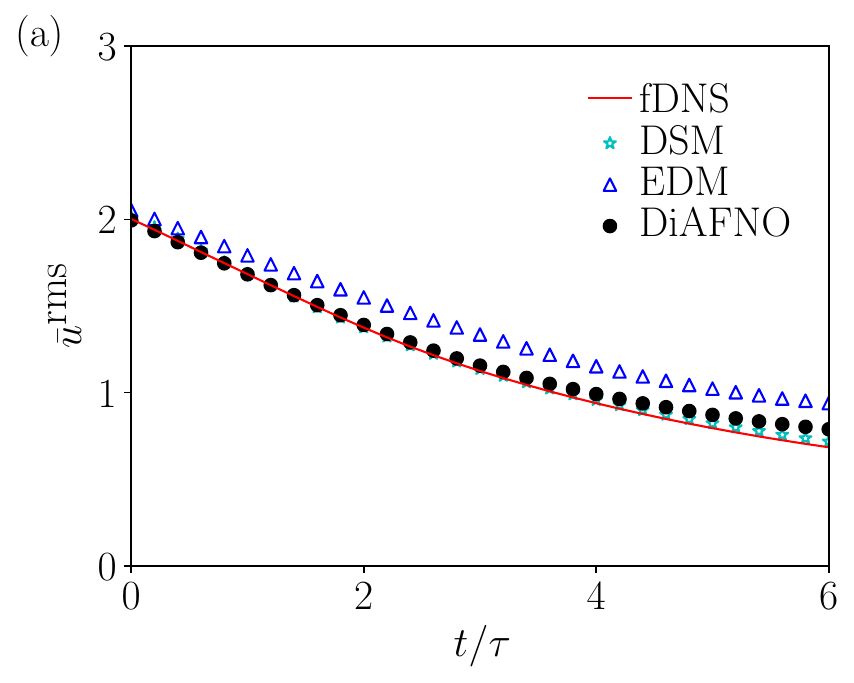}}
    \subfloat{
    \includegraphics[width=6.4cm,height = 4.8cm]{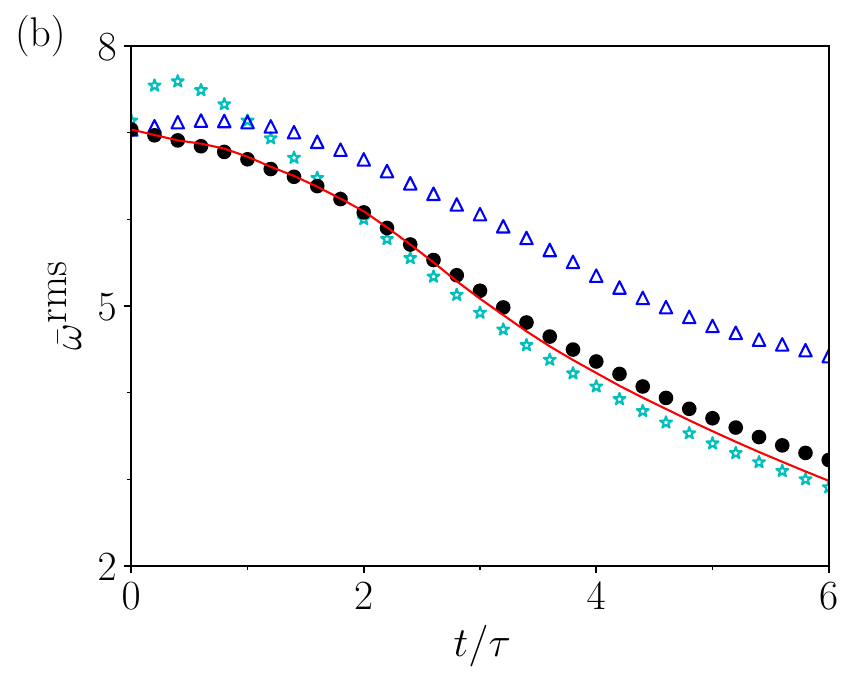}}
    \caption{\label{fig:dHITrms} Temporal evolutions of (a) the velocity rms value and (b) vorticity rms value of various models in the decaying HIT.}
\end{figure*}

\begin{figure*}[!h]
    \centering
    \subfloat{
    \includegraphics[width=6.4cm,height = 4.8cm]{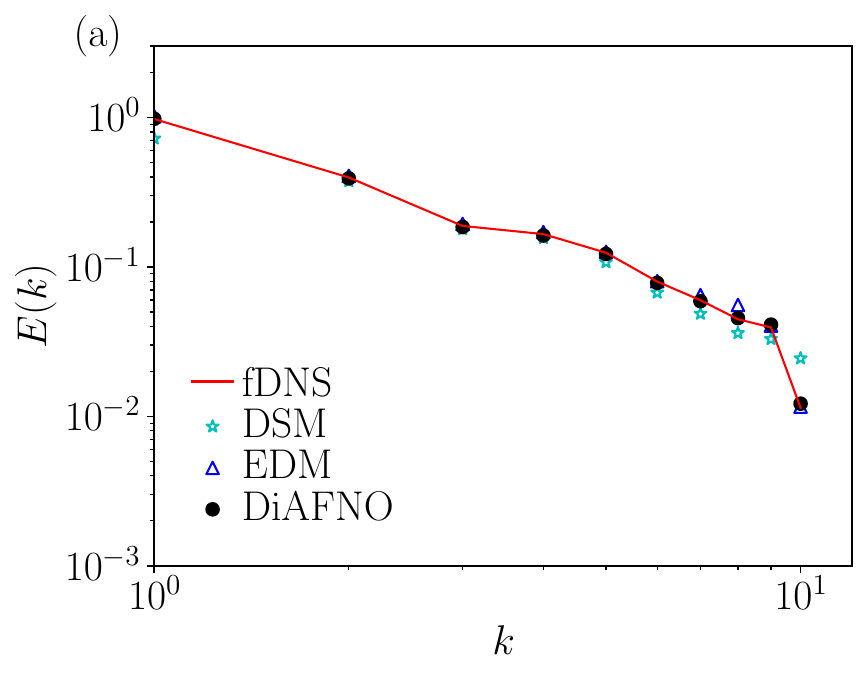}}
    \subfloat{
    \includegraphics[width=6.4cm,height = 4.8cm]{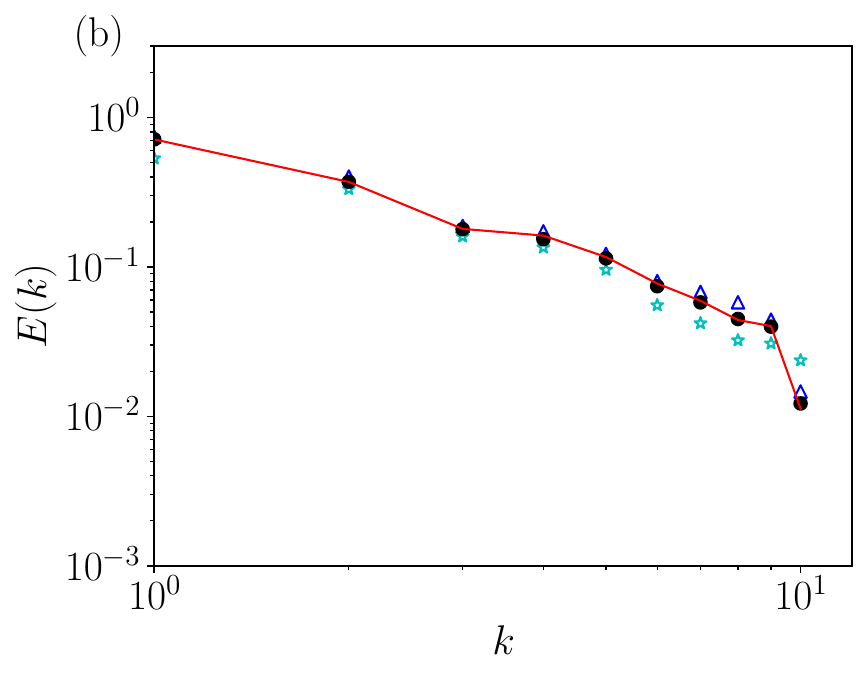}}
    \\
    \subfloat{
    \includegraphics[width=6.4cm,height = 4.8cm]{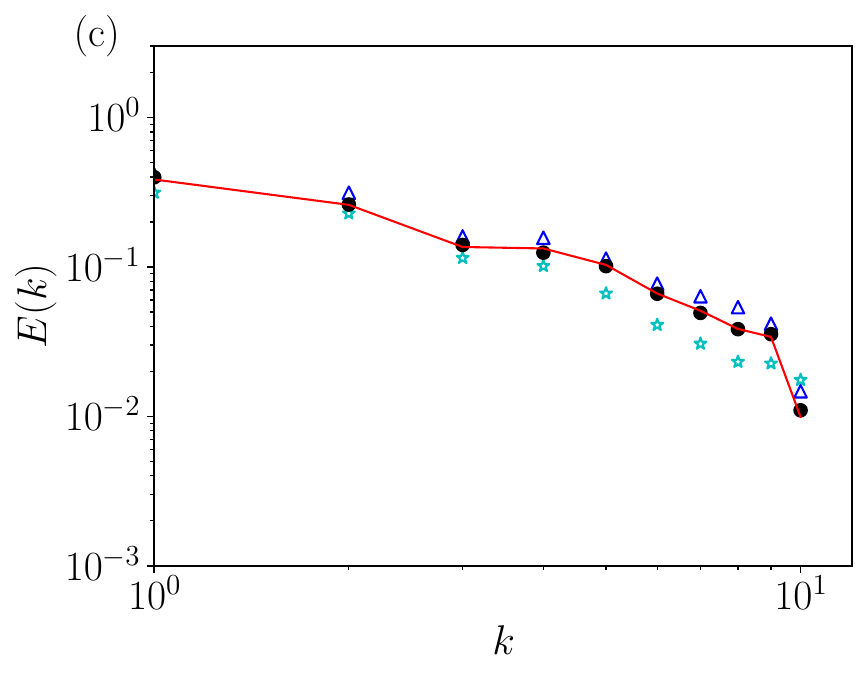}}
    \subfloat{
    \includegraphics[width=6.4cm,height = 4.8cm]{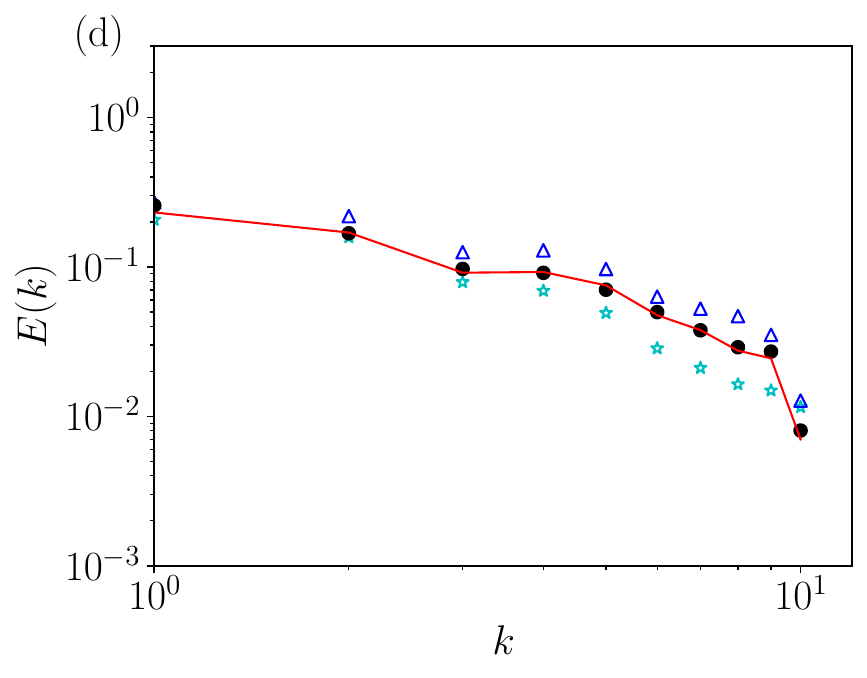}}
    \caption{\label{fig:dHITspec} The velocity spectra of various models in the decaying HIT at different time instants: (a) $t/\tau\approx 1.0$; (b) $t/\tau\approx 2.0$; (c) $t/\tau\approx 4.0$; (d) $t/\tau\approx 6.0$.}
\end{figure*}

\begin{figure*}[!h]
    \centering
    \subfloat{
    \includegraphics[width=6.4cm,height = 4.8cm]{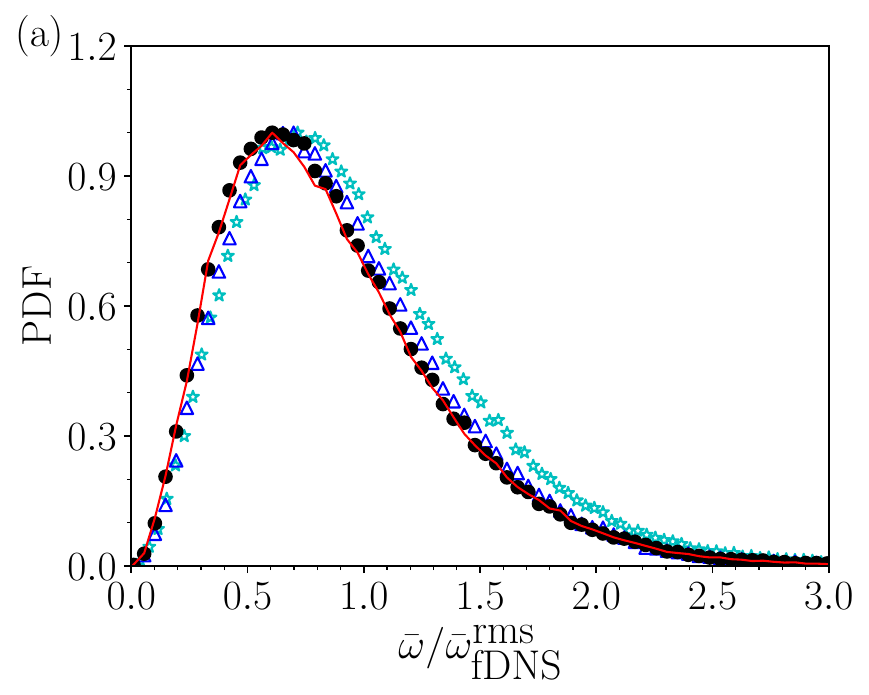}}
    \subfloat{
    \includegraphics[width=6.4cm,height = 4.8cm]{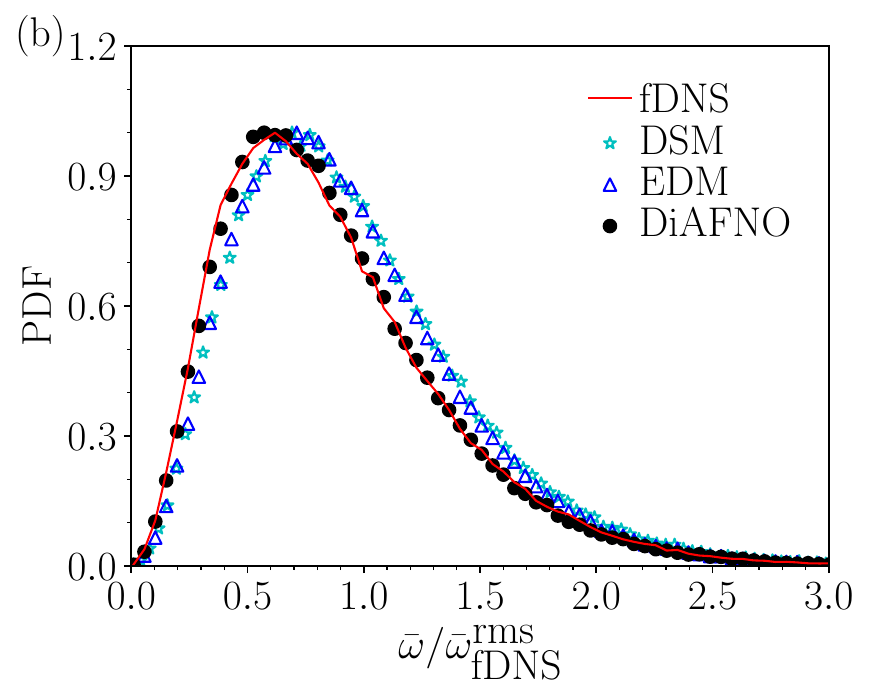}}
    \\
    \subfloat{
    \includegraphics[width=6.4cm,height = 4.8cm]{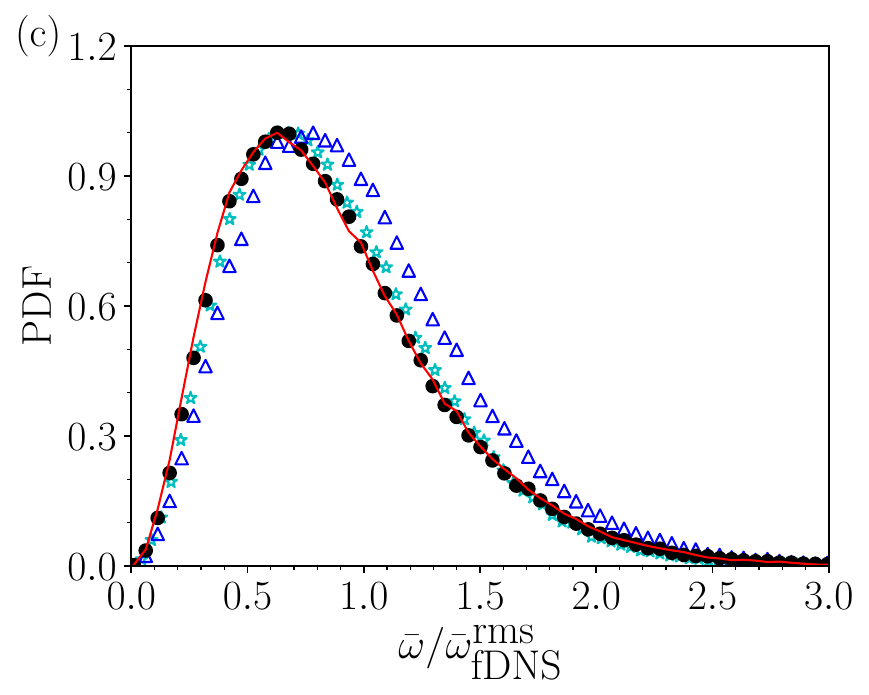}}
    \subfloat{
    \includegraphics[width=6.4cm,height = 4.8cm]{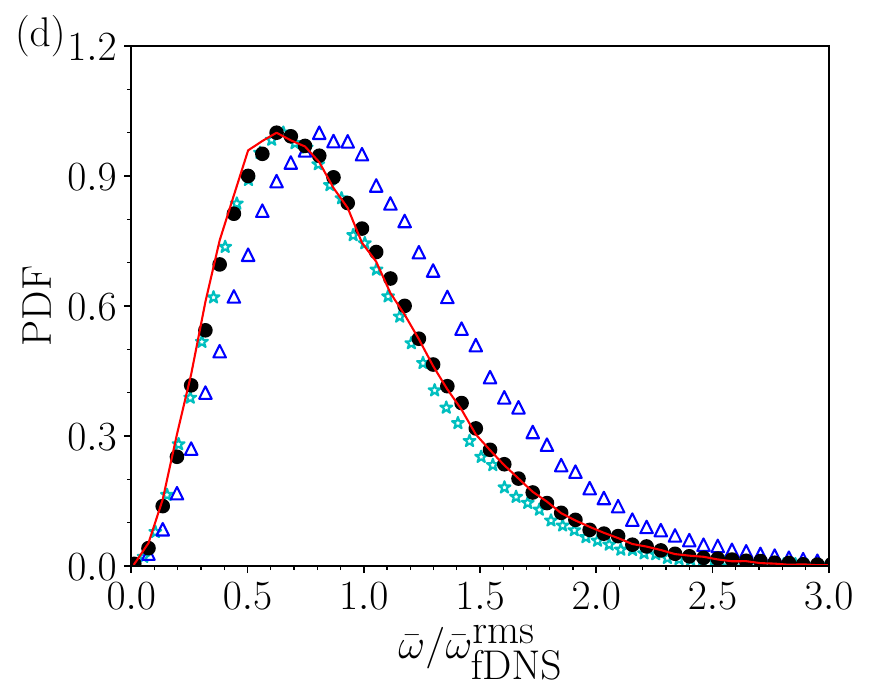}}
    \caption{\label{fig:dHITvor} The PDFs of the normalized vorticity $\bar{\omega} / \bar{\omega}^{\textrm{rms}}_{\textrm{fDNS}}$ of various models in the decaying HIT at different time instants: (a) $t/\tau\approx 1.0$; (b) $t/\tau\approx 2.0$; (c) $t/\tau\approx 4.0$; (d) $t/\tau\approx 6.0$.}
\end{figure*}

\begin{figure*}[!h]
    \centering
    \subfloat{
    \includegraphics[width=3cm,height=3cm]{ModelStructure/hit.pdf}}
    \subfloat{
    \includegraphics[width=12cm,height=8cm]{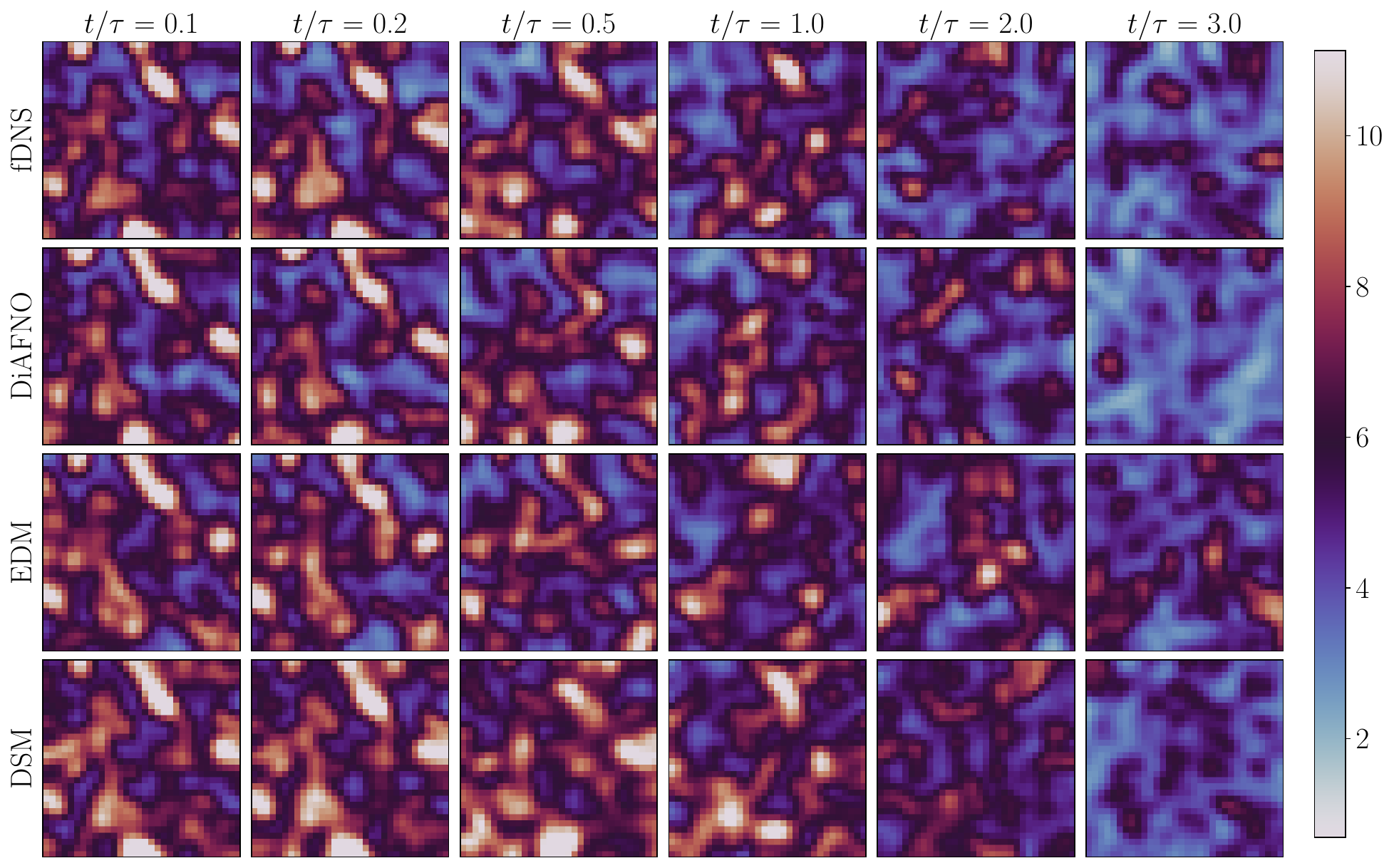}}
    \caption{\label{fig:dHITContour} Contour of vorticity $\bar{\omega}$ on the xy-plane in the middle of the z-axis at different time instants for decaying HIT.}
\end{figure*}
\begin{multicols}{2}


\subsubsection{Decaying homogeneous isotropic turbulence}

The data-driven models use five different initial flow fields to predict 60 steps for each initial field, which is corresponding to 6 large-eddy turnover times. The physical statistics obtained from post-processing these five sets of different prediction data are ensemble averaged.


We present the temporal evolutions of the rms values of velocity and vorticity of various models in the decaying HIT in Fig.~\ref{fig:dHITrms}. From Fig.~\ref{fig:dHITrms}(a), it is observed that both DiAFNO and DSM predict velocity rms value very accurately before $t/\tau\leq 2.0$. When $t/\tau> 2.0$, DiAFNO exhibits a slight deviation while DSM continues to provide accurate predictions. The rms velocity and rms vorticity predicted by EDM exhibit significant deviations as shown in Fig.~\ref{fig:dHITrms}(a)(b), indicating that EDM performs the worst. As shown in Fig.~\ref{fig:dHITrms}(b), DiAFNO gives very accurate results for vorticity rms value when $t/\tau\leq2.4$. However, DiAFNO exhibits another slight deviation when $t/\tau> 3.0$. DSM predicts an increase in vorticity rms value at the initial prediction steps. As the flow field evolves over time ($t/\tau\geq 1.8$), DSM gradually achieves accurate predictions.


In Fig.~\ref{fig:dHITspec}, we present the velocity spectra of various models in the decaying HIT at different time instants. It can be observed that DiAFNO consistently provides accurate prediction results, while DSM's predicted values tend to be slightly lower than the benchmark. When $t/\tau\leq 2.0$, EDM exhibits inaccurate predictions only at a few high wavenumbers. However, for $t/\tau\geq 4.0$, the prediction by EDM become overestimated for the entire wavenumber range.


In Fig.~\ref{fig:dHITvor}, the PDFs of the normalized vorticity $\bar{\omega} / \bar{\omega}^{\textrm{rms}}_{\textrm{fDNS}}$ predicted by various models at different time instants are presented. Here, the vorticity is normalized by the rms values of the vorticity calculated by the fDNS data. These results are consistent with those in Fig.~\ref{fig:dHITrms}(b), where DiAFNO consistently provides accurate predictions while DSM initially shows overestimation. Meanwhile, EDM exhibits a pronounced rightward shift for the PDFs, indicating that its prediction results are inaccurate.


The prediction behaviors of various models are also shown by vorticity contours in Fig.~\ref{fig:dHITContour}. When $t/\tau\leq 1.0$, DSM's results show more red areas ($7\leq\bar{\omega}\leq10$) compared to both data-driven models. When $t/\tau= 3.0$, both DiAFNO and DSM align with fDNS, predominantly displaying blue areas ($2\leq\bar{\omega}\leq5$), while EDM still exhibits a significant number of red regions. Furthermore, the overall similarity between the contour generated by DiAFNO and fDNS is the highest.


\subsubsection{Turbulent channel flow}

\begin{figure*}[!h]
    \centering
    \subfloat{\hspace{-3mm}
    \includegraphics[width=6.4cm,height = 4.8cm]{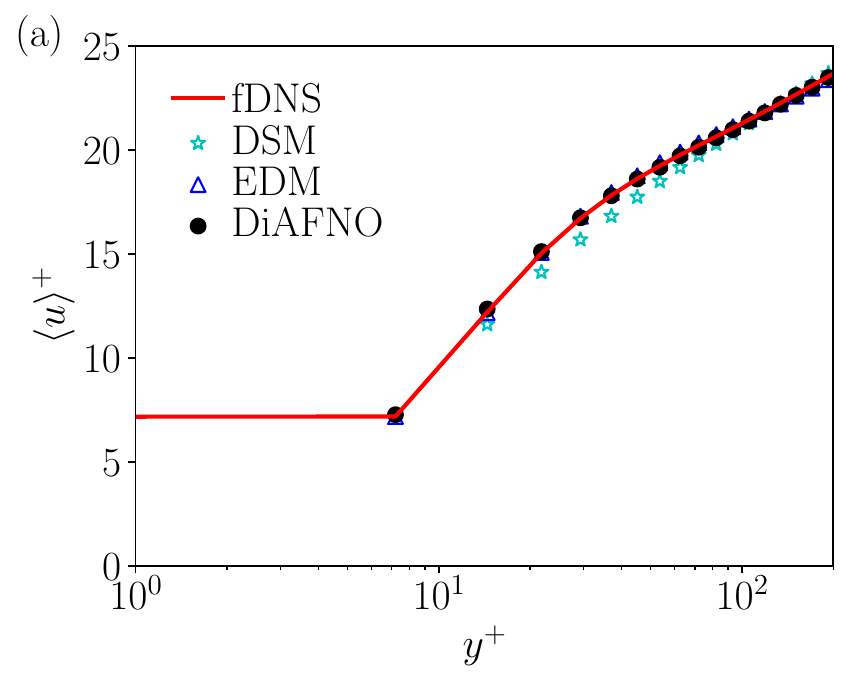}}
    \subfloat{
    \includegraphics[width=6.4cm,height = 4.8cm]{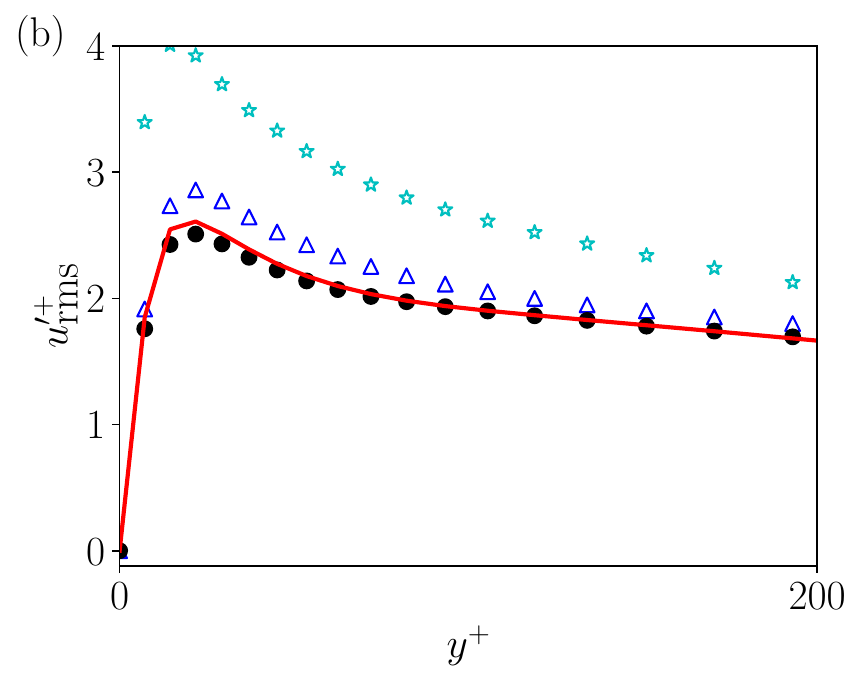}}
    \\
    \subfloat{
    \includegraphics[width=6.4cm,height = 4.8cm]{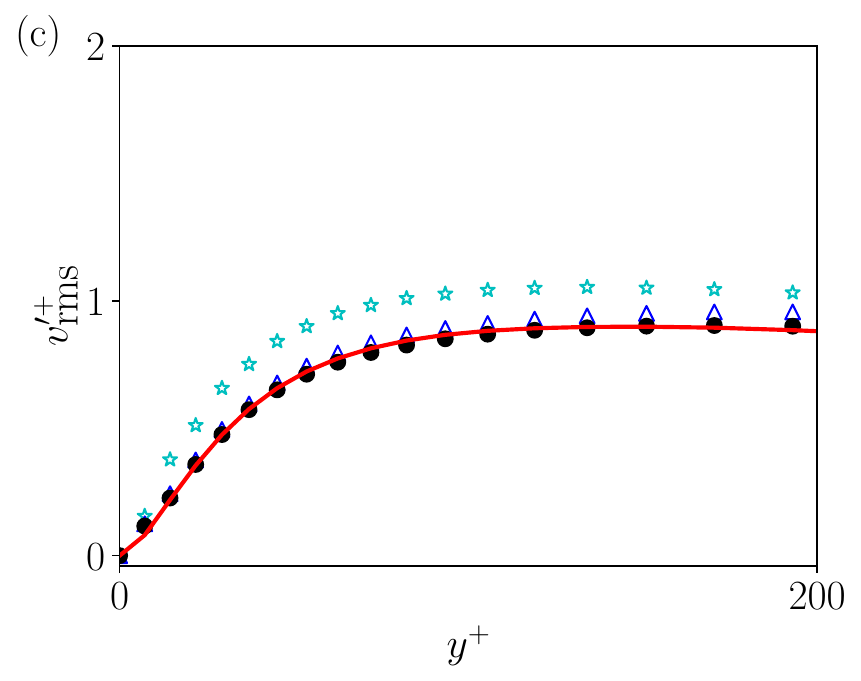}}
    \subfloat{
    \includegraphics[width=6.4cm,height = 4.8cm]{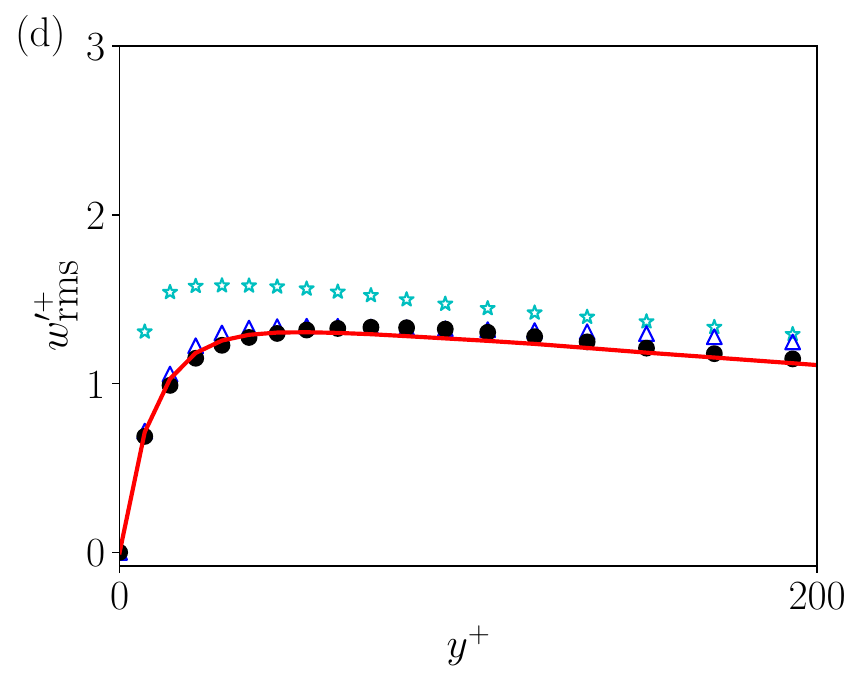}}
    \caption{\label{fig:395vel} The mean streamwise velocity and rms fluctuating velocities at $Re_{\tau}\approx395$: (a) mean  streamwise velocity; (b) rms fluctuation of streamwise velocity; (c) rms fluctuation of transverse  velocity; (d) rms fluctuation of spanwise velocity.}
\end{figure*}

\begin{figure*}[!h]
    \centering
    \subfloat{\hspace{-3mm}
    \includegraphics[width=6.4cm,height = 4.8cm]{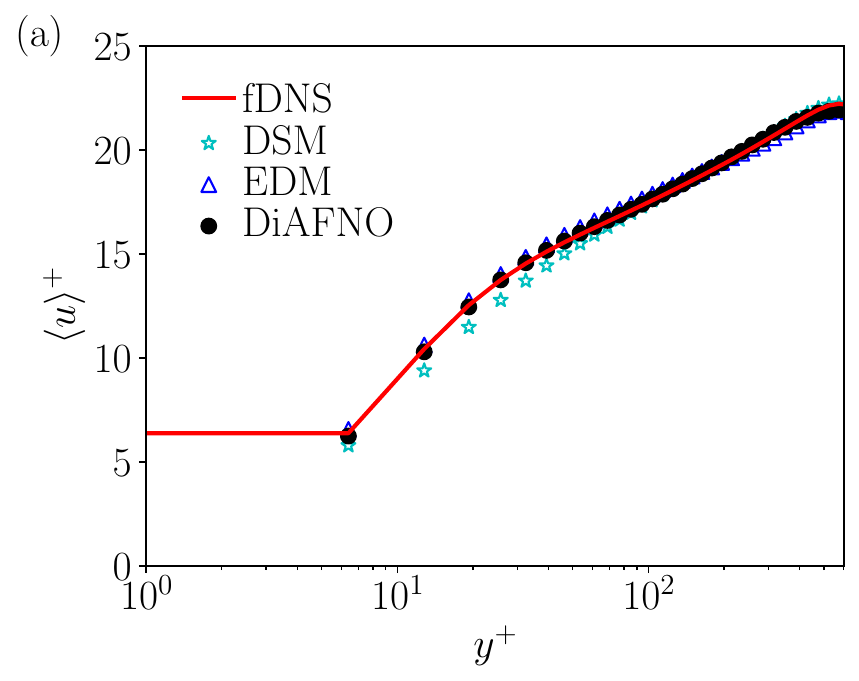}}
    \subfloat{
    \includegraphics[width=6.4cm,height = 4.8cm]{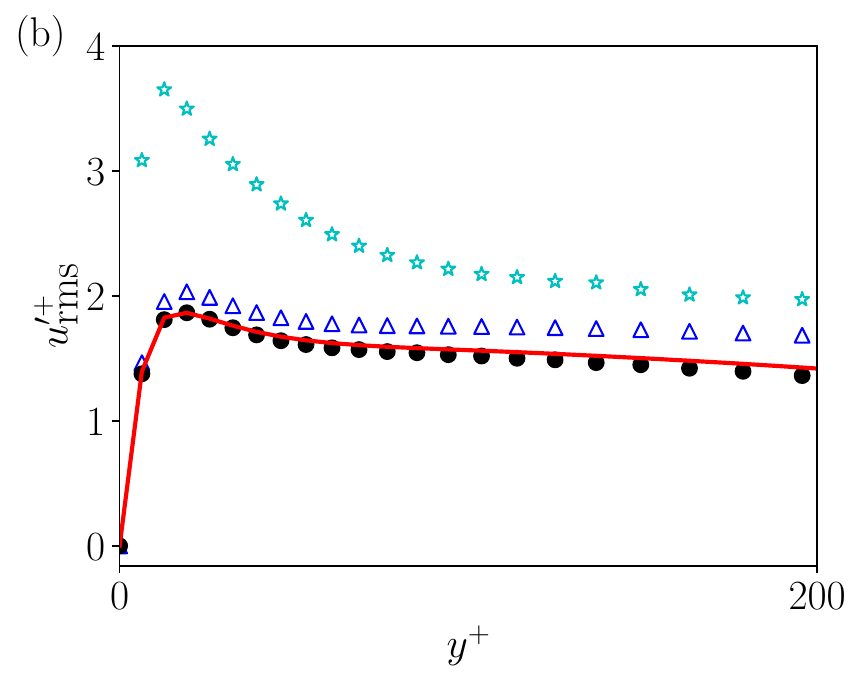}}
    \\
    \subfloat{
    \includegraphics[width=6.4cm,height = 4.8cm]{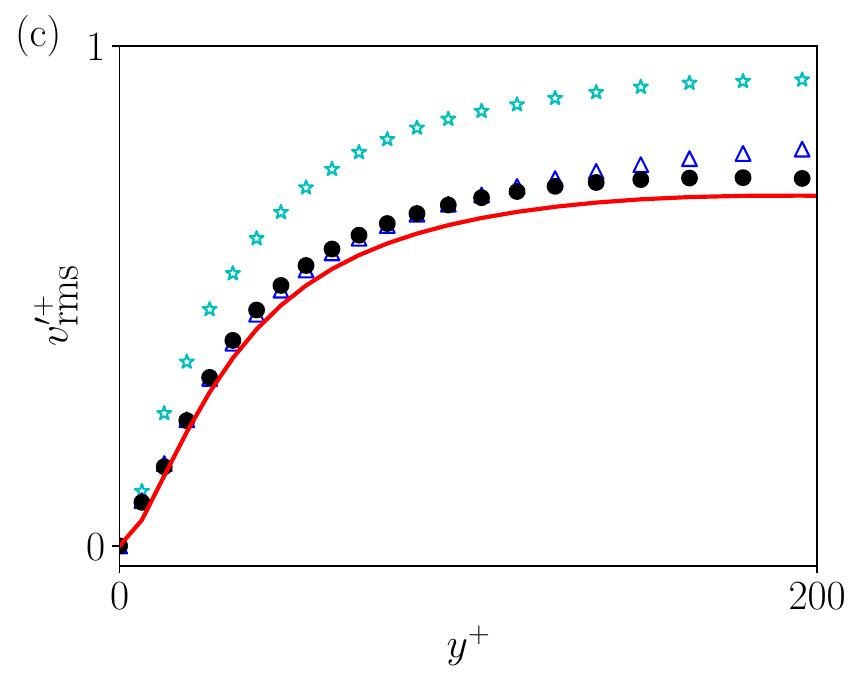}}
    \subfloat{
    \includegraphics[width=6.4cm,height = 4.8cm]{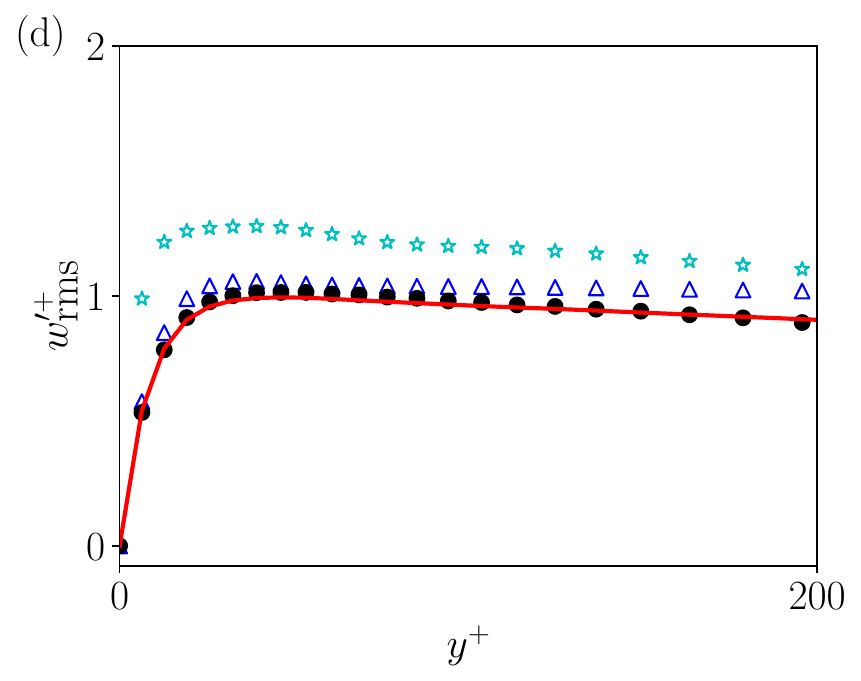}}
    \caption{\label{fig:590vel} The mean streamwise velocity and rms fluctuating velocities at $Re_{\tau}\approx590$: (a) mean  streamwise velocity; (b) rms fluctuation of streamwise velocity; (c) rms fluctuation of transverse  velocity; (d) rms fluctuation of spanwise velocity.}
\end{figure*}

In this subsection, we present the results for turbulent channel flow at two friction Reynolds numbers ($Re_{\tau}\approx395$ and $Re_{\tau}\approx590$). Fig.~\ref{fig:395vel} and Fig.~\ref{fig:590vel} show the mean streamwise velocity and rms fluctuating velocities which all have been time-averaged over the entire prediction period ($t\in[0,400]$) at $Re_{\tau}\approx395$ and $Re_{\tau}\approx590$, respectively.

As shown in Fig.~\ref{fig:395vel}(a) and Fig.~\ref{fig:590vel}(a), the DSM exhibits inaccurate predictions of the mean streamwise velocity near the wall surface ($y^+\in[10,80]$) while both two data-driven models give accurate predictions at any position. Meanwhile, the rms fluctuating velocity in three directions predicted by various models for $Re_{\tau}\approx395$ and $Re_{\tau}\approx590$ are shown in Fig.~\ref{fig:395vel}(b)(c)(d) and Fig.~\ref{fig:590vel}(b)(c)(d), respectively. It can be observed that the predicted rms values of velocity by DiAFNO are in good agreement with fDNS results at both friction Reynolds numbers, while EDM gives moderately accurate results, and DSM produces the least accurate outcomes.

\begin{figure*}[!h]
    \centering
    \subfloat{
    \includegraphics[width=6.4cm,height = 4.8cm]{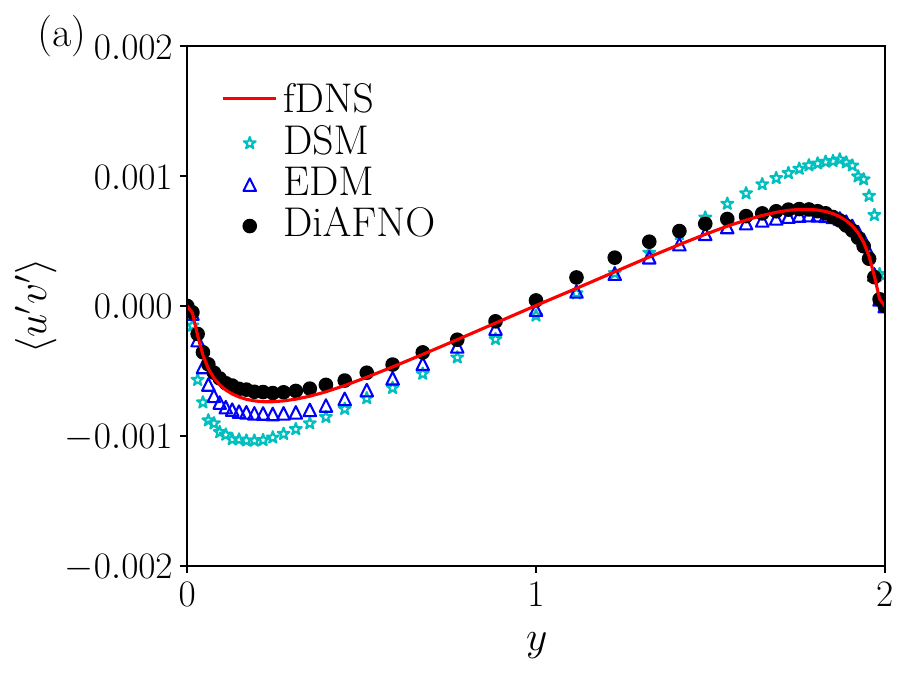}}
    \subfloat{
    \includegraphics[width=6.4cm,height = 4.8cm]{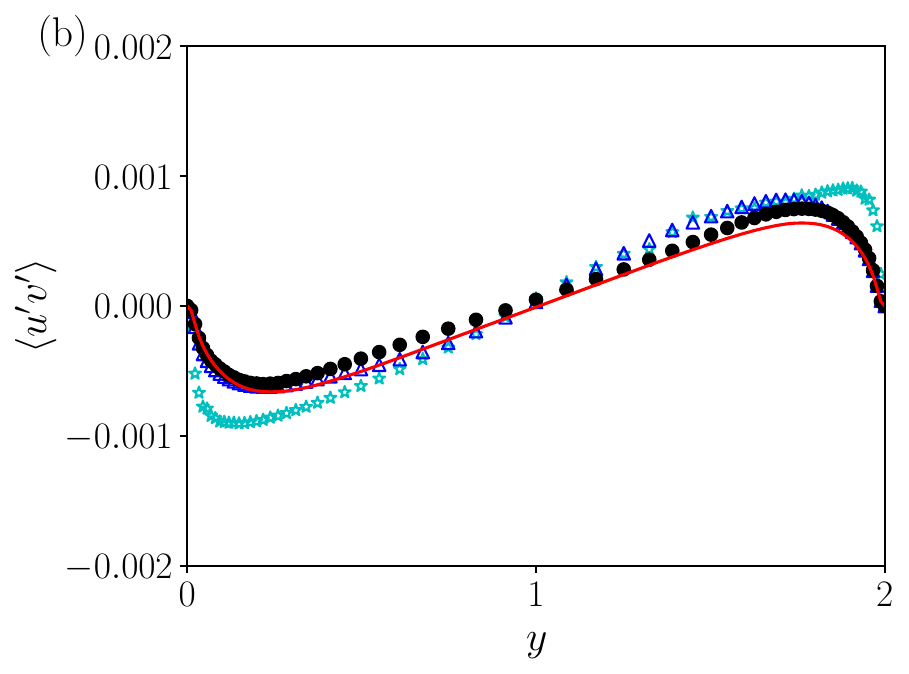}}
    \caption{\label{fig:rnss} The variation of Reynolds shear stress $\langle u'v' \rangle$ at (a) $Re_{\tau}\approx395$; (b) $Re_{\tau}\approx590$.}
\end{figure*}

The Reynolds shear stresses predicted by the DSM and the two data-driven models at $Re_{\tau}\approx395$ and $Re_{\tau}\approx590$ are shown in Fig.~\ref{fig:rnss}(a) and (b), respectively. The Reynolds shear stresses are time-averaged over the entire prediction period ($t\in[0,400]$). The maximum Reynolds shear stresses are observed to be located near the upper and lower walls, where both mean shear effects and velocity fluctuations are pronounced \cite{wang2024prediction}. Between these two peaks, the Reynolds shear stress exhibits an approximate linear dependence on the transverse coordinate which is consistent with the literature \cite{kim1987turbulence}. As shown in Fig.~\ref{fig:rnss}, DiAFNO provides the most accurate results, but some deviations occur near the two extreme points, particularly around the maximum value at $Re_{\tau}\approx590$. For EDM, its prediction results are slightly worse than those of DiAFNO, while DSM exhibits a more significant deviation in a wider range.

\begin{figure*}[!h]
    \centering
    \subfloat{
    \includegraphics[width=6.4cm,height = 4.8cm]{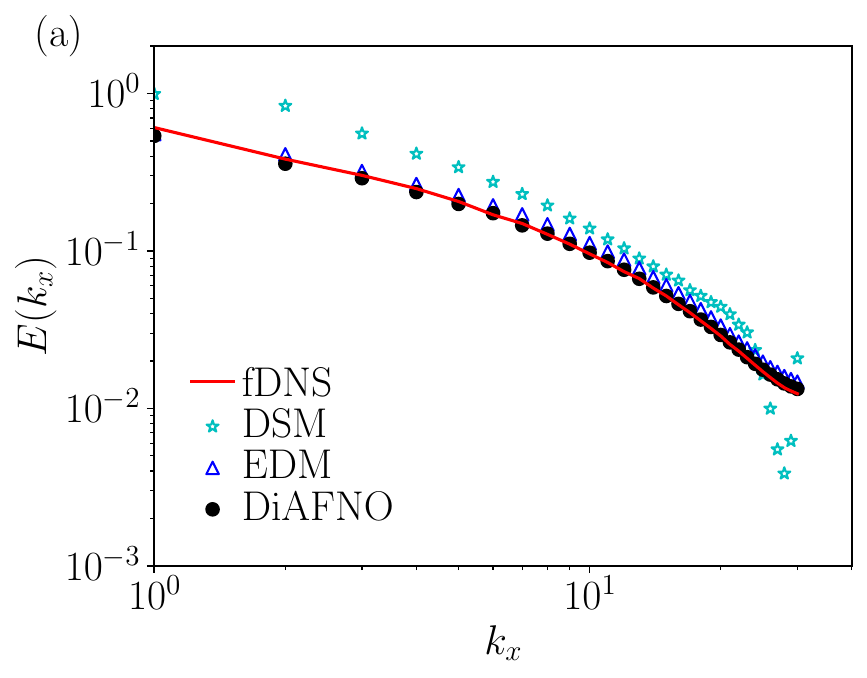}}
    \subfloat{
    \includegraphics[width=6.4cm,height = 4.8cm]{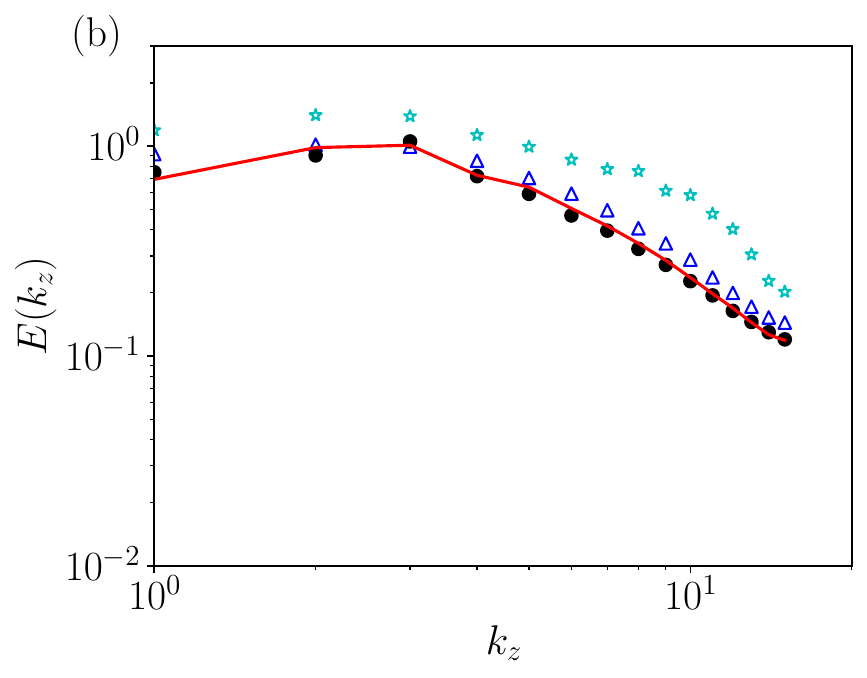}}
    \caption{\label{fig:395spec} Energy spectrum at $Re_{\tau}\approx395$: (a) streamwise spectrum; (b) spanwise spectrum.}
\end{figure*}
\begin{figure*}[!h]
    \centering
    \subfloat{
    \includegraphics[width=6.4cm,height = 4.8cm]{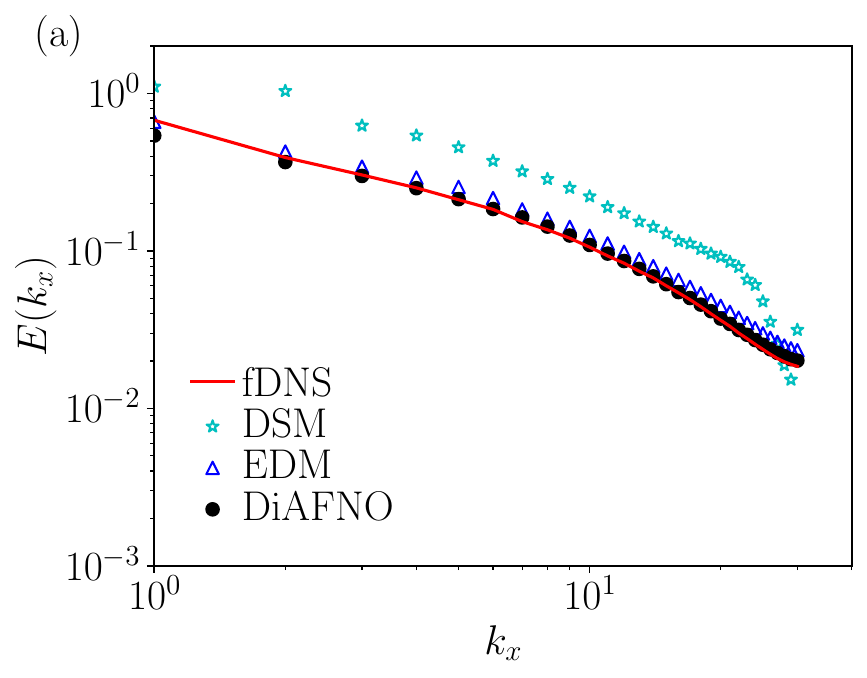}}
    \subfloat{
    \includegraphics[width=6.4cm,height = 4.8cm]{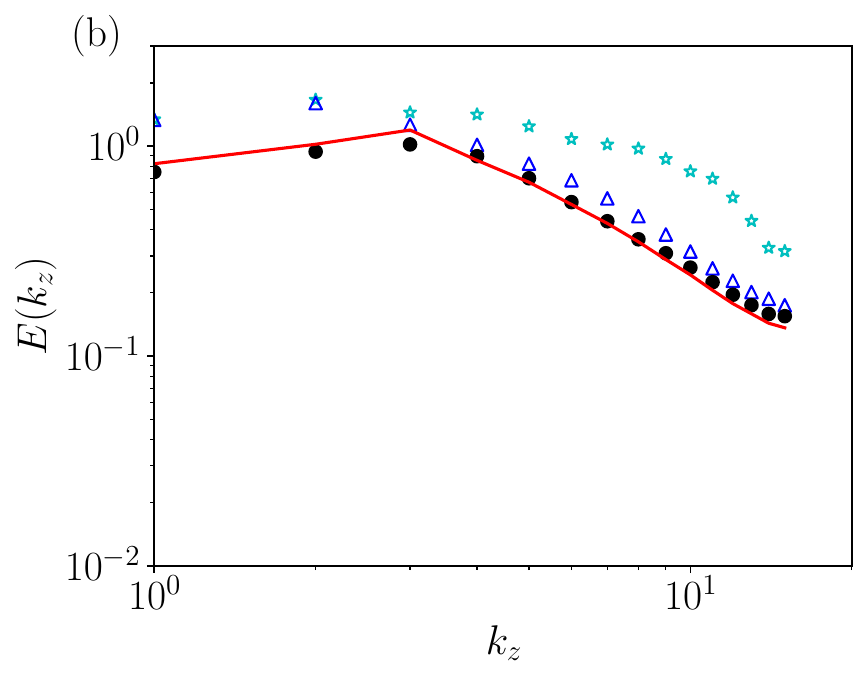}}
    \caption{\label{fig:590spec} Energy spectrum at $Re_{\tau}\approx590$: (a) streamwise spectrum; (b) spanwise spectrum.}
\end{figure*}

To further explore the performance of these models on the prediction ability of the energy distribution, we calculate the kinetic energy spectrum in the streamwise and spanwise directions for $Re_{\tau}\approx395$ and $Re_{\tau}\approx590$ as shown in Fig.~\ref{fig:395spec} and Fig.~\ref{fig:590spec}, respectively. It can be observed that the results from DSM show a significant deviation, while both data-driven models perform well. Upon closer examination of the prediction results from the two data-driven models, it is evident that EDM consistently shows deviations of varying magnitudes, whereas DiAFNO closely aligns with the benchmark. Hence, DiAFNO gives the best results.



\begin{figure*}[!h]
    \centering
    \subfloat{
    \includegraphics[width=3cm,height=3cm]{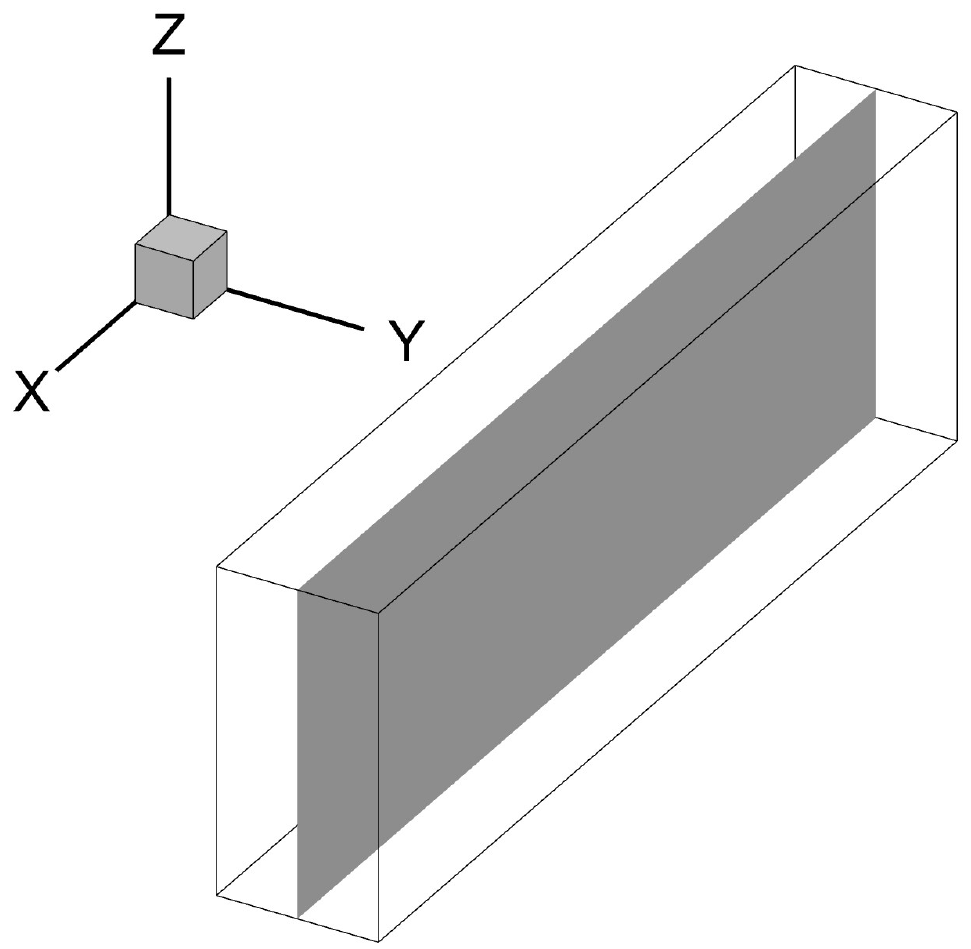}}
    \subfloat{
    \includegraphics[width=12cm,height=8cm]{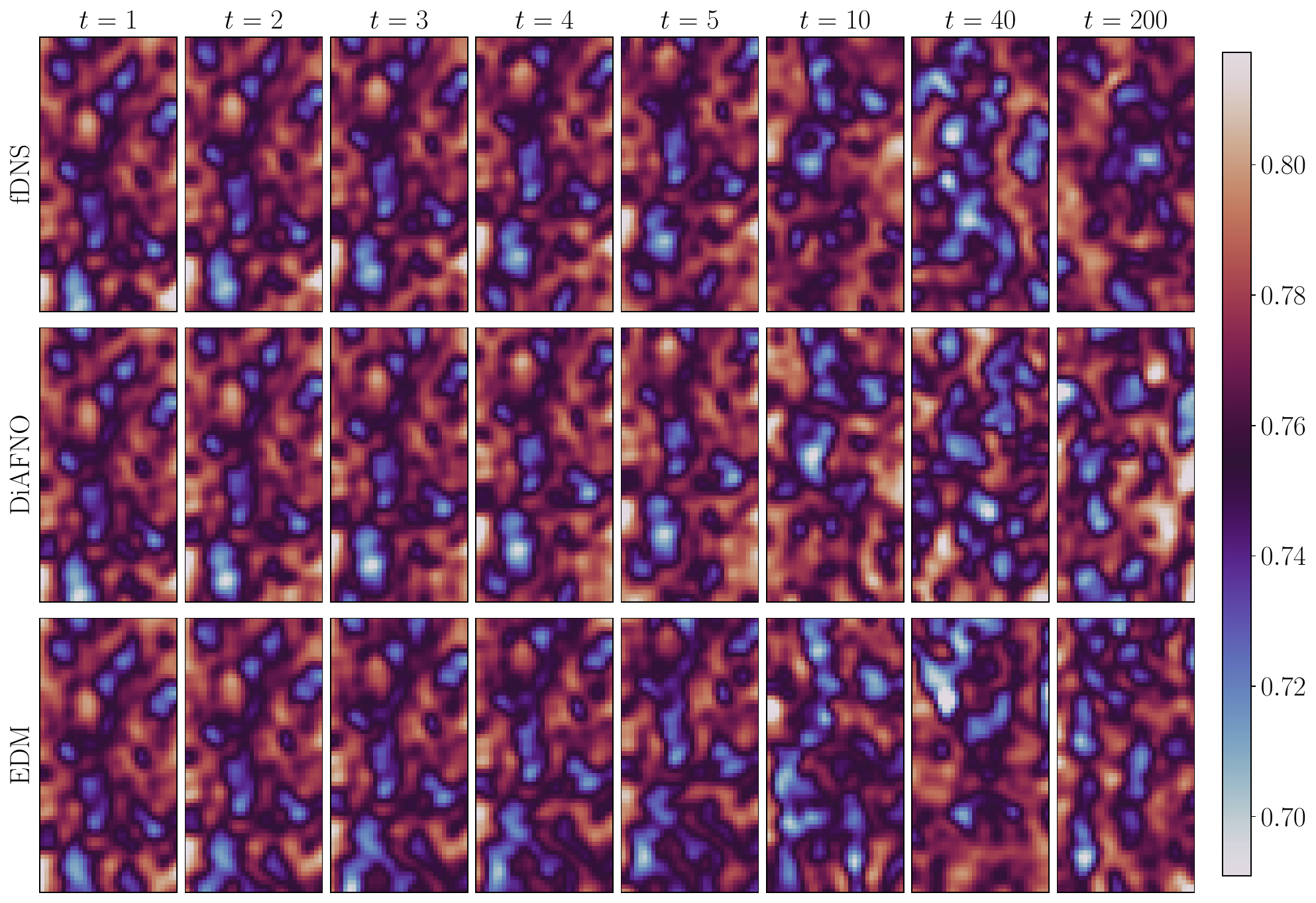}}
    \caption{\label{fig:395Contour} Contour of streamwise velocity $u$ on the zx-plane in the middle of the y-axis at different time instants for turbulent channel flow at $Re_{\tau}\approx395$.}
\end{figure*}

\begin{figure*}[!h]
    \centering
    \subfloat{
    \includegraphics[width=3cm,height=3cm]{ModelStructure/cf.pdf}}
    \subfloat{
    \includegraphics[width=12cm,height=8cm]{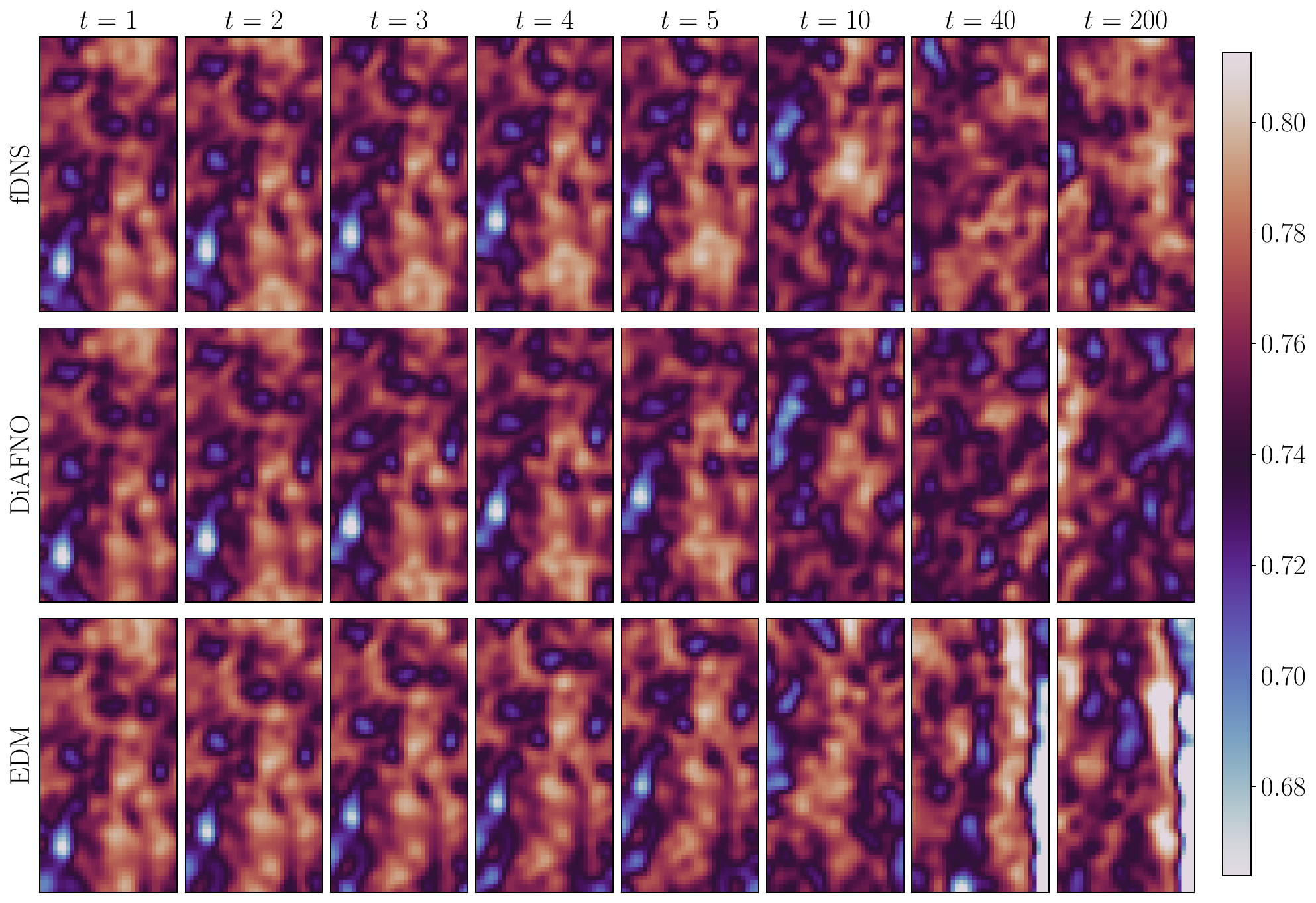}}
    \caption{\label{fig:590Contour} Contour of streamwise velocity $u$ on the zx-plane in the middle of the y-axis at different time instants for turbulent channel flow at $Re_{\tau}\approx590$.}
\end{figure*}

Given that DSM performed poorly in previous comparisons of physical statistics, we only compare DiAFNO, EDM and fDNS in contour results. We present the contour of streamwise velocity $u$ on the zx-plane in the middle of the y-axis at different time instants for turbulent channel flow at $Re_{\tau}\approx395$ and at $Re_{\tau}\approx590$ in Fig.~\ref{fig:395Contour} and Fig.~\ref{fig:590Contour}, respectively. By examining the prediction results for the first five consecutive time steps in both figures, we can observe that both data-driven models successfully predict overall spatial distribution of streamwise velocity. Further examination of the lower end of the EDM results reveals that as time progresses, its divergence from fDNS gradually increases, whereas DiAFNO maintains a better consistency with fDNS. Moreover, when $t\geq40$ in Fig.~\ref{fig:590Contour}, a non-physical results occur at the right hand side of the predictions from EDM.


\end{multicols}

\begin{multicols}{2}

\subsection{Computational costs}

We present the computational costs for the two data-driven models along with the DSM in Tab.~\ref{tab:cost}, including the number of parameters, corresponding GPU memory occupation and the inference time cost of 50 prediction steps (i.e. 10000 DNS advance steps) for forced HIT, 10 prediction steps (i.e. 1000 DNS advance steps) for decaying HIT, 50 prediction steps (i.e. 10000 DNS advance steps) for turbulent channel flow at both $Re_{\tau}\approx395$ and $Re_{\tau}\approx590$. All data-driven models are trained and tested on a NVIDIA A800 80G PCIe GPU (exclusively occupies the entire GPU with no other tasks running concurrently), while the CPU used for loading model parameters and data is Hygon(R) C86 7375 CPU @2.00 GHz. The LES simulations are implemented on a computing cluster, where the type of CPU is Intel(R) Xeon(R) Gold 6148 with 64 cores each @2.40 GHz. Depending on the grid resolution, we employ different numbers of cores for DSM calculations on the computing cluster. For details, please refer to the information in the first column of Tab.~\ref{tab:cost}.

In terms of model parameter counts for these test cases, DiAFNO has fewer parameters than EDM. However, EDM uses less GPU memory than DiAFNO, though both models exhibit very low GPU memory occupation. For the computational efficiency in term of inference time cost, DiAFNO and EDM performed similarly in the two HIT cases. In the two turbulent channel flow cases, EDM is significantly slower than DiAFNO. Since a direct comparison between traditional numerical methods and machine learning models is not fair, the inference time cost we provide only demonstrates that DiAFNO is more efficient than EDM and that the well-trained machine learning models offer efficiency advantages.

\end{multicols}
\begin{table}[!h]
\centering
\caption{\label{tab:cost}Computing costs and computational efficiency of different approaches on different types of turbulent flow.}
\renewcommand{\arraystretch}{1.5}
\begin{tabular}{cccc}
\hline\hline
\mbox{Turbulent flow / Model}&\mbox{Number of parameters ($\times10^6$)}&\mbox{GPU memory occupation (MB)}&\mbox{Inference (s)}\\
\hline
\multicolumn{4}{c}{Forced homogeneous isotropic turbulence}\\
\hline
\mbox{DSM$_{(\times16~\mathrm{cores})}$}&\mbox{N/A}&\mbox{N/A}&\mbox{65.31}\\
\mbox{EDM}&\mbox{6.388}&\mbox{2138}&\mbox{19.84}\\
\mbox{DiAFNO}&\mbox{2.318}&\mbox{2679}&\mbox{19.82}\\
\hline
\multicolumn{4}{c}{Decaying homogeneous isotropic turbulence}\\
\hline
\mbox{DSM$_{(\times16~\mathrm{cores})}$}&\mbox{N/A}&\mbox{N/A}&\mbox{9.340}\\
\mbox{EDM}&\mbox{6.388}&\mbox{2138}&\mbox{3.945}\\
\mbox{DiAFNO}&\mbox{2.318}&\mbox{2679}&\mbox{3.965}\\
\hline
\multicolumn{4}{c}{Turbulent channel flow at $Re_{\tau}\approx395$}\\
\hline
\mbox{DSM$_{(\times32~\mathrm{cores})}$}&\mbox{N/A}&\mbox{N/A}&\mbox{287.2}\\
\mbox{EDM}&\mbox{6.388}&\mbox{4921}&\mbox{162.0}\\
\mbox{DiAFNO}&\mbox{3.884}&\mbox{6843}&\mbox{72.93}\\
\hline
\multicolumn{4}{c}{Turbulent channel flow at $Re_{\tau}\approx590$}\\
\hline
\mbox{DSM$_{(\times64~\mathrm{cores})}$}&\mbox{N/A}&\mbox{N/A}&\mbox{315.3}\\
\mbox{EDM}&\mbox{6.388}&\mbox{6645}&\mbox{212.0}\\
\mbox{DiAFNO}&\mbox{4.621}&\mbox{8625}&\mbox{90.41}\\
\hline\hline
\end{tabular}
\end{table}
\begin{multicols}{2}


\section{Conclusion}
\label{conclusion}

In this work, we propose the DiAFNO model with an autoregressive framework for accurate long-term predictions of 3D turbulence. The proposed DiAFNO is validated by comparing with the EDM and traditional DSM in the large-eddy simulations of three types of 3D turbulence, including forced homogeneous isotropic turbulence (HIT), decaying HIT and turbulent channel flow. The numerical results demonstrate that:

1. Under identical training epochs and sampling efficiency, DiAFNO achieves significantly lower losses than EDM. Here, the DiAFNO employs identical hyperparameters across several distinct numerical tests.

\textcolor{red}{2. DiAFNO yields optimal predictions for most of the analyzed statistics, except for the temporal evolution of velocity rms in decaying HIT.}

\textcolor{red}{3. Ignoring the training costs, our DiAFNO outperforms EDM and DSM-based LES in inference efficiency.}

Our main contributions can be concluded as:

1. We introduce the SOTA elucidated diffusion model's framework to enhance the prediction accuracy of advanced neural operator IAFNO and enable autoregressive prediction via diffusion models based on conditional generation.

2. We replace the U-Net backbone with more advanced neural operator IAFNO, which effectively improves the accuracy and efficiency of the denoising process in diffusion models.

While our proposed DiAFNO achieves stable, accurate and autoregressive long-term prediction of 3D turbulent flows in this study, one notable limitation lies in its heavy reliance on data. In recent years, physics-informed approaches such as PINNs \cite{raissi2019physics} become very popular, and have been employed to enhance operator learning performance and reduce data dependence by embedding PDEs into loss functions. Such methods include physics-informed DeepONets \cite{wang2021learning,wang2023long}, physics-informed Fourier neural operators \cite{li2024physics,zanardi2023adaptive} and physics-informed transformers \cite{lorsung2024physics,zhao2023pinnsformer}. However, since diffusion models require the introduction of noise addition and removal processes, it is quite challenging to use physics-informed constraints to reduce the reliance on data. However, the performance of diffusion models can be enhanced through physics-informed approaches: Soni et al. improved the performance of diffusion models by introducing a weighted physics-informed loss during model training \cite{soni2025physics}. Zeng et al. introduced a physics-informed conditional embedding module in diffusion process, which ensures the training consistency with physical laws \cite{zeng2025physicsinformed}.

Moreover, we trained separate models for each dataset while keeping the hyperparameters constant. At high Reynolds numbers, complex flow structures requires high-resolution data to ensure accurate analysis of the flow. Consequently, for all machine learning models, the higher the Reynolds number, the more difficult it is to accurately predict turbulence, and the models exhibit weaker long-term stability and lack generalization ability. In addition, the vast majority of machine learning models require retraining for different turbulent flows. Endowing models with generalization capabilities is a topic that requires separate research. The DiAFNO model we propose has not undergone such research or testing. Thus, generalization capabilities will be one of the important research directions in the future.

Furthermore, we only validated the DiAFNO model on simple flows, whereas engineering applications often involve diverse complex geometries and irregular boundaries. Generating sufficient accurate data for training diffusion models under such complex flows also presents a great challenge. Thus, it is crucial to test and enhance machine learning models’ capability to handle relatively complex flow fields with parameterized boundary conditions, variable geometries and diverse Reynolds numbers \cite{gao2021phygeonet,li2023fourier,wu2024transolver,gao2026awavelet,meng2026research}.

\appendix
\section{The Fourier neural operator}
\label{app1}
\setcounter{equation}{0}
\renewcommand{\theequation}{A.\arabic{equation}}

\begin{figure*}[htb]
    \centering
    \includegraphics[width=16cm,height=9cm]{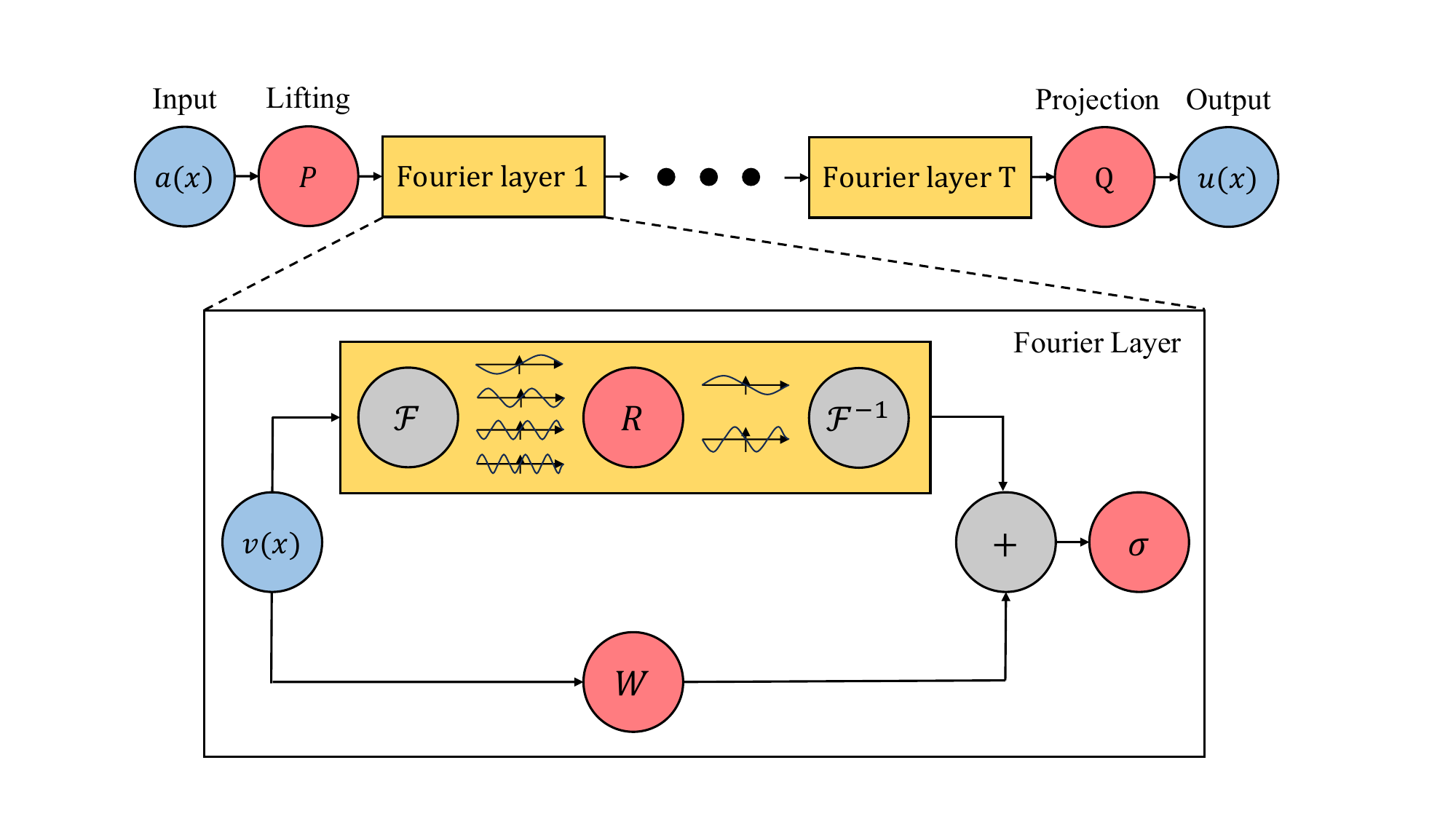}
    \caption{\label{fig:ModelStructureFNO} The architecture of FNO.}
\end{figure*}

The FNO learns a non-linear mapping between two infinite dimensional spaces from a finite collection of observed input-output pairs \cite{li2020fourier,kovachki2023neural}:

\begin{equation}
G: \mathcal{A}\times\Theta\rightarrow\mathcal{U},~~\text{or equivalently}~~G_{\theta}:\mathcal{A}\rightarrow\mathcal{U},~~\theta\in\Theta, \label{eq:FNO}
\end{equation} where $\mathcal{A}=\mathcal{A}(D;\mathbb{R}^{d_a})$ and $\mathcal{U}=\mathcal{U}(D;\mathbb{R}^{d_u})$ are separable Banach spaces consisting of functions valued in $\mathbb{R}^{d_a}$ and $\mathbb{R}^{d_u}$ respectively, with $D\subset\mathbb{R}^d$  being a bounded open set. This nonlinear mapping is parameterized by $\theta\in\Theta$, enabling Fourier neural operators to learn an approximation of the operator mapping $\mathcal{A}\rightarrow\mathcal{U}$. The optimal parameters $\theta\dagger\in\Theta$ are obtained via a data-driven approximation \cite{li2023long}. The FNO architecture is shown in Fig.~\ref{fig:ModelStructureFNO} which consists of three main steps \cite{li2020fourier}:


(1) The input $a\in\mathcal{A}$ is lifted to a higher dimension space by a fully connected layer: $v_0(x)=P(a(x))$.

(2) The calculations in higher dimension channel space is given by:

\begin{equation}
v_{t+1}:=\sigma\left( Wv_t(x)+\left( \mathcal{K}(a;\phi)v_t \right)(x) \right) ,~~\forall x\in D, \label{eq:iterFNO}
\end{equation} where $\mathcal{K}:\mathcal{A}\times\Theta_{\mathcal{K}}\rightarrow\mathcal{L}(\mathcal{U}(D;\mathbb{R}^{d_v}),\mathcal{U}(D;\mathbb{R}^{d_v}))$ maps to bounded linear operators on $\mathcal{U}(D;\mathbb{R}^{d_v})$ and is parameterized by $\phi \in \Theta_{\mathcal{K}}$. Here, $W:\mathbb{R}^{d_v}\rightarrow\mathbb{R}^{d_v}$ is a linear transformation, and $\sigma: \mathbb{R}\rightarrow\mathbb{R}$ is a component-wise non-linear activation function. The kernel integral operator mapping in Eq.~\ref{eq:iterFNO} is defined by \cite{li2020fourier}: 

\begin{equation}
\left( \mathcal{K}(a;\phi)v_t \right)(x):=\int_D\kappa\left( x,y,a(x),a(y);\phi \right)v_t(y)\mathrm{d}y, \label{eq:kernelFNO}
\end{equation} where $\kappa_{\phi}:=\mathbb{R}^{2(d+d_a)}\rightarrow\mathbb{R}^{d_v\times d_v}$ is a neural network parameterized by $\phi \in \Theta_{\mathcal{K}}$. Here, $\kappa_{\phi}(x,y)=\kappa_{\phi}(x-y)$ is imposed, now Eq.~\ref{eq:kernelFNO} becomes a convolution operator. Let $\mathcal{F}$ denotes the Fourier transform of a function $f:D\rightarrow\mathbb{R}^{d_v}$ and $\mathcal{F}^{-1}$ its inverse, we have \cite{li2020fourier}:

\begin{equation}
\left( \mathcal{K}(a;\phi)v_t \right)(x):=\mathcal{F}^{-1}\left( 
\mathcal{F}(\kappa_{\phi})\cdot\mathcal{F}(v_t) \right)(x). \label{eq:fftFNO}
\end{equation}

We then parameterize $\kappa_{\phi}$ in Fourier space:

\begin{equation}
v_{t+1}:=\sigma\left( Wv_t(x)+\mathcal{F}^{-1}\left( 
R_{\phi}\cdot\mathcal{F}(v_t) \right)(x) \right), \label{eq:fullFNO}
\end{equation} where $R_{\phi}$ denotes the Fourier transform of a periodic function $\kappa:\bar{D}\rightarrow\mathbb{R}^{d_v\times d_v}$ parameterized by $\phi \in \Theta_{\mathcal{K}}$, with $k\in \mathbb{Z}^d$ representing the frequency mode. Finite-dimensional parametrization is achieved by truncating the Fourier series at a maximum number of modes $k_{\mathrm{max}}=|Z_{k_{\mathrm{max}}}|=|\{k\in \mathbb{Z}^d:|k_j|\leq k_{\mathrm{max},j},\mathrm{for}j=1,\dots,d\}|$ \cite{li2020fourier}.

For a function $v_t\in \mathbb{R}^{n\times d_v}$, its Fourier transform $\mathcal{F}(v_t)\in \mathbb{C}^{n\times d_v}$ can be obtained by discretizing domain $D$ with $n\in \mathbb{N}$ points. By simply truncating the higher modes,  $\mathcal{F}(v_t) \in \mathbb{C}^{k_{\mathrm{max}}\times d_v}$ can be obtained, here $\mathbb{C}$ is the complex space. $R_{\phi}$ is parameterized as complex-valued-tensor $(k_{\mathrm{max}}\times d_v \times d_v)$ containing a collection of truncated Fourier modes  $R_{\phi} \in \mathbb{C}^{k_{\mathrm{max}}\times d_v\times d_v}$ \cite{li2020fourier}. Therefore, by multiplying $R_{\phi}$ and $\mathcal{F}(v_t)$, it can be derived that:

\begin{equation}
\left( 
R_{\phi}\cdot\mathcal{F}(v_t) \right)_{k,l}=\sum_{j=1}^{d_v}R_{\phi k,l,j}(\mathcal{F}v_t)_{k,j}, \label{eq:discretFNO}
\end{equation} where $k=1,\dots,k_{\mathrm{max}},~~j=1,\dots,d_v$.

(3) The output $u\in \mathcal{U}$ is obtained by $u(x) = Q(v_T(x))$, where $Q:\mathbb{R}^{d_v}\rightarrow\mathbb{R}^{d_u}$ is the projection of $v_T$ and it is parameterized by a fully connected layer.

\end{multicols}
\begin{multicols}{2}

\section{The adaptive Fourier neural operator}
\label{app2}
\setcounter{equation}{0}
\renewcommand{\theequation}{B.\arabic{equation}}

To start with, AFNO introduces the self-attention mechanism, which is defined by \cite{tsai2019transformer,kovachki2023neural,guibas2021adaptive}:

\begin{equation}
\textbf{q}=XW_q,~\textbf{k}=XW_k,~\textbf{v}=XW_v ~, \label{eq:selfattention1}
\end{equation}
\begin{equation}
\mathrm{Att}(X):=\mathrm{softmax}\left( \frac{\textbf{q}\textbf{k}^{\mathrm{T}}}{\sqrt{d}} \right)\textbf{v} ~, \label{eq:selfattention2}
\end{equation} where $\textbf{q}, \textbf{k}, \textbf{v} \in \mathbb{R}^{N\times d}$ are the query, key and value vectors, respectively. $W_q, W_k, W_v \in \mathbb{R}^{d\times d}$ are corresponding weight matrices.

\begin{figure*}[!h]
    \centering
    \includegraphics[width=16cm,height=9cm]{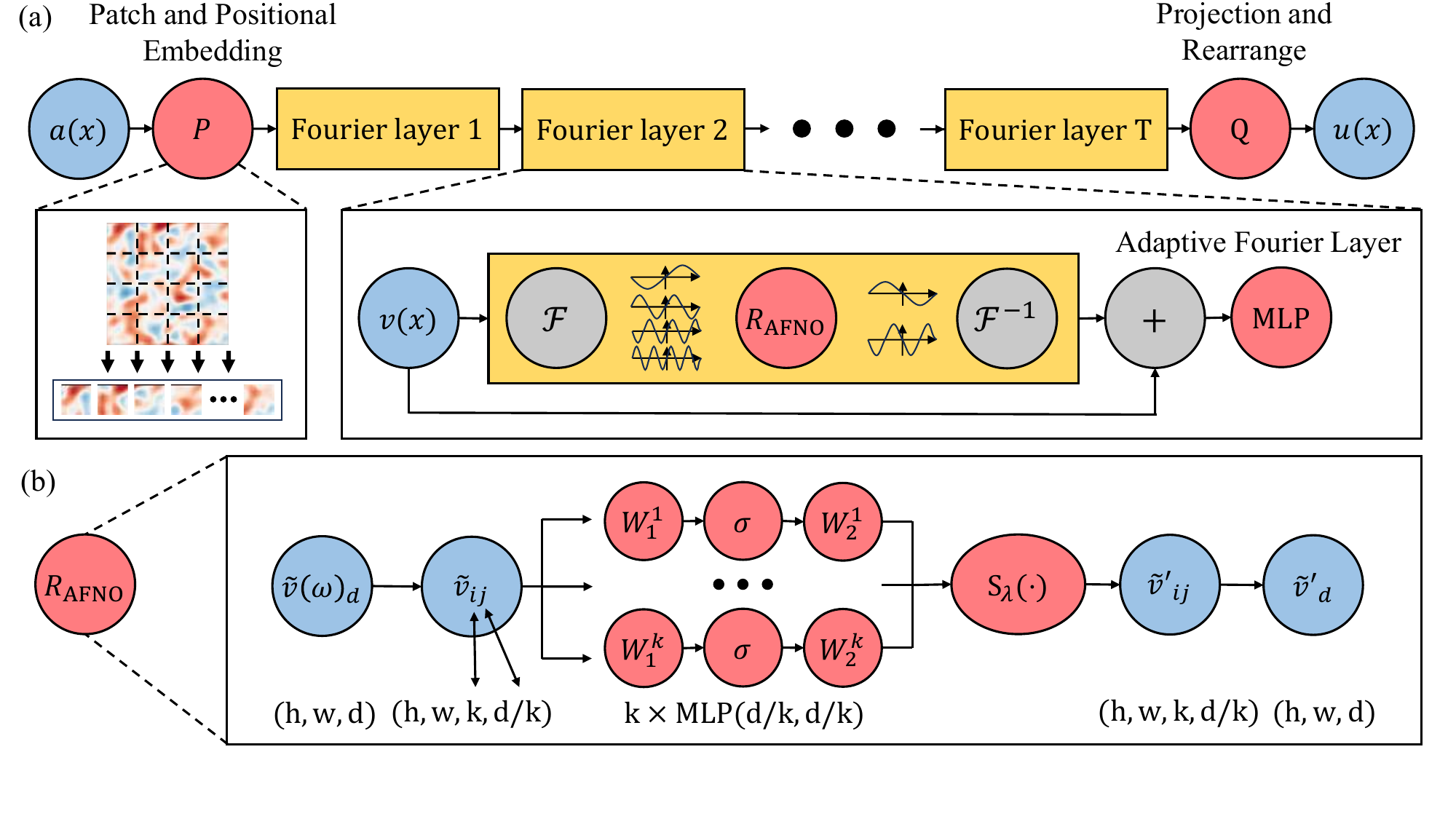}
    \caption{\label{fig:ModelStructureAFNO} The architecture of AFNO.}
\end{figure*}

As shown in Fig.~\ref{fig:ModelStructureAFNO}, input tensor $X$ can be denoted as $\tilde{v}(\omega)$ with a tensor size of $[h,w,d]$. $h$ is the height and $w$ is the width of an image, and $d$ is the channel width. We define $N:=hw$ and index the array (tensor) $X$ as a token sequence which has the form of $X[s]=X[n_s,m_s]$ for some discrete index $s$ and $t$ where $s,t\in[hw]$. Therefore $X\in \mathbb{R}^{N\times d}$ \cite{jiang2026animplicit}. Hence, the Eq.~\ref{eq:selfattention2} represents a kernel integration of $\mathbb{R}^{N\times d}\rightarrow\mathbb{R}^{N\times d}$ \cite{guibas2021adaptive}.


Furthermore, we define $K:=\mathrm{softmax}(\langle XW_q, XW_k \rangle/\sqrt{d})$ as the $N\times N$ score matrix, where $\langle \cdot,\cdot\rangle$ denotes the inner product in $\mathbb{R}^d$. We then formulate self-attention as an asymmetric matrix-valued kernel $\kappa:[N]\times[N]\rightarrow\mathbb{R}^{d\times d}$ parameterized as $\kappa[s,t]=K[s,t]\cdot W_v$ (where $K[s,t]$ is scalar valued and ``$\cdot$'' is scalar-matrix multiplication) \cite{jiang2026animplicit}. Thus, the self-attention mechanism can be re-expressed in the alternative kernel summation form \cite{guibas2021adaptive} as:

\begin{equation}
\mathrm{Att}(X)[s]:=\sum_{t=1}^{N}X[t]\kappa[s,t],~~\forall s\in [N] ~, \label{eq:SAsumform}
\end{equation} where $t$, as a discrete index, can be extended to a continuous infinitesimal change $\mathrm{d}t$ in integral computations. Following this discrete-to-continuous extension, the input tensor $X$ is no longer a finite-dimensional vector in the Euclidean space $X\in \mathbb{R}^{N\times d}$, but rather a spatial function in the function space $X\in (D,\mathbb{R}^{d})$ defined on domain $D\subset \mathbb{R}^2$ \cite{jiang2026animplicit}. In this continuum framework, the neural network is transformed into an operator acting on input functions, and thus the kernel integral operator $\mathcal{K}:(D,\mathbb{R}^{d})\rightarrow(D,\mathbb{R}^{d})$ is defined as \cite{guibas2021adaptive}:

\begin{equation}
\mathcal{K}(X)(s)=\int_{D}\kappa(s,t)X(t)\mathrm{d}t,~~\forall s\in D ~, \label{eq:SAintegralform}
\end{equation} with a continuous kernel function $\kappa:D\times D\rightarrow\mathbb{R}^{d\times d}$ \cite{li2020neural}. For Green’s kernel $\kappa(s,t)=\kappa(s-t)$, the integral leads to global convolution defined \cite{guibas2021adaptive}:

\begin{equation}
\mathcal{K}(X)(s)=\int_{D}\kappa(s-t)X(t)\mathrm{d}t,~~\forall s\in D ~. \label{eq:SAconvolutionform}
\end{equation} With FFT and inverse FFT, Eq.~\ref{eq:SAconvolutionform} becomes:

\begin{equation}
\mathcal{K}(X)(s)=\mathcal{F}^{-1}\left(\mathcal{F}(\kappa)\cdot\mathcal{F}(X)\right)(s),~~\forall s\in D ~, \label{eq:SAfftform}
\end{equation} which coincides with Eq.~\ref{eq:fftFNO}.

We now introduce the block-diagonal structure on $W$ and formulated the multiply frequency process shown in Fig.~\ref{fig:ModelStructureAFNO}(b) as:

\begin{equation}
\tilde{v}'_{d}=\tilde{v}_{ij}^{'(l)}=R_{\mathrm{AFNO}}\cdot\tilde{v}_{d}:=S_{\lambda}\left[W_2^{(l)}\sigma\left(W_1^{(l)}\tilde{v}_{ij}^{(l)}\right)\right], \label{eq:SAmf}
\end{equation} where $l=1,...,k$ and the weight matrix $W$ is divided into $k$ weight blocks of size $d/k\times d/k$. This block-diagonal weight methodology enables computational parallelism, where each block can be regarded as a head in multi-head self-attention that projects the data into a subspace \cite{guibas2021adaptive}. Accordingly, the dimension of $W_1$ is scaled from $(d/k,d/k)$ to $(d/k,f\times d/k)$, while the dimension of $W_2$ is scaled from $(d/k,d/k)$ to $(f\times d/k,d/k)$, $f$ is a scaling parameter. To sparsify the tokens, soft-thresholding (shrinkage) operation: $S_{\lambda}[x]=\mathrm{sign}(x)\mathrm{max}\{|x|-\lambda,0\}$ is employed, where $\lambda$ is a tuning parameter regulating the level of sparsity \cite{tibshirani1996regression,guibas2021adaptive}.

Hence, the AFNO architecture can be described as:

\begin{equation}
v_{t+1}=\mathrm{MLP}\left[ v_{t}+\mathcal{F}^{-1}\left( 
R_{\mathrm{AFNO}}\cdot\mathcal{F}(v_t) \right)(s) \right] ,~~\forall s\in D ~. \label{eq:fullAFNO}
\end{equation}

\end{multicols}
\begin{multicols}{2}

\section{Further analysis}
\label{app3}

\subsection{Comparison between IAFNO and DiAFNO}
\label{app31}

We conducted additional tests of IAFNO on all datasets with consistent training data, time-step interval, and grid resolution. \textcolor{red}{All models have been fine-tuned with the same effort to ensure optimal hyperparameter settings. We have not changed any of the IAFNO model’s settings and the IAFNO results presented in the paper are essentially consistent with those presented in \cite{jiang2026animplicit}.} The point-to-point errors for flow field predictions on the test sets are shown in Tab.~\ref{tab:AppClosses}.

\end{multicols}
\begin{table}[htb]
\centering
\caption{\label{tab:AppClosses}Comparison of minimum testing loss (point-to-point error) of different data-driven models in three types of turbulent flows.}
\begin{tabular}{cccc}
\hline\hline
\mbox{~~}&\mbox{IAFNO~~}&\mbox{EDM~~}&\mbox{DiAFNO}\\
\hline
\mbox{HIT}&\mbox{0.1754}&\mbox{0.0384}&\mbox{\textbf{0.0288}}\\
\mbox{dHIT}&\mbox{0.2012}&\mbox{0.0235}&\mbox{\textbf{0.0169}}\\
\mbox{CF395}&\mbox{0.1758}&\mbox{0.0501}&\mbox{\textbf{0.0410}}\\
\mbox{CF590}&\mbox{0.2170}&\mbox{0.0555}&\mbox{\textbf{0.0485}}\\
\hline\hline
\end{tabular}
\end{table}

\begin{figure*}[!h]
    \centering
    \subfloat{\hspace{-3mm}
    \includegraphics[width=6.4cm,height = 4.8cm]{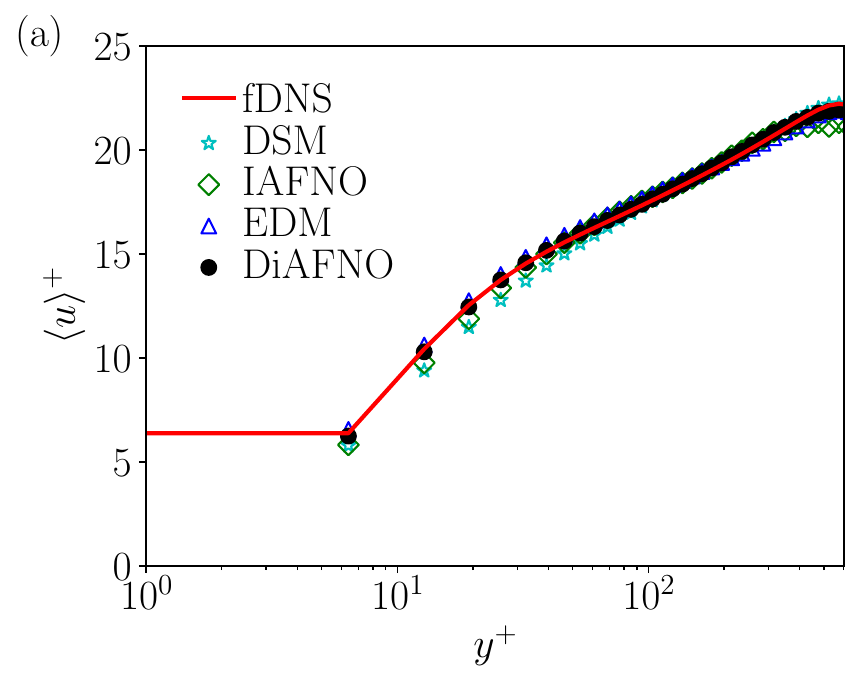}}
    \subfloat{
    \includegraphics[width=6.4cm,height = 4.8cm]{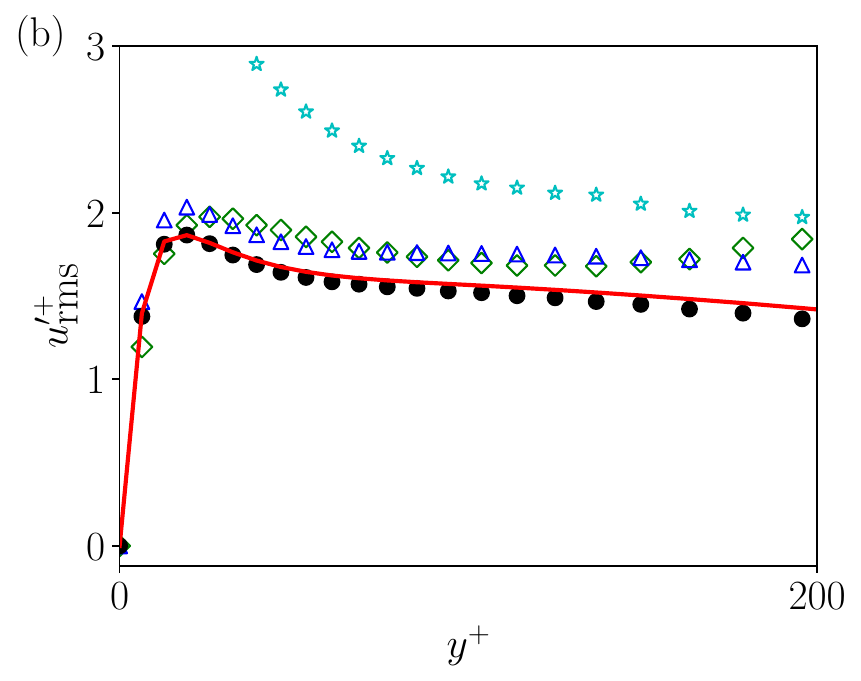}}
    \\
    \subfloat{
    \includegraphics[width=6.4cm,height = 4.8cm]{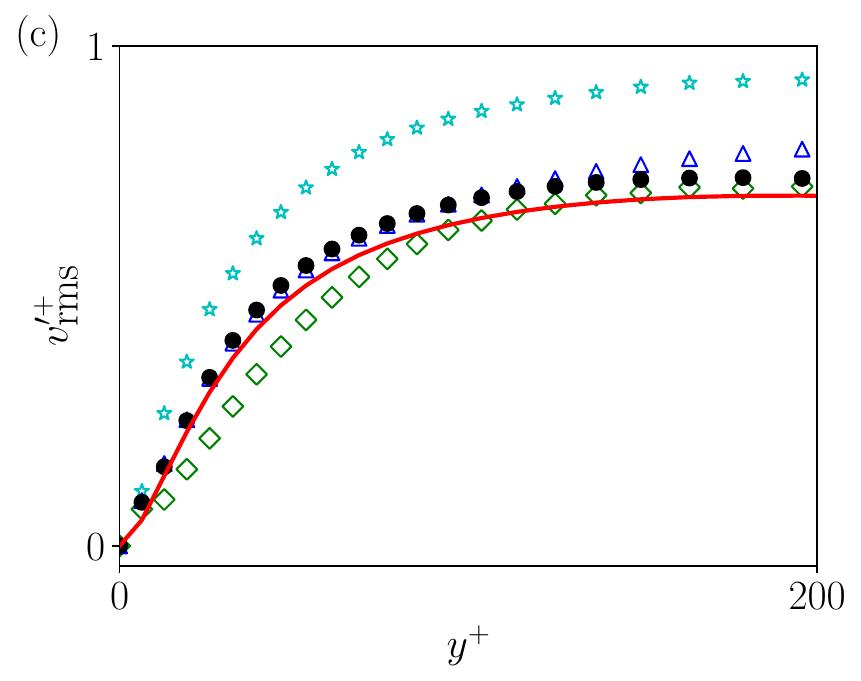}}
    \subfloat{
    \includegraphics[width=6.4cm,height = 4.8cm]{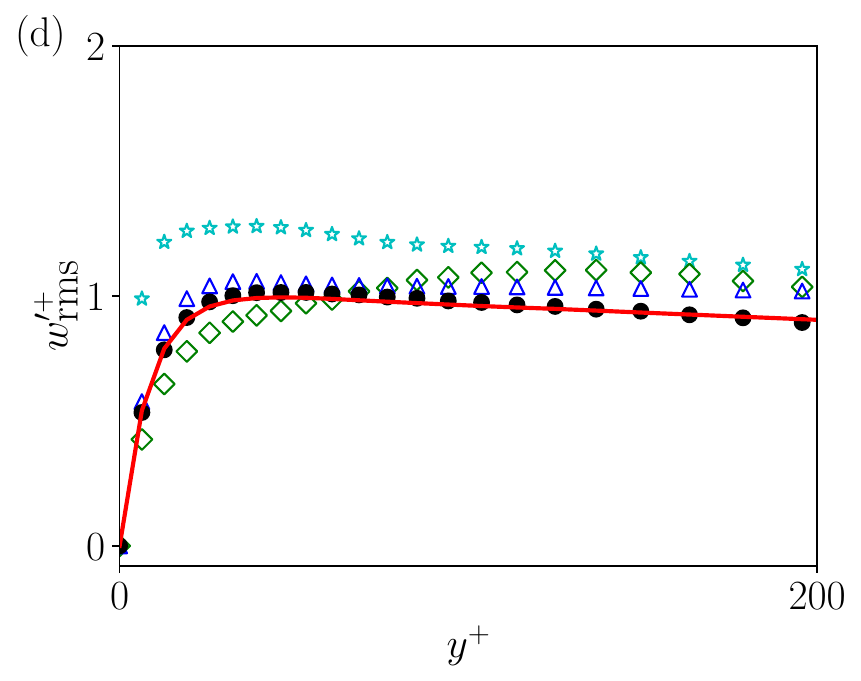}}
    \caption{\label{fig:AppC590vel} The mean streamwise velocity and rms fluctuating velocities at $Re_{\tau}\approx590$: (a) mean  streamwise velocity; (b) rms fluctuation of streamwise velocity; (c) rms fluctuation of transverse  velocity; (d) rms fluctuation of spanwise velocity.}
\end{figure*}

\begin{figure*}[!h]
    \centering
    \subfloat{
    \includegraphics[width=6.4cm,height = 4.8cm]{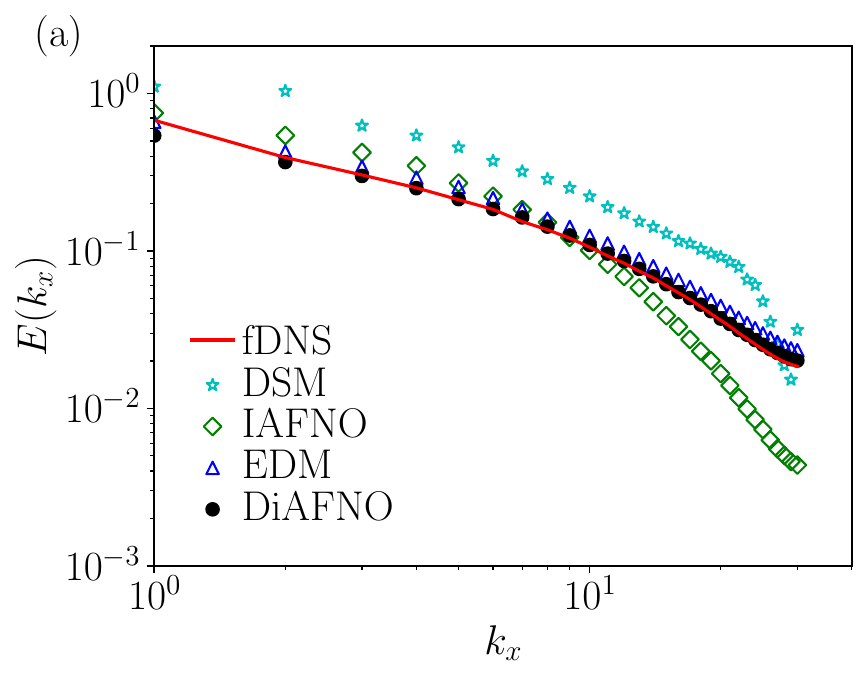}}
    \subfloat{
    \includegraphics[width=6.4cm,height = 4.8cm]{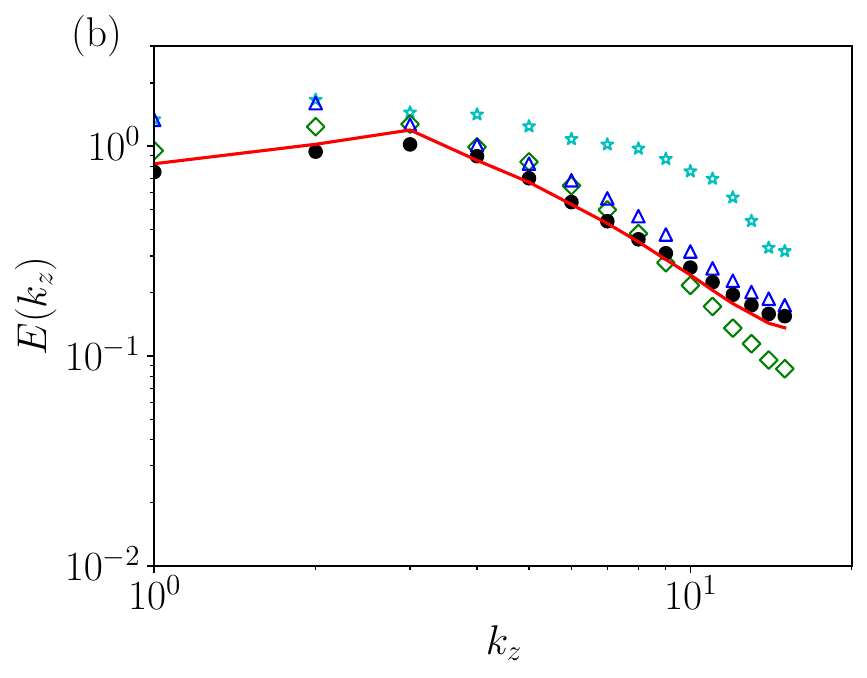}}
    \caption{\label{fig:AppC590spec} Energy spectrum at $Re_{\tau}\approx590$: (a) streamwise spectrum; (b) spanwise spectrum.}
\end{figure*}
\begin{multicols}{2}

As shown in Tab.~\ref{tab:AppClosses}, the IAFNO model’s prediction accuracy on the test set is clearly inferior to those of the two diffusion models EDM and DiAFNO. However, comparing performance based solely on error metrics is neither sufficient nor intuitive. Therefore, we conducted a series of tests on the CF590 dataset, which is considered the most challenging to predict.



As can be seen in Fig.~\ref{fig:AppC590vel} and Fig.~\ref{fig:AppC590spec}, the IAFNO model performs much worse than the two diffusion models on multiple metrics. However, it is important to note that DiAFNO took 90.41s to complete the inference process. It is 12 times longer than IAFNO (7.48s) under identical inference steps. Given that EDM took 212 seconds to complete the inference process, we can conclude that:

1. The framework of diffusion models can significantly improve the predictive accuracy of neural operator models. However, the improvements come at the cost of reduced efficiency.

2. As a relatively advanced neural operator model, IAFNO can improve the predictive accuracy and inference efficiency of diffusion models.

\subsection{Statistical convergence analysis}
\label{app32}
\setcounter{equation}{0}
\renewcommand{\theequation}{C.\arabic{equation}}

\textcolor{red}{We performed a statistical convergence analysis on the energy spectrum results for the CF590 case, providing the mean values and upper and lower bounds (maximum and minimum values) for 400 generated samples and their corresponding true values.}

\begin{figure*}[htb]
    \centering
    \subfloat{
    \includegraphics[width=6.4cm,height = 4.8cm]{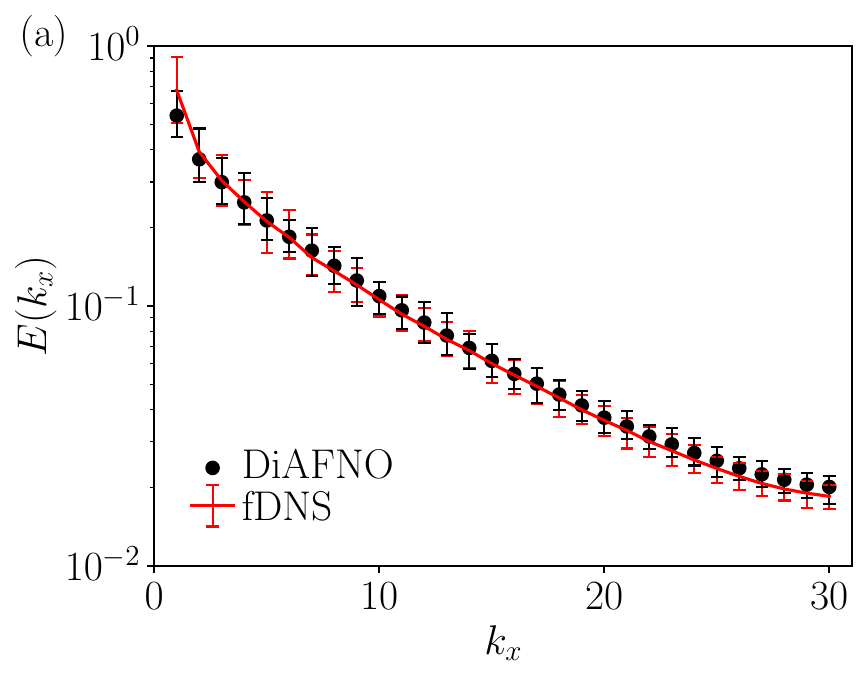}}
    \subfloat{
    \includegraphics[width=6.4cm,height = 4.8cm]{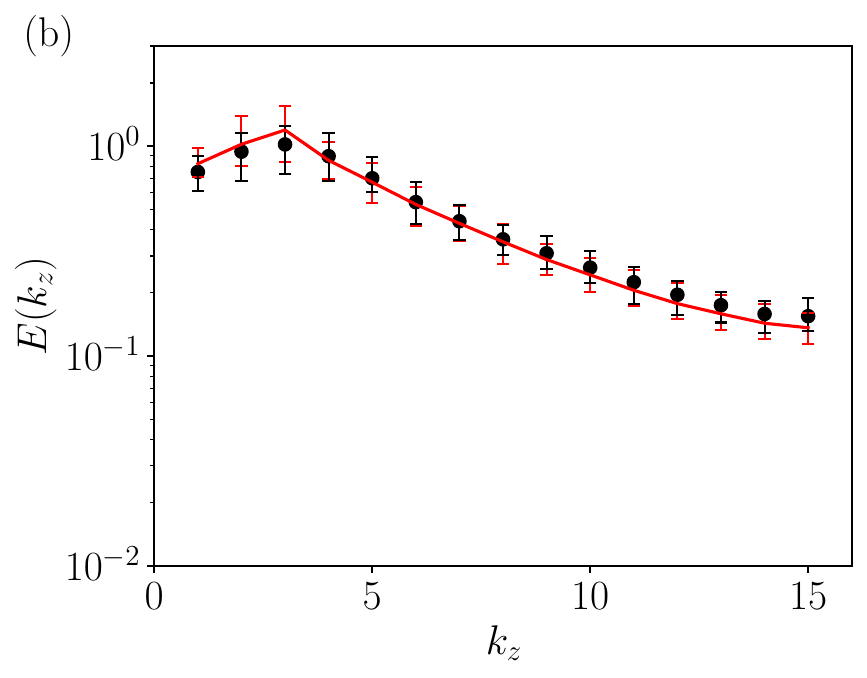}}
    \caption{\label{fig:AppC590specErr} Energy spectrum at $Re_{\tau}\approx590$ with upper and lower bounds of error: (a) streamwise spectrum; (b) spanwise spectrum.}
\end{figure*}

As shown in Fig.~\ref{fig:AppC590specErr}, the range of the true steady-state energy spectrum values converges relatively well while preserving some differences between samples. The range of our model’s predicted values is very close to that of the true values, indicating that our model exhibits statistical convergence. \textcolor{red}{Moreover, since the other physical quantities we reported are all derived from post-processing of the predicted velocity field, we performed the same analysis directly on the predicted velocity field.}

\begin{figure*}[htb]
    \centering
    \subfloat{\hspace{-3mm}
    \includegraphics[width=6.4cm,height = 4.8cm]{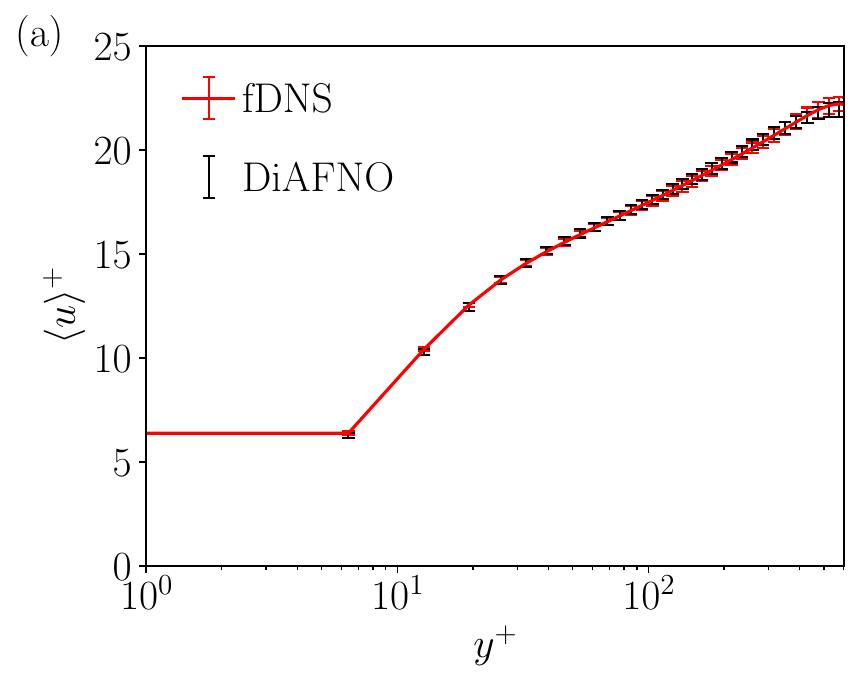}}
    \subfloat{
    \includegraphics[width=6.4cm,height = 4.8cm]{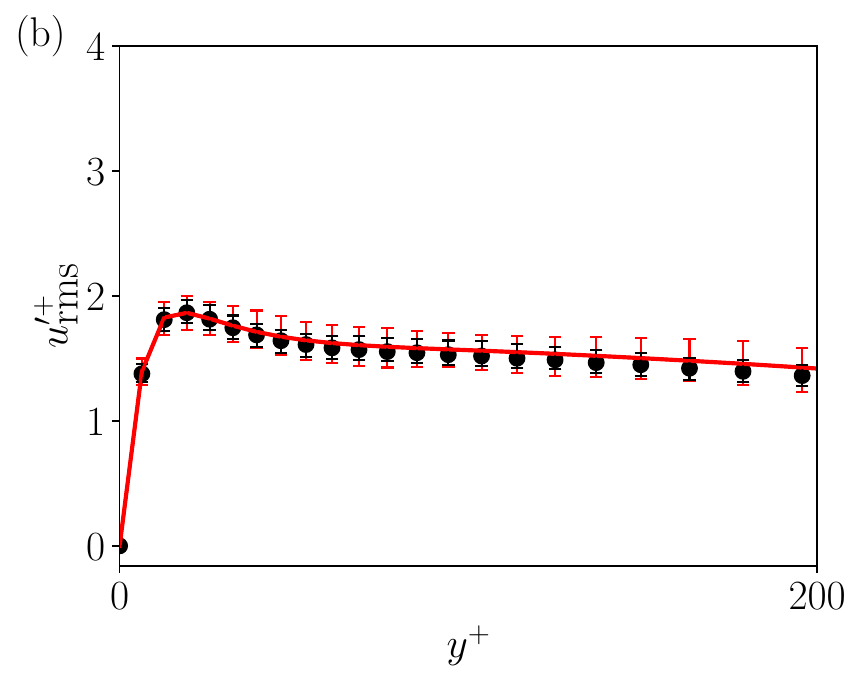}}
    \\
    \subfloat{
    \includegraphics[width=6.4cm,height = 4.8cm]{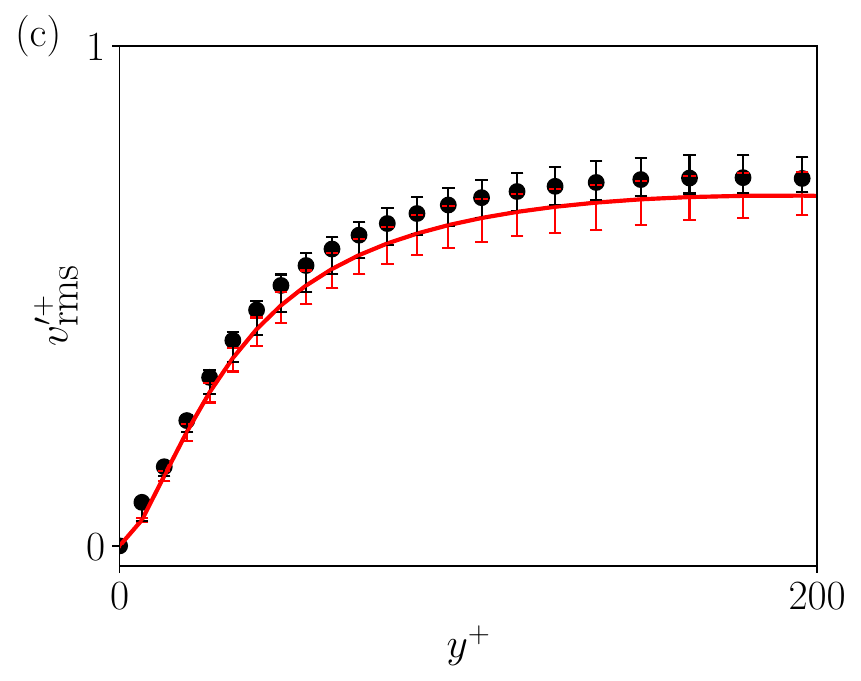}}
    \subfloat{
    \includegraphics[width=6.4cm,height = 4.8cm]{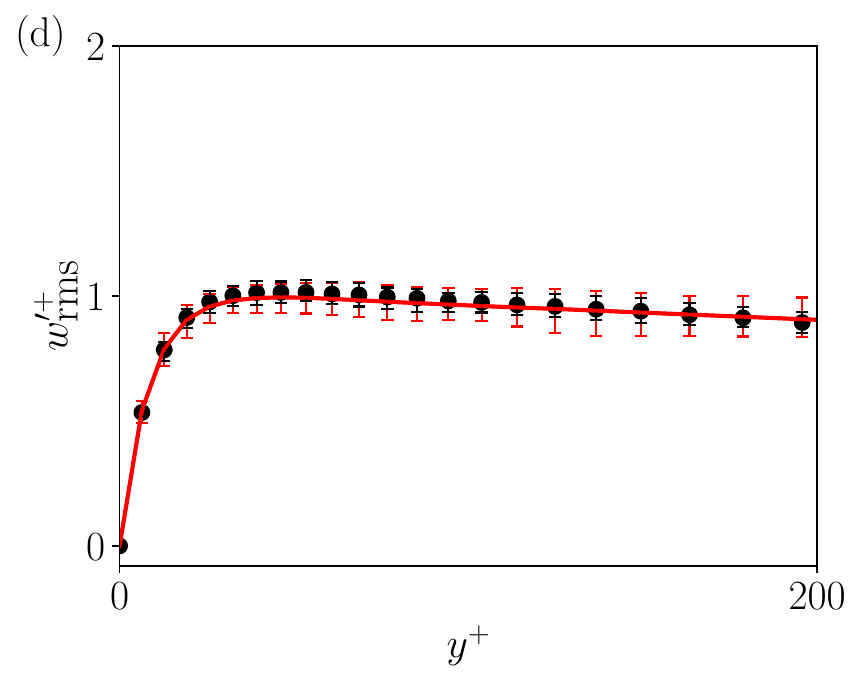}}
    \caption{\label{figvel_err} The mean streamwise velocity and rms fluctuating velocities at $Re_{\tau}\approx590$: (a) mean  streamwise velocity; (b) rms fluctuation of streamwise velocity; (c) rms fluctuation of transverse  velocity; (d) rms fluctuation of spanwise velocity.}
\end{figure*}

\textcolor{red}{It is shown in Fig.~\ref{figvel_err} that the uncertainty of the average field is very small, while the fluctuating field exhibits notable variability and uncertainty. Furthermore, we analyzed the temporal correlations among our 400 samples for the velocity fluctuation. We used the Pearson correlation coefficient for the analysis, namely:}

\begin{equation}
\textcolor{red}{r(t)=\frac{\sum_i[u_i(t)-\bar{u}(t)][u_i(1)-\bar{u}(1)]}{\sqrt{\sum_i[u_i(t)-\bar{u}(t)]^2\sum_i[u_i(1)-\bar{u}(1)]^2}}},
\end{equation} \textcolor{red}{where $u_i(t)=\tilde{U_i}(t)$ represents the fluctuating velocity component $i$ at sample $t$, $\bar{u}(t)$ represents the average fluctuating velocity at sample $t$ and $-1\leq r\leq1$. When $r$ approaches 1, it indicates a strong correlation; when $r$ approaches 0, it indicates weak correlation. We select the first sample as the base field and calculate $r$ for every sample relative to the first sample.}

\begin{figure*}[htb]
    \centering
    \includegraphics[width=10cm,height = 7.5cm]{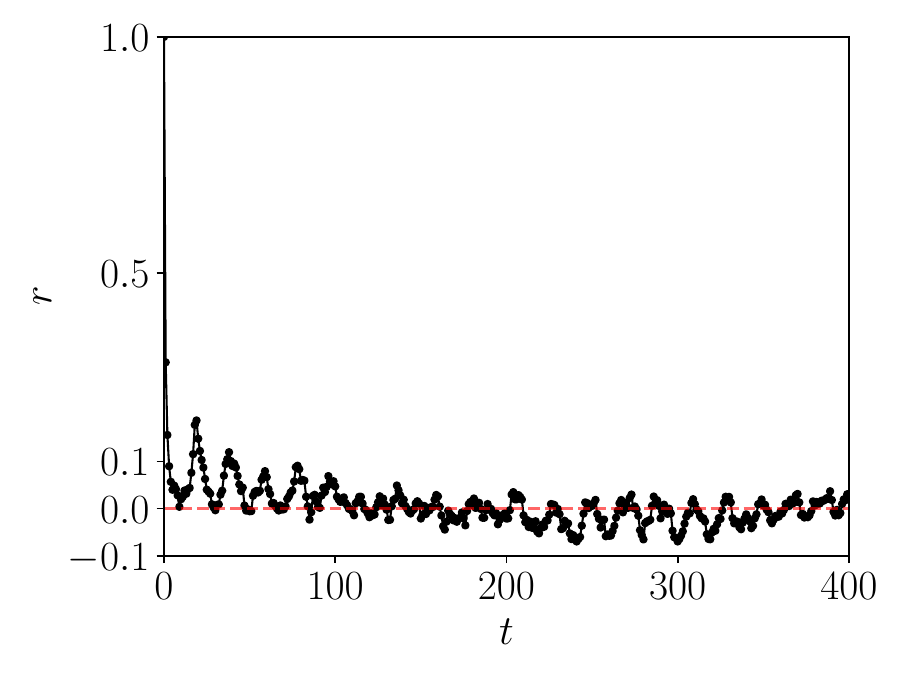}
    \caption{\label{figpearson} Pearson correlation coefficient $r$ with respect to sample time $t$.}
\end{figure*}

\textcolor{red}{As can be seen in Fig.~\ref{figpearson}, the sample correlation for the fluctuating field decays quickly to zero for time lags greater than a few steps.}

\textcolor{red}{Since most machine learning models are trained and validated using a fixed manual random seed, we followed the same approach (our manual random seed used in DiAFNO is 123). However, we supplemented another five sets of different random seeds and predicted the flow field for 400 time steps using the same model parameters. We calculated the point-to-point errors between the predictions at the 400th time step and the corresponding fDNS fields to assess our model's sensitivity to different random seeds. Based on the results shown in Tab.~\ref{tab:randomseed}, all errors are close to 0.14 and the maximum difference compared to the baseline is less than 3\%. Therefore, we can conclude that our model is not sensitive to random seeds.}

\end{multicols}

\begin{table}[htb]
\centering
\caption{\label{tab:randomseed}Point-to-point errors of different random seeds.}
\begin{tabular}{ccccccc}
\hline\hline
\mbox{Random seed:}&\mbox{123 (baseline)}&\mbox{1}&\mbox{2}&\mbox{3}&\mbox{4}&\mbox{5}\\
\hline
\mbox{Point-to-point error:}&\mbox{0.1400~}&\mbox{0.1399~}&\mbox{0.1394~}&\mbox{0.1403~}&\mbox{0.1439~}&\mbox{0.1374}\\
\hline\hline
\end{tabular}
\end{table}

\begin{multicols}{2}

\subsection{Analysis of long-term stability of point-wise error}
\label{app33}

\begin{figure*}[htb]
    \centering
    \includegraphics[width=10cm,height = 7.5cm]{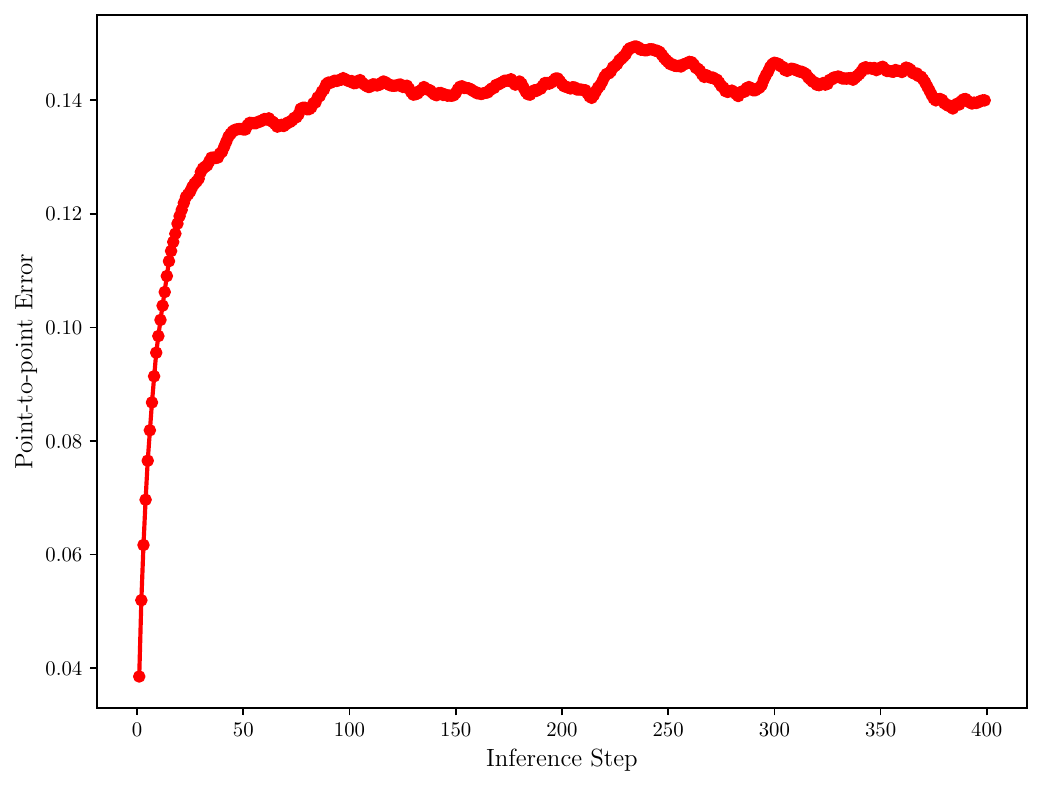}
    \caption{\label{fig:AppCErr} Point-to-point errors per step.}
\end{figure*}

We plotted the point-to-point error between the predicted and actual values for each step on the $Re_{\tau}\approx590$ dataset. As shown in Fig.~\ref{fig:AppCErr}, the error accumulates rapidly in the early stages and eventually converges to around 0.14. This indicates that the distribution learned by the model is an approximation of the true distribution, but the convergence of the error suggests that this approximate distribution is numerically stable. However, point-wise error is not a suitable metric for turbulence simulations. Even for two direct numerical simulation cases starting from identical initial conditions, the relative error between them will grow dramatically over time due to numerical round-off errors and the inherent chaotic nature of turbulence.

Additionally, the results in Fig.~\ref{fig:AppCErr} also indicate that our model actually learned a distribution that approximates the true distribution. The statistical results presented in the paper demonstrate that the approximate distribution accurately predicts key statistical features of the true distribution, thereby validating the effectiveness of our model. However, it must be acknowledged that our model does not employ any specific strategies to constrain or optimize this distribution shift, which provides inspiration for developing more advanced models in the future.

\end{multicols}
\begin{multicols}{2}

\subsection{\textcolor{red}{Quantitative testings of hyperparameters}}
\label{app34}

\textcolor{red}{We fix all other hyperparameters and vary only one at a time to independently quantify the impact of each hyperparameter. To maintain the conciseness and preciseness of this study, we report the testing loss (as shown in Eq.~\ref{eq:lossfunction}) when the model is fully converged and the corresponding training time cost per epoch. The baseline hyperparameter settings are shown in Tab.~\ref{tab:configs} with the retained Fourier modes = 100\% and the soft-thresholding parameter $\lambda=0.01$. The results presented in Tab.~\ref{tab:hyperparams} confirm the validity of our previous parameter settings.}

\end{multicols}
\begin{table}[h]
\centering
\caption{\label{tab:hyperparams}Testing loss and time cost per epoch of different hyperparameter settings.}
\begin{tabular}{ccc}
\hline\hline
\mbox{}&\mbox{~~Testing loss~~}&\mbox{~~Time cost(s)~~}\\
\hline
\mbox{~~\textbf{Baseline}}&\mbox{\textbf{0.0485}}&\mbox{\textbf{844.5}}\\
\hline
\mbox{~~Implicit layers = 2}&\mbox{0.0493}&\mbox{439.1}\\
\mbox{~~Implicit layers = 8}&\mbox{0.0483}&\mbox{1651}\\
\hline
\mbox{~~Explicit layers = 1}&\mbox{0.0508}&\mbox{439.0}\\
\mbox{~~Explicit layers = 4}&\mbox{0.0471}&\mbox{1785}\\
\hline
\mbox{~~Patch size = (4,2,4)}&\mbox{0.0580}&\mbox{403.8}\\
\mbox{~~Patch size = (1,2,1)}&\mbox{0.0454}&\mbox{3177}\\
\hline
\mbox{~~Number of blocks = 2}&\mbox{0.0492}&\mbox{870.4}\\
\mbox{~~Number of blocks = 4}&\mbox{0.0503}&\mbox{852.1}\\
\hline
\mbox{~~Retained Fourier modes = 75\%}&\mbox{0.0513}&\mbox{736.3}\\
\hline
\mbox{~~Soft-thresholding $\lambda$ = 0.02}&\mbox{0.0488}&\mbox{946.2}\\
\hline
\mbox{~~Sampling steps = 16}&\mbox{0.0533}&\mbox{495.1}\\
\hline\hline
\end{tabular}
\end{table}
\begin{multicols}{2}

\Acknowledgements{\textbf{Conflict of interest} \hspace{1em} The authors declare that they have no known competing financial interests or personal relationships that could have appeared to influence the work reported in this paper.\\ \\
\textbf{Author contributions} \hspace{1em} 
\textbf{Yuchi Jiang}: Conceptualization, Methodology, Investigation, Coding, Validation, Writing - draft preparation, Writing - reviewing and editing. \textbf{Yunpeng Wang}: Conceptualization, Investigation, Writing - reviewing and editing. \textbf{Huiyu Yang}: Conceptualization, Investigation, Writing - reviewing and editing. \textbf{Jianchun Wang}: Conceptualization, Methodology, Investigation, Supervision, Writing - reviewing and editing, Project administration, Funding acquisition.\\ \\
\textbf{Acknowlegdements} \hspace{1em} This work was supported by the National Natural Science Foundation of China (NSFC) (Grant Nos. 12588301, and 12302283); the NSFC Excellence Research Group Program for ‘Multiscale Problems in Nonlinear Mechanics’ (No. 12588201), by the Shenzhen Science and Technology Program (Grant Nos. SYSPG20241211173725008, and KQTD20180411143441009), and by Department of Science and Technology of Guangdong Province (Grant Nos. 2019B21203001, 2020B1212030001, and 2023B1212060001). Additional support was provided by the Innovation Capability Support Program of Shaanxi (Program No. 2023-CXTD-30) and the Center for Computational Science and Engineering of Southern University of Science and Technology.}


\bibliographystyle{spmpsci} 

\end{multicols}


\end{document}